\documentclass[12pt]{iopart}

\usepackage{graphicx}
\usepackage{amssymb}
\usepackage{dcolumn}
\usepackage{bm}

\begin{document}

\title{Orbital ordering phenomena in $d$- and $f$-electron systems}

\author{Takashi Hotta}

\address{Advanced Science Research Center,
Japan Atomic Energy Agency \\
Tokai, Ibaraki 319-1195, Japan}
\ead{hotta.takashi@jaea.go.jp}

%
%
\begin{abstract}
In recent decades, novel magnetism of $d$- and $f$-electron compounds
has been discussed very intensively both in experimental and theoretical
research fields of condensed matter physics.
It has been recognized that those material groups are in the same category
of strongly correlated electron systems, while the low-energy physics of
$d$- and $f$-electron compounds has been separately investigated rather
in different manners.
One of common features of both $d$- and $f$-electron systems is certainly
the existence of active orbital degree of freedom,
but in $f$-electron materials, due to the strong spin-orbit interaction
in rare-earth and actinide ions, the physics seems to be quite different
from that of $d$-electron systems.
In general, when the number of internal degrees of freedom and relevant
interactions is increased, it is possible to obtain rich phase diagram
including large varieties of magnetic phases by using several kinds of
theoretical techniques.
However, we should not be simply satisfied with the reproduction of
rich phase diagram.
It is believed that more essential point is to seek for a simple principle
penetrating complicated phenomena in common with $d$- and $f$-electron
materials, which opens the door to a new stage in orbital physics.
In this sense, it is considered to be an important task of
this article to explain common features of magnetism
in $d$- and $f$-electron systems from a microscopic viewpoint,
using a key concept of orbital ordering, in addition to the review
of the complex phase diagram of each material group.
As a typical $d$-electron complex material exhibiting orbital order,
first we focus on perovskite manganites,
in which remarkable colossal magneto-resistance effect has been
intensively studied.
The manganites provide us a good stage to understand
that a simple mechanism works for the formation of
complex spin, charge, and orbital ordering.
We also explain intriguing striped charge ordering
on the orbital-ordered background in nickelates
and the effect of orbital ordering to resolve spin frustration
in geometrically frustrated $e_{\rm g}$ electron systems.
Note that orbital ordering phenomena are also found
in $t_{\rm 2g}$ electron systems.
Here we review recent advances in the understanding of
orbital ordering phenomenon in Ca$_2$RuO$_4$.
Next we discuss another spin-charge-orbital complex system
such as $f$-electron compound.
After the detailed explanation of the construction of microscopic models
on the basis of a $j$-$j$ coupling scheme,
we introduce a $d$-electron-like scenario to understand
novel magnetism in some actinide compounds with the HoCoGa$_5$-type
tetragonal crystal structure.
Finally, we show that complicated multipole order can be
understood from the spin-orbital model on the basis of the
$j$-$j$ coupling scheme.
As a typical material with multipole order,
we pick up NpO$_2$ which has been believed
to exhibit peculiar octupole order.
Throughout this review, it is emphasized that
the same orbital physics works both
in $d$- and $f$-electron complex materials
in spite of the difference between $d$ and $f$ orbitals.
\end{abstract}

\maketitle
\tableofcontents
\pagestyle{headings}
\markboth{Orbital ordering phenomena in $d$- and $f$-electron systems}{}

%
%
\section{Introduction}

It has been widely recognized that orbital degree of freedom plays
a key role in understanding of novel magnetism observed in
transition metal oxides
\cite{Imada1998,Tokura2000a,Dagotto2001,Hotta2004a}.
A typical material of such spin-charge-orbital complex is the
manganese oxide, exhibiting remarkable colossal magneto-resistance
(CMR) phenomena \cite{Tokura2000b}.
In the recent decade, the study of manganites has been one of
the most important areas of research in condensed matter physics.
In one word, the CMR effect is considered to occur
when the manganite ground-state changes from insulating to
ferromagnetic (FM) metallic, after a small magnetic field is applied.
Based on the concept of two-phase competition,
the CMR behavior has been successfully qualitatively reproduced
in computational simulations, for instance,
employing resistor-network models \cite{Dagotto2002}.
In the two phases, the appearance of the FM metallic phase in manganites
has been usually rationalized by the so-called double-exchange (DE)
mechanism \cite{Zener1951},
based on a strong Hund's rule coupling between
mobile $e_{\rm g}$ electrons and localized $t_{\rm 2g}$ spins.
On the other hand, the insulating phase in manganites is basically
understood by the coupling between degenerate $e_{\rm g}$ electrons
and Jahn-Teller (JT) distortions of the MnO$_6$ octahedra
\cite{Dagotto2001,Hotta2004a},
leading to the various types of charge and/or orbital orders observed
experimentally.

The rich phase diagram of manganites has been revealed due to competition
and interplay among spin, charge, and orbital degrees of freedom,
but a recent trend is to unveil further new phases
both from experimental and theoretical investigations.
A typical example can be found in the undoped perovskite manganite,
RMnO$_3$ with rare earth ion R, which is the mother compound of
CMR manganites.
For R=La, it has been understood clearly that
the A-type antiferromagnetic (AF) phase appears
\cite{Wollan1955,Matsumoto1970}
with the C-type ordering of $(3x^2$$-$$r^2)$- and $(3y^2$$-$$r^2)$-orbitals
\cite{Murakami1998a}.
Here ``A-type'' denotes a layered antiferro structure with
ferro-order in the $ab$ plane and antiferro-order along the $c$ axis,
while ``C-type'' indicates a chain-type antiferro structure with
antiferro-order in the $ab$ plane and ferro-order along the $c$ axis
\cite{Wollan1955}.
See Fig.~\ref{fig7}(a) for each structure.
Theoretically, the A-type ordering has been explained by
using several kinds of techniques
\cite{Mizokawa1995,Mizokawa1996,Solovyev1996a,Solovyev1996b,Koshibae1997,
Shiina1997,Ishihara1997,Sawada1997,Feinberg1998,Maezono1998,Feiner1999,
Horsch1999,Brink1999,Benedetti1999,Hotta1999,Allen1999,Capone2000}.
By substituting La by alkaline earth ions such as Sr and Ca,
holes are effectively doped into $e_{\rm g}$-electron band and
due to the DE mechanism, the FM metallic phase appears
with its concomitant CMR effect.
Most of the discussion in manganites has centered on the many phases
induced by doping with holes the A-type AF state,
at different values of their bandwidths.
In this framework, it is implicitly assumed that the undoped material
is always in the A-type AF state.

However, recently, a new AF phase has been reported as the ground state
in the undoped limit for R=Ho \cite{Munoz2001,Kimura2003}.
This phase is called the ``E-type'' spin structure
following the standard notation in this context \cite{Wollan1955}.
See Fig.~\ref{fig8}(b) for the E-type spin structure.
It is surprising that a new phase can be still found even in the
undoped material, previously considered to be well understood.
In addition, the nature of the states obtained by lightly doping
this E-phase is totally unknown, and new phenomena may be unveiled
experimentally in the near future.
This is believed to open an exciting new branch of investigations
in manganites \cite{Hotta2003a,Hotta2003b},
since novel phases appear to be hidden
in the vast parameter space of these compounds.
A clear example has been recently provided by the prediction of
a FM charge-ordered (CO) phase at x=1/2 \cite{Yunoki2000,Hotta2001},
which may have been found experimentally \cite{Loudon2002,Mathur2003}.
These facts indicate the importance of both experimental and
theoretical efforts to unveil new phases in manganites,
in addition to the explanation of the complex phases already observed.
Such efforts have also been made to find new phases in
other transition metal oxides, for instance,
ruthenates and nickelates, as we will see later in this article.
Concerning RMnO$_3$ with hexagonal structure, quite recently,
``multiferroics'' has been another keyword to understand
exotic magnetic phenomena emerging from
the multi-phase competition \cite{multiferroics}.
We believe that it is useful to review the nature of
spin, charge, and orbital ordered phases of manganites and
other transition metal oxides from a unified viewpoint,
even though it is true that more work remains to be done to
fully understand transition metal oxides, in particular,
unusual magneto-transport properties of manganese oxides
and appearance of unconventional superconductivity.

A trend to seek for new magnetic as well as superconducting phases
has been also found in the $f$-electron system, which is another type
of spin-charge-orbital complex \cite{f-review1,f-review2}.
Among so many kinds of $f$-electron materials, in recent years,
$f$-electron compounds with HoCoGa$_5$-type tetragonal
crystal structure, frequently referred to as ``115'',
have been intensively investigated both in experimental and theoretical
research fields of condensed matter physics.
Such vigorous activities are certainly motivated by ``high''
temperature superconductivity observed in some 115 compounds.
First, unconventional superconductivity has been found in
Ce-based 115 compounds, CeTIn$_5$ (T=Rh, Ir, and Co).
A surprising point is that CeCoIn$_5$ exhibits the superconducting
transition temperature $T_{\rm c}$=2.3K \cite{Ce115-1},
which was the highest among yet observed for heavy fermion materials
at ambient pressure when it was discovered.
On the other hand, CeIrIn$_5$ shows $T_{\rm c}$=0.4K \cite{Ce115-2}
which is much less than that of CeCoIn$_5$.
Note that CeRhIn$_5$ is antiferromagnet with a N\'eel temperature
$T_{\rm N}$=3.8K at ambient pressure, while under high pressure,
it becomes superconducting with $T_{\rm c}$=2.1K \cite{Ce115-3}.

After the discovery of superconductivity in Ce-115,
the rapid expansion of the research frontier to transuranium
systems has been accelerated by the discovery of superconductivity
of Pu-based 115 compounds, PuTGa$_5$ (T=Co and Rh).
It has been reported that $T_{\rm c}$ of PuCoGa$_5$ is 18.5K
\cite{Sarrao,SarraoASR},
which is amazingly high value even compared with
other well-known intermetallic compounds.
The coefficient of electronic specific heat $\gamma$ is estimated
as $\gamma$=77mJ/mol$\cdot$K$^2$, moderately enhanced relative to
that for normal metals, suggesting that PuCoGa$_5$ should be
heavy-fermion superconductor.
In PuRhGa$_5$, superconductivity has been also found \cite{Wastin}.
Although the value of $T_{\rm c}$=8.7K is lower than that of
PuCoGa$_5$, it is still high enough compared with
other heavy-fermion superconductors.
Quite recently, high quality single crystal PuRhGa$_5$ has been
synthesized \cite{Haga} and the Ga-NQR measurement has revealed
that $d$-wave superconductivity is realized in PuRhGa$_5$ \cite{Sakai}.
The Ga-NMR measurement of PuCoGa$_5$ is consistent with this conclusion
\cite{Curro}.
PuIrGa$_5$ has been also synthesized, but it is considered
to be paramagnetic at ambient pressure.
At least up to now, there is no indication of superconductivity
even under the pressure of 9.5GPa down to 1.4K \cite{Griveau}.

Besides such high temperature superconductivity of Ce-115 and Pu-115
compounds, interesting magnetic properties have been reported for
UTGa$_5$, where T is a transition metal ion.
For several transition metal ions T, UTGa$_5$ are AF metals
or Pauli paramagnets \cite{U115-1a,U115-1b,U115-1c,U115-1d,
U115-2,U115-3,U115-4,U115-5,U115-6,U115-7,U115-8,U115-9}.
Among them, neutron scattering experiments have revealed that
UNiGa$_5$ exhibits the G-type AF phase, while UPdGa$_5$ and UPtGa$_5$
have the A-type AF state \cite{U115-5,U115-9}.
See Fig.~\ref{fig6-1} for the magnetic structure.
Note that G-type indicates a three-dimensional N\'eel state \cite{Wollan1955}.
On the other hand, for UTGa$_5$ with T=Co, Rh, Ir, Fe, Ru, and Os,
magnetic susceptibility is almost independent of temperature,
since these are Pauli paramagnets.
It is quite interesting that the magnetic structure is different
for U-115 compounds which differ only by the substitution
of transition metal ions.

Recently, Np-115 compounds NpTGa$_5$ (T=Fe, Co, Rh, and Ni) have been
also synthesized and several kinds of physical quantities have been
successfully measured \cite{Colineau,Aoki-Ni,Aoki-Co1,Aoki-Co2,
Aoki-Rh,Honda,Yamamoto,Homma,Metoki,Jonen}.
In particular, the de Haas-van Alphen (dHvA) effect has been observed
in NpNiGa$_5$ \cite{Aoki-Ni}, which is the first observation of
dHvA signal in transuranium compounds.
Quite recently, dHvA measurement has been also successfully performed
in plutonium compound PuIn$_3$ over beyond several kinds
of difficulties \cite{Haga2}.
For NpCoGa$_5$ and NpRhGa$_5$, the dHvA oscillations have been also
detected and plural number of cylindrical Fermi surfaces are
found \cite{Aoki-Co2,Aoki-Rh}.
For NpFeGa$_5$, the magnetic moment at Fe site has been suggested
in neutron scattering experiments \cite{Honda} and it has been
also detected by $^{57}$Fe M\"ossbauer spectroscopy \cite{Homma}.
The magnetic structure of Np-115 compounds also depends sensitively
on transition metal ion \cite{Honda,Metoki,Jonen}:
G-AF for NpNiGa$_5$, A-AF for NpCoGa$_5$ and NpRhGa$_5$,
and C-AF for NpFeGa$_5$.
See Fig.~\ref{fig6-1} for the magnetic structure.
Note also that in the neutron scattering experiment for NpNiGa$_5$,
the signal suggesting the G-AF grows,
after the FM transition occurs \cite{Honda}.
This G-AF structure is due to canted magnetic moments of Np ions.
It is characteristic of U-115 and Np-115 compounds that
the magnetic properties are sensitive to the choice of
transition metal ions.

The appearance of several kinds of AF states in U-115 and Np-115
compounds reminds us of the magnetic phase diagram of CMR manganites.
Thus, we envisage a scenario to understand the complex magnetic
structure of actinide compounds based on an orbital degenerate model
similar to that of manganites.
However, one must pay due attention to the meanings of
``spin'' and ``orbital'' in $f$-electron systems.
Since they are tightly coupled with each other through
a strong spin-orbit interaction, distinguishing them is not
straightforward in comparison with $d$-electron systems.
This point can create serious problems when we attempt to
understand microscopic aspects of magnetism and
superconductivity in $f$-electron compounds.
Thus, it is necessary to carefully define the terms ``orbital''
and ``spin'' for $f$ electrons in a microscopic discussion of
magnetism and superconductivity in $f$-electron compounds.

From a conceptual viewpoint,
in general, $f$ electrons are more localized in comparison with
$d$ electrons, but when we turn our attention from $4f$ to $5f$,
electronic properties are changed gradually
from localized to itinerant nature, leading to rich phenomena
which have been recently investigated intensively.
In this sense, it has been also highly requested to push forward
the microscopic research on $f$ electron systems.
However, in sharp contrast to $d$ electron systems,
the existence of strong spin-orbit coupling has been
a problem to develop a theoretical study in the same
level as $d$-electron research.

In order to overcome such problems, it has been recently proposed
to employ a $j$-$j$ coupling scheme to discuss microscopic aspects
of magnetism and superconductivity of $f$-electron systems
\cite{Hotta2003,Hotta2003c},
on the basis of the relativistic band-structure
calculation results \cite{Maehira2002,Maehira2003a,Maehira2003b,
Maehira2003c,Maehira2003d,Maehira2003e,Maehira2004,Maehira2005}.
There are a couple of advantages of the $j$-$j$ coupling scheme.
First, it is quite convenient for the inclusion of many-body effects
using standard quantum-field theoretical techniques,
since individual $f$-electron states are clearly defined.
In contrast, in the $LS$ coupling scheme we cannot use such standard
techniques, since Wick's theorem does not hold.
Second we can, in principle, include the effects of valence fluctuations.
In some uranium compounds, the valence of the uranium ion is neither
definitely U$^{3+}$ nor U$^{4+}$, indicating that the $f$-electron
number takes a value between 2 and 3.
In the $j$-$j$ coupling scheme this is simply regarded
as the average number of $f$ electron per uranium ion.

As we will discuss later in detail,
with the use of the $j$-$j$ coupling scheme,
it is possible to establish the microscopic Hamiltonian
for $f$-electron systems.
In particular, under crystalline electric field of cubic symmetry,
the $f$-electron model on the basis of the $j$-$j$ coupling scheme
can be reduced to the orbital degenerate model,
which is equivalent to the $e_{\rm g}$ electron Hubbard model
\cite{Hotta2003}.
The common microscopic model for $d$- and $f$-electron compounds
is a simple explanation for the appearance of magnetic structure
in common with manganites and actinide 115 compounds,
as actually analyzed by using numerical calculations
\cite{Hotta2004,Onishi2004}.
In order to understand unconventional superconductivity
of Ce-115 and Pu-115 materials,
the orbital degenerate model has been also analyzed with
the use of a fluctuation-exchange approximation to include
effectively spin and orbital fluctuations
\cite{Takimoto2002,Takimoto2003,Takimoto2004}.
A possible scenario for odd-parity triplet superconductivity
induced by the Hund's rule interaction has been discussed based on
the orbital degenerate model on the non-Bravais lattice
such as honeycomb lattice \cite{Hotta2004b,Hotta2004c}.
Novel magnetism and exotic superconductivity of filled skutterudite
materials \cite{Sato2003,Aoki2005}
have been discussed by using the microscopic model
on the basis of the $j$-$j$ coupling scheme
\cite{Hotta2005a,Hotta2005b,Hotta2005c,Hotta2005d}.
We would like to emphasize that the same orbital physics should work
between $d$- and $f$-electron systems,
after the application of the $j$-$j$ coupling scheme to
$f$-electron materials.
This is an important clue to establish the unified picture, which
penetrates complex phenomena both in $d$- and $f$-electron compounds.

Another advantage of the model on the basis of the $j$-$j$ coupling
is that it is possible to develop a microscopic theory
for multipole ordering of $f$-electron materials.
Recently, ordering of high-order multipole such as octupole has
been intensively discussed for Ce$_x$La$_{1-x}$B$_6$
\cite{Kuramoto,Kusunose,Sakakibara2002a,Sakakibara2002b,Sakakibara2004,
Kubo2003,Kubo2004,Kobayashi,Iwasa,Suzuki}
and for NpO$_2$
\cite{Santini2000,Santini2002,Paixao,caciuffo,Lovesey,Kiss,Tokunaga,
Sakai2,Kubo2005a}
to reconcile experimental observations which seem to contradict
one another at first glance.
Very recently, a possibility of octupole ordering has been also
proposed for SmRu$_4$P$_{12}$ \cite{Yoshizawa,Hachitani}.
Here we note that spin and orbital degrees of freedom correspond
to dipole and quadrupole moment, respectively.
The microscopic aspects of octupole ordering has not been discussed
satisfactorily in the context of electronic models.
Rather, phenomenological theories have been developed
under the assumption that octupole ordering occurs.
Note that direct detection of octupole ordering is very difficult,
since the octupole moment directly couples to neither a magnetic
field nor lattice distortions.
However, those phenomenological theories have successfully
explained several experimental facts consistently, e.g.,
induced quadrupole moments in octupole ordered states in
Ce$_x$La$_{1-x}$B$_6$ \cite{Kusunose,Kubo2003,Kubo2004}
and NpO$_2$ \cite{Paixao,Kiss}.
On the other hand, it has been highly required to proceed to
microscopic theory, in order to understand the origin of
multipole ordering in $f$-electron systems
over beyond the phenomenological level.

Concerning this issue, recently, it has been clearly shown
that the model based on the $j$-$j$ coupling scheme also
works for the explanation of octupole ordering
\cite{Kubo2005a,Kubo2005b,Kubo2005c,Kubo2005d,Kubo2005e}.
It is possible to obtain the multipole interaction from
the $f$-electron model, by applying the same procedure
to derive the orbital-dependent superexchange interaction
in $d$-electron systems.
Namely, we can provide the microscopic basis for multipole interaction.
It is also possible to show the stability of octupole ordered phase
depending on the lattice structure.
In this scenario, octupole is the combined degree of freedom of
spin and orbital.
In another word, octupole is considered to be characterized
by the anisotropic spin densities.
Thus, the difference in anisotropic up- and down-spin densities
naturally provide the magnetic moment of octupole.
It is one of progresses in orbital physics that higher-order
multipole can be also understood in the context of spin and
orbital ordering phenomena.

In this article, in Sec.~2, we review the construction
of the microscopic models for $d$-electron systems in detail.
Then, we arrive at three kinds of Hamiltonians,
depending on the active orbitals, which are
(i) $e_{\rm g}$-electron doubly degenerate model,
(ii) $t_{\rm 2g}$-electron triply degenerate model,
and (iii) $e_{\rm g}$-orbital double-exchange model coupled with
$t_{\rm 2g}$ localized spins.
In Sec.~3, in order to explain the orbital ordering in manganites,
we focus on the theoretical results on the model (iii).
In particular, the topological aspects of orbital ordering
will be discussed in detail.
Then, we explain characteristic features of orbital ordering
depending on hole doping x in subsections for
x=0.0, x=0.5, x$>$0.5, and x$<$0.5.
In Sec.~4, we discuss the results of models (i) and (ii),
by introducing nickelates and ruthenates as typical materials,
respectively.
In Sec.~5, we will move on to the model construction for
$f$-electron materials with the use of the $j$-$j$ coupling scheme.
We will explain the spirit of the $j$-$j$ coupling scheme
in detail, in comparison with the $LS$ coupling scheme.
Then, we establish the orbital degenerate model
for $f$-electron systems.
In Sec.~6, we will show theoretical results on the microscopic
model for $f$-electron systems.
We will explain possible orbital ordering of 115 compounds and
octupole ordering of NpO$_2$ from a microscopic viewpoint.
Finally, in Sec.~7, we will summarize this article.

%
%
\section{Model Hamiltonian for $d$-electron systems}

Before proceeding to the description of theoretical results
on orbital ordering phenomena in $d$-electron compounds,
first it is necessary to define the model Hamiltonian
for $d$-electron systems.
The model should be composed of three parts as
\begin{equation}
   H = H_{\rm kin} + H_{\rm loc} + H_{\rm inter-site},
\end{equation}
where $H_{\rm kin}$ expresses the kinetic term of $d$ electrons,
$H_{\rm loc}$ denotes the local term for $d$ electrons,
and $H_{\rm inter-site}$ indicates the inter-site interaction
among $d$ electrons, which is not fully included by
the combination of $H_{\rm kin}$ and $H_{\rm loc}$.
Among them, the local term $H_{\rm loc}$ includes
three important ingredients, written as
\begin{equation}
   H_{\rm loc} = H_{\rm CEF} + H_{\rm el-el} + H_{\rm el-ph},
\end{equation}
where $H_{\rm CEF}$ is the crystalline electric field (CEF) term,
$H_{\rm el-el}$ denotes the Coulomb interactions among $d$ electrons,
and $H_{\rm el-ph}$ indicates the coupling term between $d$ electrons
and lattice distortions.
The full Hamiltonian includes several competing tendencies and couplings,
but as shown below, the essential physics can be obtained using
relatively simple models, deduced from the complicated full Hamiltonian.

\subsection{Crystalline electric field effect}

In order to construct the model Hamiltonian for $d$-electron systems,
let us start our discussion at the level of the atomic problem,
in which just one electron occupies a certain orbital
in the $3d$ shell of a transition metal ion.
In the next subsection, we will include the effect of
Coulomb interactions among $d$ electrons.
For an isolated ion, a five-fold degeneracy exists for the
occupation of the $3d$ orbitals,
but this degeneracy is partially lifted by the CEF potential
from anions surrounding the transition metal ion.
Since it is the electrostatic potential field acting on one electron state
even in the complicated crystal structure,
the CEF potential should be, in any case,
given in the second-quantized form as
\begin{equation}
   H_{\rm CEF} = \sum_{{\bf i},\sigma,m,m'}
   A_{m,m'} d^{\dag}_{{\bf i}m \sigma} d_{{\bf i} m' \sigma},
\end{equation}
where $d_{{\bf i} m \sigma}$ is the annihilation operator for
a $d$-electron with spin $\sigma$ in the $m$-orbital at site ${\bf i}$,
$m$(=$-2,\cdots,2$) is the $z$-component of angular momentum $\ell$(=2),
and $A_{m,m'}$ is the coefficient of the CEF potential,
depending on the crystal structure and the angular momentum.

The explicit form of $A_{m,m'}$ has been analyzed
in detail by the ligand field theory.
Here we briefly explain the derivation of $A_{m,m'}$
by following the procedure of Hutchings \cite{Hutchings,Syokabo}.
The CEF potential is given by the sum of electro-static potential
from anions surrounding the transition metal ion, written by
\begin{equation}
   V_{\rm CEF} ({\bf r})=
   \sum_i \frac{Z e^2}{|{\bf R}_i-{\bf r}|},
\end{equation}
where $Z$ is the valence of anion,
$e$ is elementary electric charge,
${\bf R}_i$ denotes the position of $i$-th anion,
and ${\bf r}$ indicates the position of $3d$ electron
around the
nucleus of transition metal ion.
Then, the CEF coefficient is evaluated by
\begin{equation}
  A_{m,m'} = \int d{\bf r} \Psi^*_{n \ell m}({\bf r})
  V_{\rm CEF} ({\bf r}) \Psi_{n \ell m'}({\bf r}),
\end{equation}
where $\Psi_{n \ell m}({\bf r})$ is the wavefunction
of $d$ electron, expressed as
\begin{equation}
  \Psi_{n \ell m}({\bf r})=R_{n\ell}(r)Y_{\ell m}(\theta,\varphi),
\end{equation}
with ${\bf r}$=$(r,\theta,\varphi)$ in the polar coordinate.
Here $R_{n\ell}(r)$ is the radial wavefunction and
$Y_{\ell m}$ is the spherical harmonics,
which are obtained by solving the Schr\"odinger equation for
the hydrogen-like potential problem.
Note that $n$ denotes the principal quantum number.
In the actual situation, $|{\bf r}|$ is in the order of Bohr radius,
while $|{\bf R}_i|$ is related to the lattice constant
which is in the order of several angstroms.
Thus, it is convenient to expand $V_{\rm CEF}$ in terms of
$|{\bf r}|/|{\bf R}_i|$ as
\begin{equation}
  V_{\rm CEF} ({\bf r})=
  \sum_i \frac{Ze^2}{R_i} \sum_{k=0}^{\infty} \sum_{m=-k}^{k}
  \frac{4\pi}{2k+1} \Bigl( \frac{r}{R_i} \Bigr)^k
  Y_{km}(\theta,\varphi) Y^*_{km}(\theta_i,\varphi_i),
\end{equation}
where ${\bf R}_i$=$(R_i,\theta_i,\varphi_i)$ in the polar coordinate.

\begin{figure}[t]
\begin{center}
\includegraphics[width=0.4\textwidth]{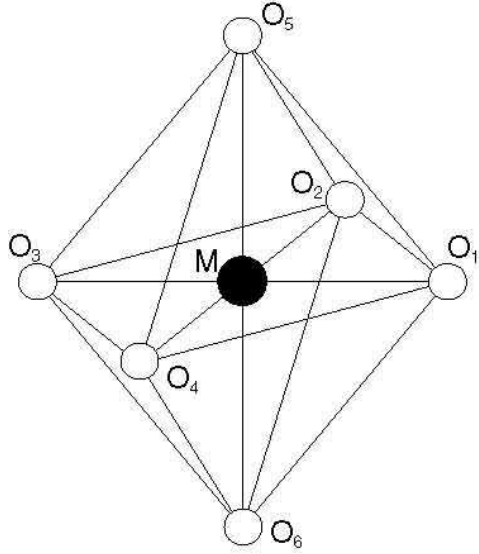}
\end{center}
\caption{%
MO$_6$ octahedron at site ${\bf i}$.
The position of the $i$-th oxygen ion labeled by O$_i$ is ${\bf R}_i$.
The size of the octahedron is specified by $a$ and $b$,
which are given by $a$=$|{\bf R}_1-{\bf R}_3|$=$|{\bf R}_2-{\bf R}_4|$
and  $b$=$|{\bf R}_5-{\bf R}_6|$, respectively.
See the maintext for the definitions of ${\bf R}_i$.
}
\label{fig1}
\end{figure}

For further calculations, it is necessary to set
the actual crystal structure.
As a typical example, in this review article, we pick up
the perovskite structure composed of MO$_6$ octahedron,
in which transition metal ion $M$ is surrounded by six oxygen ions,
as shown in Fig.~\ref{fig1}.
To make the situation general, we set the different values,
$a$ and $b$, for the size of octahedron in the $xy$ plane and
along the $z$ axis, respectively.
The positions of oxygen ions are, then, given by
${\bf R}_1$=$(a/2,0,0)$,
${\bf R}_2$=$(0,-a/2,0)$,
${\bf R}_3$=$(-a/2,0,0)$,
${\bf R}_4$=$(0,a/2,0)$,
${\bf R}_5$=$(0,0,b/2)$,
and
${\bf R}_6$=$(0,0,-b/2)$.
For the case of $a=b$, the cubic symmetry is maintained,
while for the case of $a \ne b$,
the system is in the tetragonal symmetry.
After some algebraic calculations using the explicit form of
the wavefunction for $d$ electrons,
a general form for $A_{m,m'}$ is given by
\begin{eqnarray}
  A_{m,m'} =
  &-& \frac{\langle r^2 \rangle}{a^2} \frac{2Ze^2}{a}
  \Bigl[ 1-\Bigl(\frac{a}{b}\Bigr)^3 \Bigr]
  c^{(2)}(2m,2m') \delta_{m,m'} \nonumber \\
  &+& \frac{\langle r^4 \rangle}{a^4} \frac{2Ze^2}{a}
  \Bigl[ \frac{3}{4}+\Bigl( \frac{a}{b} \Bigr)^5 \Bigr]
  c^{(4)}(2m,2m') \delta_{m,m'} \nonumber \\ 
  &+& \frac{\langle r^4 \rangle}{a^4} \frac{Ze^2}{a}
  \sqrt{\frac{35}{8}} c^{(4)}(2m,2m') \delta_{m-m',\pm 4},
\end{eqnarray}
where $\delta_{m,m'}$ is the Kronecker's delta,
$\langle r^k \rangle$ is given by
\begin{eqnarray}
  \langle r^k \rangle = \int_0^{\infty} r^k |R_{3d}(r)|^2 r^2 dr,
\end{eqnarray}
and $c^{(k)}(\ell m, \ell' m')$ is the so-called Gaunt coefficient
\cite{Gaunt,Racah1942},
defined by
\begin{equation}
  c^{(k)}(\ell m, \ell' m')
  \!=\! \sqrt{\frac{4\pi}{2k+1}}
  \! \int \! \sin \theta d\theta d\varphi Y^*_{\ell m}(\theta,\varphi)
  Y_{k m-m'}(\theta,\varphi)Y_{\ell' m'}(\theta,\varphi).
\end{equation}
The non-zero values for the Gaunt coefficients have been tabulated
in the standard textbook \cite{Slater1960}.
By consulting with the table for the Gaunt coefficient,
we explicitly obtain $A_{m,m'}$ as
\begin{eqnarray}
  \begin{array}{l}
    A_{2,2}=A_{-2,-2}=2A_{20}+A_{40}, \\
    A_{1,1}=A_{-1,-1}=-A_{20}-4A_{40}, \\
    A_{0,0}=-2A_{20}+6A_{40}, \\
    A_{2,-2}=A_{-2,2}= A_{44},
  \end{array}
\end{eqnarray}
where
\begin{eqnarray}
  &&A_{20}=\frac{2}{7}\frac{Ze^2}{a}\frac{\langle r^2 \rangle}{a^2}
  \Bigl[ 1-\Bigl( \frac{a}{b} \Bigr )^3 \Bigr], \\ \nonumber
  &&A_{40}=\frac{1}{6}\frac{Ze^2}{a}\frac{\langle r^4 \rangle}{a^4}
  \Bigl[ \frac{3}{7} + \frac{4}{7} \Bigl( \frac{a}{b} \Bigr )^5 \Bigr],\\
  &&A_{44}=\frac{5}{6}\frac{Ze^2}{a}\frac{\langle r^4 \rangle}{a^4}.\nonumber
\end{eqnarray}
Note that $A_{20}$=0 and $A_{44}$=$5A_{40}$
for the cubic case ($a$=$b$).

For instructive purpose, here we have shown a part of tedious calculations
to determine CEF potential, but for actual purpose, it is more convenient
to consult with the table of Hutchings,
in which all possible CEF parameters have been listed
for arbitrary angular momentum $J$ (both integer and half-integer).
For the $d$-electron case, we can simply refer the results for $J$=2
in the Hutchings table \cite{Hutchings}.
For instance, in the tetragonal CEF as discussed above,
we easily obtain
\begin{eqnarray}
  \begin{array}{l}
    A_{2,2}=A_{-2,-2}=6B_{2}^{0}+12B_{4}^{0}, \\
    A_{1,1}=A_{-1,-1}=-3B_{2}^{0}-48B_{4}^{0}, \\
    A_{0,0}=-6B_{2}^{0}+72B_{4}^{0}, \\
    A_{2,-2}=A_{-2,2}=12B_{4}^{4},
  \end{array}
\end{eqnarray}
where $B_{n}^{m}$ with integers $n$ and $m$
is the so-called CEF parameter,
traditionally used in the ligand field theory.
It is easy to express the CEF parameters $B_n^m$ by using ours
as $B_2^0=A_{20}/3$, $B_4^0=A_{40}/12$, and $B_4^4=A_{44}/12$.
Note again that $B_2^0$=0 and $B_4^4$=$5B_4^0$ for the cubic case.
In actuality, we do not estimate the CEF parameters purely theoretically,
but they are determined from the fitting of experimental results
such as magnetic susceptibility and specific heat
at high temperatures.

\begin{figure}[t]
\begin{center}
\includegraphics[width=1.0\textwidth]{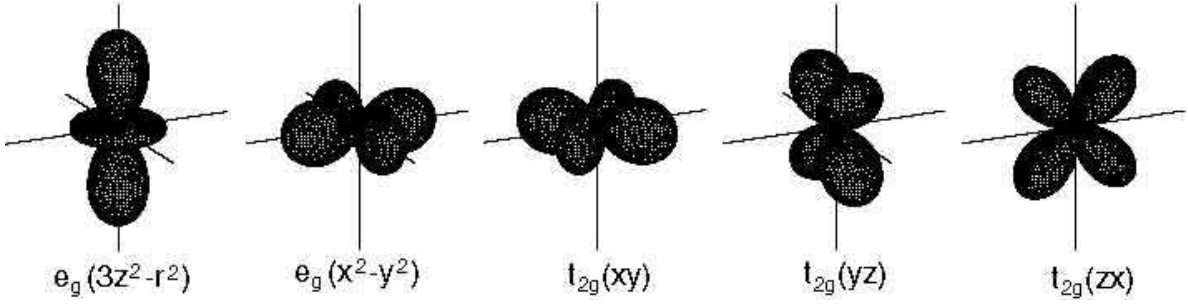}
\end{center}
\caption{%
Views for $d$-electron orbitals.
}
\label{fig2}
\end{figure}

For a cubic case ($a$=$b$),
the five-fold degeneracy in $d$ orbital is lifted into
doubly-degenerate $e_{\rm g}$-orbitals
($d_{x^2-y^2}$ and $d_{3z^2-r^2}$)
and triply-degenerate $t_{\rm 2g}$-orbitals
($d_{xy}$, $d_{yz}$, and $d_{zx}$).
The shape of each orbital is illustrated in Fig.~\ref{fig2}.
The eigenenergy for $e_{\rm g}$-orbitals is given by
\begin{equation}
  \varepsilon_{3z^2-r^2}=\varepsilon_{x^2-y^2}
  =6A_{40}=72B_4^0=\frac{Ze^2}{a}\frac{\langle r^4 \rangle}{a^4},
\end{equation}
while for $t_{\rm 2g}$ orbitals, we obtain
\begin{equation}
  \varepsilon_{xy}=\varepsilon_{yz}=\varepsilon_{zx}
  =-4A_{40}=-48B_4^0
  =-\frac{2}{3}\frac{Ze^2}{a}\frac{\langle r^4 \rangle}{a^4}.
\end{equation}
Then, the CEF term can be written as
\begin{equation}
  \label{Eq:dCEF}
  H_{\rm CEF} = \sum_{{\bf i},\sigma,\gamma}
  \varepsilon_{\gamma}
  d^{\dag}_{{\bf i}\gamma \sigma} d_{{\bf i} \gamma \sigma},
\end{equation}
where $d_{{\bf i}\gamma\sigma}$ is the annihilation operator
for a $d$-electron with spin $\sigma$
in the $\gamma$-orbital at site ${\bf i}$.
The energy difference between those two levels is usually expressed as
$10Dq$ in the traditional notation in the ligand field theory.
It is explicitly written as
\begin{equation}
  10Dq=10A_{40}=120B_{4}^0=\frac{5}{3} \frac {Ze^2}{a}
  \frac{\langle r^4 \rangle}{a^4}.
\end{equation}
Let us try to estimate theoretically the value of $10Dq$.
For the purpose, it is convenient to transform the above equation into
\begin{equation}
  10Dq=\frac{10Z}{3} \frac {e^2}{2a_{\rm B}}
  \Big(\frac{a_{\rm B}}{a}\Bigr)^5
  \frac{\langle r^4 \rangle}{a_B^4},
\end{equation}
where $a_{\rm B}$ denotes the Bohr radius ($a_{\rm B}$=0.529\AA).
Note that $Z$=2 for O$^{2-}$ ion and $e^2/(2a_{\rm B})$=1 Ryd.=13.6 eV.
By using the solution of the hydrogen-like potential problem,
we obtain $\langle r^4 \rangle/a_B^4$=$25515/Z_M^4$,
where $Z_M$ denotes the atomic number of transition metal atom.
For instance, if we simply set $Z_M$=25 for manganese and
$a$=$2{\rm \AA}$ as a typical value for perovskite structure,
we obtain $10Dq$=7.7meV, which is very small compared with
the observed value, since $10Dq$ is found to be
in the order of eV in actual transition metal oxides.
In fact, the estimation by Yoshida suggests that $10Dq$ is
about 10000-15000cm$^{-1}$
(remember that 1 eV = 8063 cm$^{-1}$) \cite{Yoshida}.
The electrostatic contribution obtained in the above naive estimation
is considered to be much smaller than the observed CEF splitting.

Here note that the energy level for the $t_{\rm 2g}$-orbitals is lower 
than that for $e_{\rm g}$-orbitals for perovskite structure.
Qualitatively this can be intuitively understood as follows:
The energy difference originates in the Coulomb interaction
between the $3d$ electrons and the oxygen ions surrounding
transition metal ion.
As shown in Fig.~\ref{fig2},
while the wave-functions of the $e_{\rm g}$-orbitals
are extended along the direction of the bond between transition metal ion
and oxygen ions, those in the $t_{\rm 2g}$-orbitals avoid this direction.
Thus, an electron in $t_{\rm 2g}$-orbitals is not heavily influenced
by the Coulomb repulsion due to the negatively charged oxygen ions, 
and the energy level for $t_{\rm 2g}$-orbitals is lower than that
for $e_{\rm g}$-orbitals.

For a tetragonal case with $a<b$, as observed in cuprates,
we find the splitting of each degenerate orbital.
The $e_{\rm g}$ orbital is split into $a_{\rm 1g}$ ($3z^2-r^2$)
and $b_{\rm 1g}$ ($x^2-y^2$) orbitals, of which eigen energies
are, respectively, given by
\begin{equation}
  \varepsilon_{3z^2-r^2}=6{\tilde D}q
  -\frac{4}{7}\frac{Ze^2}{a}\frac{\langle r^2 \rangle}{a^2}
  \Bigl[ 1- \Bigl( \frac{a}{b} \Bigr)^3 \Bigr]
  -\frac{5}{21}\frac{Ze^2}{a}\frac{\langle r^4 \rangle}{a^4}
  \Bigl[ 1- \Bigl( \frac{a}{b} \Bigr)^5 \Bigr],
\end{equation}
and
\begin{equation}
  \varepsilon_{x^2-y^2}=6{\tilde D}q
  +\frac{4}{7}\frac{Ze^2}{a}\frac{\langle r^2 \rangle}{a^2}
  \Bigl[ 1- \Bigl( \frac{a}{b} \Bigr)^3 \Bigr]
  +\frac{5}{21}\frac{Ze^2}{a}\frac{\langle r^4 \rangle}{a^4}
  \Bigl[ 1- \Bigl( \frac{a}{b} \Bigr)^5 \Bigr],
\end{equation}
where ${\tilde D}q$ is defined as
\begin{equation}
  {\tilde D}q=
  \frac{1}{6}\frac{Ze^2}{a}\frac{\langle r^4 \rangle}{a^4}
  \Bigl[ \frac{2}{3} + \frac{1}{3}\Bigl( \frac{a}{b} \Bigr)^5 \Bigr].
\end{equation}
On the other hand, $t_{\rm 2g}$ orbital is split into $e_{\rm g}$
degenerate ($yz$ and $zx$) and $b_{\rm 2g}$ ($xy$) orbitals,
of which eigen energies are, respectively, given by
\begin{equation}
  \varepsilon_{yz}=\varepsilon_{zx}=-4{\tilde D}q
  -\frac{2}{7}\frac{Ze^2}{a}\frac{\langle r^2 \rangle}{a^2}
  \Bigl[ 1- \Bigl( \frac{a}{b} \Bigr)^3 \Bigr]
  +\frac{10}{63}\frac{Ze^2}{a}\frac{\langle r^4 \rangle}{a^4}
  \Bigl[ 1- \Bigl( \frac{a}{b} \Bigr)^5 \Bigr],
\end{equation}
and
\begin{equation}
  \varepsilon_{xy}=-4{\tilde D}q
  +\frac{4}{7}\frac{Ze^2}{a}\frac{\langle r^2 \rangle}{a^2}
  \Bigl[ 1- \Bigl( \frac{a}{b} \Bigr)^3 \Bigr]
  -\frac{20}{63}\frac{Ze^2}{a}\frac{\langle r^4 \rangle}{a^4}
  \Bigl[ 1- \Bigl( \frac{a}{b} \Bigr)^5 \Bigr].
\end{equation}
Note that also for the tetragonal case,
the CEF term is written as Eq.~(\ref{Eq:dCEF}).

The splitting in the tetragonal case with $a<b$ is intuitively
understood from the shape of orbitals, on the basis of the point
that the energy difference originates in the Coulomb interaction
between the $3d$ electrons and the oxygen ions surrounding
transition metal ion.
While the wave-function of the $3z^2$$-$$r^2$ orbital is extended
along the $z$-direction, $x^2$$-$$y^2$ orbital shape is extended
in the $x$-$y$ plane.
Since apical oxygens move to the $z$-direction for the case of $a<b$,
the energy loss for $3z^2$$-$$r^2$ orbital becomes small,
indicating that the $3z^2$$-$$r^2$ orbital is lower.
Concerning the splitting of $t_{\rm 2g}$ orbitals, we note that
$xy$ orbital is extended in the $xy$ plane, while $yz$ and $zx$
orbitals have some extension along $z$ axis.
Then, for $a<b$, the penalty from the electrostatic potential
would be smaller for $yz$ and $zx$, compared with that for $xy$.
Then, we intuitively consider that the energy level for $xy$ orbital
is higher than those for $yz$ and $zx$ orbitals.

\subsection{Coulomb interactions}

Now we include the effect of Coulomb interactions
among $d$ electrons in the level of atomic problem.
In the localized ion system, the Coulomb interaction term
among $d$-electrons is generally given by
\begin{eqnarray}
  H_{\rm el-el} = \frac{1}{2}
  \sum_{\bf i}\sum_{m_1,m_2,m_3,m_4}\sum_{\sigma_1 \sigma_2}
  I^d_{m_1 m_2, m_3 m_4}
  d_{{\bf i}m_1\sigma_1}^{\dag}
  d_{{\bf i}m_2\sigma_2}^{\dag}
  d_{{\bf i}m_3\sigma_2}
  d_{{\bf i}m_4\sigma_1},
\end{eqnarray}
where the Coulomb matrix element among $d$ electrons $I^d$
is expressed as
\begin{eqnarray}
  I^d_{m_1 m_2, m_3 m_4}
  = \int  \int  d{\bf r}_1 d{\bf r}_2
  \Psi_{3 2 m_1}^*({\bf r}_1) \Psi_{3 2 m_2}^*({\bf r}_2)
  g_{12} \Psi_{3 2 m_3}({\bf r}_2) \Psi_{3 2 m_4}({\bf r}_1).
\end{eqnarray}
Here $g_{12}$=$g(|{\bf r}_1-{\bf r}_2|)$
is the screened Coulomb interaction, which can be expanded
in spherical harmonics
\begin{eqnarray}
  g_{12}=g(|{\bf r}_1-{\bf r}_2|)
  =\sum_{\ell,m} g_{\ell}(r_1,r_2)Y_{\ell m}(\theta_1,\phi_1)
  Y_{\ell m}(\theta_2,\phi_2),
\end{eqnarray}
with ${\bf r}_1$=$(r_1,\theta_1,\phi_1)$ and
${\bf r}_2$=$(r_2,\theta_2,\phi_2)$ in the polar coordinate.
The complicated integrals can be partly performed and the result
is given by using the Gaunt coefficients as
\begin{eqnarray}
  I^d_{m_1 m_2, m_3 m_4}
  = \delta_{m_1+m_2,m_3+m_4}
  \sum_{k=0,2,4} c^{(k)}(m_1,m_3) c^{(k)}(m_4,m_2) F^k_d.
\end{eqnarray}
Note that the sum is limited by the Wigner-Eckart theorem to
$k$=0, 2, and 4.
Here we define $F^k_d$, which is the radial integral
for the $k$-th partial wave,
called Slater integral or Slater-Condon parameter
\cite{Slater1929,Condon1931}, given by
\begin{eqnarray}
  F^{k}_d=\int_0^{\infty} r_1^2dr_1 \int_0^{\infty} r_2^2dr_2
  R_{3d}^2(r_1)R_{3d}^2(r_2) g_{k}(r_1,r_2).
\end{eqnarray}
By using Slater-Condon parameters, it is more convenient to
define Racah parameters, given by
\begin{eqnarray}
  \begin{array}{l}
    A=F^0_d-F^4_d/9,\\
    B=(9F^2_d-5F^4_d)/441,\\
    C=5F^4_d/63,
  \end{array}
\end{eqnarray}
where $A$, $B$, and $C$ are Racah parameters \cite{Racah1942}.
All possible Coulomb integrals are expressed by these
Racah parameters \cite{Tang1998}.
In usual, the Racah parameters are determined from
the experimental results in the high-energy region.

\begin{figure}[t]
\begin{center}
\includegraphics[width=0.8\textwidth]{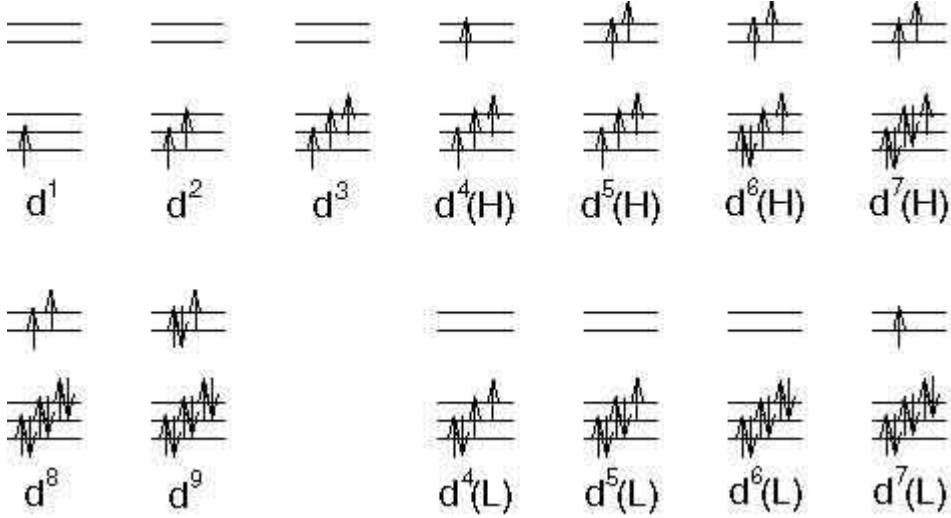}
\end{center}
\caption{%
Configurations of $d^n$ electrons for $n$=1$\sim$9.
For $n$=4$\sim$7, we show both the high- and low-spin states.
}
\label{fig3}
\end{figure}

By diagonalizing $H_{\rm CEF}$+$H_{\rm el-el}$,
let us discuss $d^n$-electron configuration under the cubic CEF,
where the superscript $n$ denotes the local $d$-electron number.
It is possible to obtain analytic result partly, but most of the
solutions can be obtained only numerically.
As summarized in Fig.~\ref{fig3},
the results are understood from the electron configuration
obtained by accommodating $n$ electrons in $e_{\rm g}$ and
$t_{\rm 2g}$ orbitals due to the competition between
the CEF splitting $10Dq$ and the Hund's rule interaction.
For $n$=1, we find 6-fold degenerate ground state, originating from
the three $t_{\rm 2g}$ orbitals with spin degree of freedom.
For $n$=2, 9-fold degeneracy is found in the ground state.
When we accommodate a couple of electrons among three $t_{\rm 2g}$
orbitals, there are three possibilities for the formation of
total spin $S$=1 between two of three orbitals.
Since each spin $S$=1 has three degenerate states with $S_z$=$-1$, $0$,
and $+1$, in total we obtain 9-fold degeneracy.
For $n$=3, 4-fold degeneracy is found in the ground state.
Now three $t_{\rm 2g}$ electrons form $S$=3/2 spin due to Hund's rule,
which have four degenerate states with $S_z$=$\pm 3/2$ and $\pm 1/2$.

For the case of $n$=4, we find two types of ground states,
depending on the balance between $10Dq$ and Coulomb interaction.
For small $10Dq$, 10-fold degeneracy is observed in the ground state,
while the degeneracy is changed to 9-fold for large $10Dq$.
These are high- and low-spin states,
labeled by ``H'' and ``L'', respectively.
In the high-spin configuration, due to the strong Hund's rule
interaction, total spin $S$=2 is formed, but we consider that
one electron is added to the $d^3$ configuration.
The fourth electron should be accommodated in $e_{\rm g}$ sector,
which provide an extra double degeneracy.
Since $S$=2 state include five degenerate states with
$S_z$=$\pm 2$, $\pm 1$, and 0, in total we obtain 10-fold degeneracy.
The eigen energy for $d^4$(H) is exactly given by
$E[d^4({\rm H})]$=$6A-21B-72B_4^0$.
On the other hand, for the low-spin state, we consider that
all four electrons are included among $t_{\rm 2g}$ orbitals
to gain the CEF potential energy.
Namely, $S$=1 spin is formed in the $t_{\rm 2g}$ sector,
leading to 9-fold degeneracy as in the case of $d^2$.
The energy for $d^4$(L) can be obtained only numerically,
but in the limit of infinite $10Dq$, the energy is asymptotically
given by $E[d^4({\rm L})]$=$6A-15B+5C-192B_4^0$.
Due to the comparison of two energies,
we obtain a rough condition for the change of the spin state as
$6B+5C > 10Dq$, which indicates clearly that the high-spin state
is chosen when the Hund's rule interaction is stronger than
the effect of CEF potential.

In the high-spin state for $n$=4, as observed in Mn$^{3+}$ ion,
three electrons are accommodated in the $t_{\rm 2g}$ orbitals,
while one electron exists in the $e_{\rm g}$ orbital.
In such a case, it is necessary to treat both $t_{\rm 2g}$
and $e_{\rm g}$ electrons in the same model Hamiltonian.
On the other hand, if the low-spin state for $n$=4 is favored
due to the large CEF potential, it is enough to consider only
the $t_{\rm 2g}$ orbital, as in the cases of $n$=1$\sim$3,
In fact, for $4d$ electron systems, the effect of CEF potential
becomes strong due to the large ion radius.
For instance, in ruthenates, Ru$^{4+}$ ion includes four $4d$
electrons, which form the low-spin state with $S$=1,
as deduced from the AF phase of Ca$_2$RuO$_4$.

For $n$=5, we find 6-fold degeneracy in the region of small $10Dq$.
This is originating from the high-spin $S$=5/2 state
with $S_z$=$\pm 5/2$, $\pm 3/2$ and $\pm 1/2$, in which each orbital
is occupied by one electron.
For very large $10Dq$, the ground state degeneracy is still six, but
the wavefunction is drastically changed in this low-spin $S$=1/2 state.
It is understood that five electrons (or one hole) occupies
the $t_{\rm 2g}$ orbitals.
Namely, there are three kinds of $S$=1/2 state, leading to
six-fold degeneracy in total.
In principle, the intermediate spin state with $S$=3/2 is possible,
but the region for the intermediate state seems
to be very limited in the parameter space.

For $n$=6, the ground state for small $10Dq$ has 15-fold degeneracy.
This is the high-spin $S$=2 state, in which two electrons in
the $e_{\rm g}$ sector, while four electrons in the $t_{\rm 2g}$ section.
As shown in the configuration of $d^6$(H) in Fig.~\ref{fig3},
there are three possibility for down spin electron
in the $t_{\rm g}$ orbitals.
Namely, there are three kinds of $S$=2 states,
leading to 15-fold degeneracy in total.
For large $10Dq$, we obtain the singlet ground state,
which is the low-spin state with $S$=0,
in which $t_{\rm 2g}$ orbitals are fully occupied by six electrons.

For $n$=7, the high-spin state in the region of small $10Dq$
has 12-fold degeneracy, by accommodating two electrons in
the $e_{\rm g}$ and five electrons in the $t_{\rm 2g}$ orbitals.
As shown in the configuration of $d^7$(H) in Fig.~\ref{fig3},
there are three possibility for one hole
in the $t_{\rm g}$ orbitals.
Thus, there are three kinds of $S$=3/2 states, leading to 
12-fold degeneracy in total.
In the low-spin state, on the other hand, the ground state
degeneracy is four.
It is easily understood, when we put one more electron in
the $e_{\rm g}$ orbital in addition to the $d^6$(L) configuration.

Both for $n$=8 and 9, since $t_{\rm 2g}$ orbitals are fully occupied
and thus, active orbital is $e_{\rm g}$.
For $n$=8, the ground state with $S$=1 composed of a couple of
$e_{\rm g}$ electrons is triply degenerate.
For $n$=9, we find one hole in the $e_{\rm g}$ orbitals,
there are four fold degeneracy in the ground state.
Note that for $n$=9, when the tetragonal CEF effect is strong enough,
as actually found in perovskite copper oxides,
only $x^2-y^2$ orbital becomes active.

In general, it is necessary to consider all five orbitals
in the Hamiltonian for the complete analysis of electronic
properties of transition metal oxides, even if
the full Hamiltonian includes several
competing tendencies and couplings.
However, the essential physics can be frequently obtained
even using relatively simple models,
by focusing on the active orbital,
emerging out of the competition between
the CEF potential and the Coulomb interaction.
In this sense, we immediately arrive at two possibilities:
One is $t_{\rm 2g}$ model for the cases of $d^1 \sim d^3$ and 
$d^4(L) \sim d^6(L)$.
Another is $e_{\rm g}$ model for the cases of $d^7$(L), $d^8$, and $d^9$.
For other cases, it is necessary to consider both $e_{\rm g}$
and $t_{\rm 2g}$ orbitals in the same Hamiltonian simultaneously.
However, for the cases of $d^4$(H), it is possible to define
a simplified model, which will be explained later for
Mn$^{3+}$ ion with the high-spin state.

For the purpose to express the above models,
it is convenient to simplify the Coulomb interaction term
in a qualitatively correct form,
since the representation using Racah parameters are exact,
but too complicated for further analysis.
Under the cubic CEF potential,
as described in the previous subsection,
two $e_{\rm g}$ and three $t_{\rm 2g}$ orbitals provide appropriate basis.
Then, the Coulomb interaction term should be expressed as
\begin{eqnarray}
  H_{\rm el-el} = \frac{1}{2} \sum_{\bf i}
  \sum_{\gamma_1 \gamma_2 \gamma_3 \gamma_4}
  \sum_{\sigma_1 \sigma_2}
  {\tilde I}_{\gamma_1 \sigma_1,\gamma_2 \sigma_2;
  \gamma_3 \sigma_2,\gamma_4 \sigma_1}
  d_{{\bf i}\gamma_1\sigma_1}^{\dag}
  d_{{\bf i}\gamma_2\sigma_2}^{\dag}
  d_{{\bf i}\gamma_3\sigma_2}
  d_{{\bf i}\gamma_4\sigma_1},
\end{eqnarray}
where the modified Coulomb matrix element ${\tilde I}$ is given by
the proper combination of the original one $I^d$.
Note that the orbital index $\gamma$ denotes $e_{\rm g}$ and $t_{\rm 2g}$
orbitals, which are defined by linear combination of the original $d$-electron
orbitals labeled by $m$.

By using the modified Coulomb matrix element,
it is useful to define the so-called ``Kanamori parameters'', 
$U$, $U'$, $J$, and $J'$ \cite{Kanamori1963}.
Among them, $U$ is the intra-orbital Coulomb interaction, given by 
\begin{eqnarray}
  U_{\gamma} = {\tilde I}_{\gamma \sigma,\gamma \sigma';
      \gamma \sigma',\gamma \sigma},
\end{eqnarray}
with $\sigma \ne \sigma'$.
$U'$ is the inter-orbital Coulomb interaction, expressed by 
\begin{eqnarray}
  U'_{\gamma,\gamma'}= {\tilde I}_{\gamma \sigma,\gamma' \sigma';
  \gamma' \sigma',\gamma \sigma},
\end{eqnarray}
with $\gamma \ne \gamma'$.
$J$ is the inter-orbital exchange interaction, leading to
the Hund's rule coupling, written as 
\begin{eqnarray}
  J_{\gamma,\gamma'} = {\tilde I}_{\gamma \sigma,\gamma' \sigma';
  \gamma \sigma',\gamma' \sigma},
\end{eqnarray}
with $\gamma \ne \gamma'$.
Finally, $J'$ is the pair-hopping amplitude between different orbitals,
given by
\begin{eqnarray}
  J'_{\gamma,\gamma'} = {\tilde I}_{\gamma \sigma,\gamma \sigma';
  \gamma' \sigma',\gamma' \sigma},
\end{eqnarray}
with $\gamma \ne \gamma'$ and $\sigma \ne \sigma'$.

In Table.~1, we list the values for possible Kanamori parameters.
Here we have four comments on the relation among Kanamori parameters.
(i) A relation $J$=$J'$ holds, which is simply due to the fact that
each of the parameters above is given by an integral of the Coulomb
interaction sandwiched with appropriate orbital wave functions.
Analyzing the form of those integrals the equality between $J$ and $J'$
can be deduced.
(ii) For completeness, we explicitly show the orbital indices
for the definition of Kanamori parameters.
In fact, as found in Table.~1, the inter-orbital Coulomb
interactions between $e_{\rm g}$ and $t_{\rm 2g}$ orbitals
depend on the kind of orbitals,
while there is no orbital dependence in Coulomb interactions
among $e_{\rm g}$ or $t_{\rm 2g}$ orbitals.
(iii) We point out that among $e_{\rm g}$ or $t_{\rm 2g}$ orbitals,
another relation $U$=$U'$+$J$+$J'$ holds in any combination of orbitals.
This relation is needed to recover the rotational invariance
in orbital space.
(iv) Since the largest energy scale among the Kanamori parameters is $U$,
the orbitals are not doubly occupied by both up- and down-spin electrons.
Thus, only one electron can exist in each orbital of the degenerate
$e_{\rm g}$ or $t_{\rm 2g}$ sector.
Furthermore, in order to take advantage of $J$, the spins of electrons
point along the same direction.
This is the so-called ``Hund's rule''.

\begin{table*}
\begin{center}
    \begin{tabular}{|c|c|c|c|}
    \hline $\gamma$ & $\gamma'$ &
    $U'_{\gamma,\gamma'}$  & $J_{\gamma,\gamma'}$ \\ \hline
    $xy,yz,zx$ & $xy,yz,zx$ &
    $A$$-$$2B$+$C$  & $3B$+$C$  \\ \hline
    $x^2$$-$$y^2$, $3z^2$$-$$r^2$ & $x^2$$-$$y^2$, $3z^2$$-$$r^2$ &
    $A$$-$$4B$+$C$  & $4B$+$C$  \\ \hline
    ${xy}$ & $x^2$$-$$y^2$ &
    $A$+$4B$+$C$  & $C$  \\ \hline
    ${xy}$ & $3z^2$$-$$r^2$ &
    $A$$-$$4B$+$C$  & $4B$+$C$  \\ \hline
    ${yz}$, ${zx}$ & $x^2$$-$$y^2$  &
    $A$$-$$2B$+$C$  & $3B$+$C$  \\ \hline
    ${yz}$, ${zx}$ & $3z^2$$-$$r^2$  &
    $A$+$2B$+$C$  & $B$+$C$  \\ \hline
    \end{tabular}
\end{center}
\caption{Expressions for $U'$ and $J$ by using
Racah parameters $A$, $B$, and $C$.
Note that $U$=$A$+$4B$+$3C$ for each orbital.
For more information, see Refs.~\cite{Tang1998} and \cite{Kanamori1963}.}
\end{table*}

Then, for the case with active $e_{\rm g}$ orbital degree of freedom,
we can define the Coulomb interaction term as
\begin{eqnarray}
  H^{e_{\rm g}}_{\rm el-el} &=&
  U \sum_{{\bf i},\gamma}
  n_{{\bf i}\gamma\uparrow} n_{{\bf i}\gamma\downarrow}
  + U' \sum_{\bf i} n_{{\bf i}{\rm a}} n_{{\bf i}{\rm b}}
  \nonumber \\
  &+& J \sum_{{\bf i},\sigma,\sigma'}
  d_{{\bf i}{\rm a}\sigma}^{\dag}d_{{\bf i}{\rm b}\sigma'}^{\dag}
  d_{{\bf i}{\rm a}\sigma'}d_{{\bf i}{\rm b}\sigma}
  + J' \sum_{\bf i}
  (d_{{\bf i}{\rm a}\uparrow}^{\dag}d_{{\bf i}{\rm a}\downarrow}^{\dag}
  d_{{\bf i}{\rm b}\downarrow}d_{{\bf i}{\rm b}\uparrow}
  +{\rm h.c.}),
\end{eqnarray}
where the subscripts, a and b, denote
$x^2-y^2$ and $3z^2-r^2$ orbitals, respectively,
$n_{{\bf i}\gamma\sigma}$=
$d_{{\bf i}\gamma\sigma}^{\dag}d_{{\bf i}\gamma\sigma}$,
and
$n_{{\bf i}\gamma}$=$\sum_{\sigma}n_{{\bf i}\gamma\sigma}$.
As easily understood from Table.~1,
we find $U'$=$A-4B+C$ and $J$=$4B+C$ for $e_{\rm g}$ orbital.

On the other hand, for the case with active $t_{\rm 2g}$ orbital,
the Coulomb interaction term is expressed by
\begin{eqnarray}
  H_{\rm el-el}^{t_{\rm 2g}} &=&
  U \sum_{{\bf i},\gamma}
  n_{{\bf i}\gamma\uparrow} n_{{\bf i}\gamma\downarrow}
  + \frac{U'}{2} \sum_{{\bf i},\gamma \ne \gamma'}
  n_{{\bf i}\gamma}n_{{\bf i}\gamma'}
  \nonumber \\
  &+& \frac{J}{2} \sum_{{\bf i},\sigma,\sigma'}\sum_{\gamma \ne \gamma'}
  d_{{\bf i}\gamma\sigma}^{\dag}d_{{\bf i}\gamma'\sigma'}^{\dag}
  d_{{\bf i}\gamma\sigma'}d_{{\bf i}\gamma'\sigma}
  + \frac{J'}{2} \sum_{{\bf i},\gamma \ne \gamma'}
  d_{{\bf i}\gamma\uparrow}^{\dag}d_{{\bf i}\gamma\downarrow}^{\dag}
  d_{{\bf i}\gamma'\downarrow}d_{{\bf i}\gamma'\uparrow},
\end{eqnarray}
where $\gamma$ and $\gamma'$ run in the $t_{\rm 2g}$ orbitals,
$U'$=$A-2B+C$, and $J$=$3B+C$.

Next we consider the more complicated case, in which the Coulomb
interaction term is not expressed only among $e_{\rm g}$ or
$t_{\rm 2g}$ orbitals.
A typical situation can be found in the high-spin state of $d^4$
configuration, which is relevant to manganites in the
cubic perovskite structure.
Due to the Hund's rule, tetravalent manganese ion includes three
$d$ electrons to form $S$=3/2 in the $t_{\rm 2g}$ orbitals.
By adding one more electron to Mn$^{4+}$ with three up-spin
$t_{\rm 2g}$-electrons, let us consider the configuration for the
Mn$^{3+}$ ion.
As mentioned above, there are two possibilities due to the balance
between the crystalline-field splitting and the Hund's rule coupling,
but in $3d$ electron system such as manganites,
the high-spin state is realized, since the Hund's rule coupling
dominates over $10Dq$.

In order to simplify the model without loss of essential physics,
it is reasonable to treat the three spin-polarized
$t_{\rm 2g}$-electrons as a localized ``core-spin'' expressed by
${\bf S}_{\bf i}$ at site ${\bf i}$, since the overlap integral
between $t_{\rm 2g}$ and oxygen $p$ orbital is small compared
with that between $e_{\rm g}$ and $p$ orbitals.
Moreover, due to the large value of the total spin $S$=$3/2$,
it is usually approximated by a classical spin.
This approximation has been tested by using computational techniques
\cite{Yunoki1998a,Dagotto1998}.
Thus, the effect of Coulomb interaction between $e_{\rm g}$ and
$t_{\rm 2g}$ electrons can be effectively included by
the strong Hund's rule coupling between the $e_{\rm g}$-electron
spin and localized $t_{\rm 2g}$-spins, given by
\begin{equation}
  H_{\rm Hund} = -J_{\rm H} \sum_{\bf i}
  {\bf s}_{\bf i} \cdot {\bf S}_{\bf j},
\end{equation}
where ${\bf s}_{\bf i}$=
$\sum_{\gamma\alpha\beta}d^{\dag}_{{\bf i}\gamma\alpha}
\sigma_{\alpha\beta}d_{{\bf i}\gamma\beta}$,
$J_{\rm H}$($>$0) is the Hund's rule coupling between
localized $t_{\rm 2g}$-spin and mobile $e_{\rm g}$-electron,
and ${\sigma}$=$(\sigma_x, \sigma_y, \sigma_z)$ are the Pauli matrices.
The magnitude of $J_{\rm H}$ is considered to be of the order of $J$.
Here note that ${\bf S}_{\bf i}$ is normalized as $|{\bf S}_{\bf i}|$=1.
Thus, the direction of the classical $t_{\rm 2g}$-spin at site
${\bf i}$ is defined as
\begin{equation}
  {\bf S}_{\bf i}=(\sin\theta_{\bf i}\cos\phi_{\bf i}, 
  \sin\theta_{\bf i}\sin\phi_{\bf i}, 
  \cos\theta_{\bf i}),
\end{equation}
by using the polar angle $\theta_{\bf i}$ and the azimuthal angle
$\phi_{\bf i}$.

Note that the effect of the Coulomb interaction is not fully
taken into account only by $H_{\rm Hund}$, since there remains the 
direct electrostatic repulsion between $e_{\rm g}$-electrons.
Then, by further adding the $e_{\rm g}$ electron interaction term,
we obtain the Coulomb interaction term for $d^4$(H) configuration as
\begin{equation}
  H^{\rm DE}_{\rm el-el} = H_{\rm Hund} + H_{\rm el-el}^{e_{\rm g}}.
\end{equation}
Here we use ``DE'' in the superscript for the Hamiltonian,
which is the abbreviation of ``double exchange''.
We will explain later the reason why we use ``DE'',
in the section
explaining the result for manganites.

\subsection{Electron-phonon interactions}

Another important ingredient in the model Hamiltonian
for transition metal oxides is the lattice distortion
coupled to $e_{\rm g}$ or $t_{\rm 2g}$ electrons.
In particular, the orbital degeneracy is
lifted by the Jahn-Teller distortion.
It is difficult to explain all possible types of
electron-phonon coupling term, but here we explain
the coupling between degenerate $e_{\rm g}$ orbital
and the distortion of the octahedron,
since later we will focus on the magnetic properties
of manganites in the perovskite structure.
The coupling of the same type of distortions of octahedron
to $t_{\rm 2g}$ orbitals is also discussed in this subsection.

The basic formalism for the study of electrons coupled to Jahn-Teller
modes has been set up by Kanamori \cite{Kanamori1960}.
He focused on cases where the electronic orbitals are degenerate in
the undistorted crystal structure, as in the case of manganese ion
in an octahedron of oxygens.
As explained by Kanamori, the Jahn-Teller effect in this context
can be simply stated as follows.
When a given electronic level of a cluster is degenerate
in a structure of high symmetry, this structure is generally
unstable, and the cluster will present a distortion
toward a lower symmetry ionic arrangement \cite{Jahn-Teller}.
In the case of Mn$^{3+}$, which is
orbitally doubly degenerate
when the crystal is undistorted,
a splitting will occur when the crystal is distorted.
The distortion of the MnO$_6$ octahedron is ``cooperative'' since once
it occurs in a particular octahedron, it will affect the neighbors. 
The basic Hamiltonian to describe the interaction between electrons
and Jahn-Teller modes was written by Kanamori.
In the adiabatic approximation, it is given in the form of
\begin{equation}
  H_{\rm JT} = g \sum_{\bf i} (Q_{2{\bf i}} \tau_{x{\bf i}} +
  Q_{3{\bf i}} \tau_{z{\bf i}})
  + k_{\rm JT} \sum_{\bf i} (Q_{2{\bf i}}^2+Q_{3{\bf i}}^2)/2,
\end{equation}
where $g$ is the coupling constant between the $e_{\rm g}$-electrons and
distortions of the MnO$_6$ octahedron,
$Q_{2{\bf i}}$ and $Q_{3{\bf i}}$ are normal modes of vibration of the 
oxygen octahedron that remove the degeneracy between the electronic
levels, and $k_{\rm JT}$ is the spring constant for the Jahn-Teller
mode distortions.
We note that the sign of $g$ depends on the kind of ion and
electron configuration.
The pseudospin operators are defined as
\begin{eqnarray}
  \tau_{x{\bf i}} = \sum_{\sigma}
  (d_{{\bf i}{\rm a}\sigma}^{\dag}d_{{\bf i}{\rm b}\sigma}
  +d_{{\bf i}{\rm b}\sigma}^{\dag}d_{{\bf i}{\rm a}\sigma}),~~
  \tau_{z{\bf i}} = \sum_{\sigma}
  (d_{{\bf i}{\rm a}\sigma}^{\dag}d_{{\bf i}{\rm a}\sigma}
  -d_{{\bf i}{\rm b}\sigma}^{\dag}d_{{\bf i}{\rm b}\sigma}).
\end{eqnarray}
In the expression of $H_{\rm JT}$, a $\tau_{y{\bf i}}$-term does
not appear for symmetry reasons, since it belongs to 
the $A_{\rm 2u}$ representation.
The non-zero terms should correspond to the irreducible
symmetric representations of $E_{\rm g}$$\times$$E_{\rm g}$,
namely, $E_{\rm g}$ and $A_{\rm 1g}$. 
The former representation is expressed by using the pseudo spin
operators $\tau_{x{\bf i}}$ and $\tau_{z{\bf i}}$ as discussed here,
while the latter, corresponding to the breathing mode,
is discussed later.

Following Kanamori, $Q_{2{\bf i}}$ and $Q_{3{\bf i}}$ are explicitly
given by
\begin{equation}
  \label{eq:q2}
  Q_{2{\bf i}}=(X_{1{\bf i}}-X_{3{\bf i}}
               -Y_{2{\bf i}}+Y_{4{\bf i}})/\sqrt{2},
\end{equation}
and 
\begin{equation}
  \label{eq:q3}
  Q_{3{\bf i}}=(2Z_{5{\bf i}}-2Z_{6{\bf i}}
               -X_{1{\bf i}}+X_{3{\bf i}}
               -Y_{2{\bf i}}+Y_{4{\bf i}})/\sqrt{6},
\end{equation}
where $X_{\mu {\bf i}}$, $Y_{\mu {\bf i}}$, and $Z_{\mu {\bf i}}$
are the displacement of oxygen ions from the equilibrium positions
along the $x$-, $y$-, and $z$-direction, respectively.
Explicitly, they are included in the positions of oxygen ions as
${\bf R}_{1 {\bf i}}$=$(a/2+X_{1 {\bf i}},0,0)$,
${\bf R}_{2 {\bf i}}$=$(0,-a/2+Y_{2 {\bf i}},0)$,
${\bf R}_{3 {\bf i}}$=$(-a/2+X_{3 {\bf i}},0,0)$,
${\bf R}_{4 {\bf i}}$=$(0,a/2+Y_{4 {\bf i}},0)$,
${\bf R}_{5 {\bf i}}$=$(0,0,a/2+Z_{5 {\bf i}})$,
and
${\bf R}_{6 {\bf i}}$=$(0,0,-a/2+Z_{6 {\bf i}})$.
Here we consider the deformation of octahedron from the cubic case.
The convention for the labeling $\mu$
of coordinates is shown in Fig.~\ref{fig1}.

In order to solve the local Jahn-Teller problem,
it is convenient to scale the phononic degrees of freedom as
\begin{equation}
  Q_{2{\bf i}}=(g/k_{\rm JT})q_{2{\bf i}},~~
  Q_{3{\bf i}}=(g/k_{\rm JT})q_{3{\bf i}},
\end{equation}
where $g/k_{\rm JT}$ is the typical length scale for the Jahn-Teller
distortion, which is of the order of 0.1 \AA, namely, a few percent
of the lattice constant.
When the Jahn-Teller distortion is expressed in the polar coordinate as 
\begin{equation}
  \label{eq:polar}
  q_{2{\bf i}} = q_{{\bf i}} \sin \xi_{\bf i},~~
  q_{3{\bf i}} = q_{{\bf i}} \cos \xi_{\bf i},
\end{equation}
the ground state is easily obtained as
$(-\sin [\xi_{\bf i}/2] {d}_{{\bf i} a\sigma}^{\dag}
+\cos [\xi_{\bf i}/2] {d}_{{\bf i} b\sigma}^{\dag})|0\rangle$
with the use of the phase $\xi_{\bf i}$. 
The corresponding eigenenergy is given by $-E_{\rm JT}$,
where $E_{\rm JT}$ is the static Jahn-Teller energy,
defined by 
\begin{equation}
  E_{\rm JT}=g^2/(2k_{\rm JT}).
\end{equation}
Note here that the ground state energy is independent of the phase
$\xi_{\bf i}$.
Namely, the shape of the deformed isolated octahedron
is not uniquely determined in this discussion.
In the Jahn-Teller crystal, the kinetic motion of $e_{\rm g}$
electrons, as well as the cooperative effect between adjacent
distortions, play a crucial role in lifting the degeneracy and fixing
the shape of the local distortion.

To complete the electron-phonon coupling term, it is necessary to
consider the breathing mode distortion,
coupled to the local electron density as
\begin{equation}
  H_{\rm br} = g \sum_{\bf i} Q_{1{\bf i}} n_{{\bf i}}
  + k_{\rm br} \sum_{\bf i} Q_{1{\bf i}}^2/2,
\end{equation}
where
$n_{\bf i}$=$\sum_{\gamma,\sigma}
d_{{\bf i}\gamma\sigma}^{\dag}d_{{\bf i}\gamma\sigma}$
and the breathing-mode distortion $Q_{1{\bf i}}$ is given by
\begin{equation}
  \label{eq:q1}
  Q_{1{\bf i}}=(X_{1{\bf i}}-X_{3{\bf i}}
  +Y_{2{\bf i}}-Y_{4{\bf i}}+Z_{5{\bf i}}-Z_{6{\bf i}})/\sqrt{3},
\end{equation}
and $k_{\rm br}$ is the associated spring constant.
Note that, in principle, the coupling constants of the $e_{\rm g}$ 
electrons with the $Q_1$, $Q_2$, and $Q_3$ modes could be 
different from one another.
For simplicity, here it is assumed that those coupling constants
take the same value.
On the other hand, for the spring constants, a different notation
for the breathing mode is introduced, since the frequency
for the breathing mode distortion has been found experimentally to be
different from that for the Jahn-Teller mode.
This point will be briefly discussed later. 
Note also that the Jahn-Teller and breathing modes are competing with
each other. As it was  shown above, the energy gain due to the
Jahn-Teller distortion is maximized when one electron exists per
site. On the other hand, the breathing mode distortion energy is
proportional to the total number of $e_{\rm g}$ electrons per site,
since this distortion gives rise to an effective on-site attraction
between electrons.

By combining the Jahn-Teller mode and breathing mode distortions,
the electron-phonon term for the $e_{\rm g}$ model is summarized as
\begin{equation}
  \label{eph-eg}
  H^{e_{\rm g}}_{\rm el-ph}=H_{\rm JT}+H_{\rm br}.
\end{equation}
This expression depends on the parameter $\beta$=$k_{\rm br}/k_{\rm JT}$,
which regulates which distortion, the Jahn-Teller or breathing mode,
plays a more important role.
This point will be discussed in a separate subsection.

When we simply ignore buckling and rotational modes and consider
only the Jahn-Teller-type distortions of octahedron,
the electron-lattice coupling for $t_{\rm 2g}$ orbitals is given by
\cite{Struge}
\begin{eqnarray}
 \label{eph-t2g}
  H^{t_{\rm 2g}}_{\rm el-ph} = g \sum_{\bf i}
   (Q_{z{\bf i}} n_{{\bf i}{\rm xy}}
   +Q_{x{\bf i}} n_{{\bf i}{\rm yz}}
   +Q_{y{\bf i}} n_{{\bf i}{\rm zx}})
   +(k_{\rm JT}/2) \sum_{\bf i} (Q_{2{\bf i}}^2 + Q_{3{\bf i}}^2),
\end{eqnarray}
where
$Q_{x{\bf i}}$=$(-1/2)Q_{3{\bf i}}$+$(\sqrt{3}/2)Q_{2{\bf i}}$,
$Q_{y{\bf i}}$=$(-1/2)Q_{3{\bf i}}$$-$$(\sqrt{3}/2)Q_{2{\bf i}}$,
and $Q_{z{\bf i}}$=$Q_{3{\bf i}}$.
Later we will consider the effect of this type of electron-phonon
coupling on the magnetic structure of $t_{\rm 2g}$ electron systems.

Let us now consider the cooperative effect.
Although we have not explicitly mentioned thus far,
the distortions at each site are $not$ independent,
since all oxygens are shared by neighboring MnO$_6$ octahedra,
as easily understood by the explicit expressions of $Q_{1{\bf i}}$,
$Q_{2{\bf i}}$, and $Q_{3{\bf i}}$ presented before.
A direct and simple way to consider such cooperative effect is
to determine the oxygen positions $X_{1{\bf i}}$, $X_{4{\bf i}}$,
$Y_{2{\bf i}}$, $Y_{5{\bf i}}$, $Z_{3{\bf i}}$, and $Z_{6{\bf i}}$,
by using computational method such as
the Monte Carlo simulations and numerical relaxation techniques.
To reduce the burden on the numerical calculations,
for instance, 
the displacements of oxygen ions are assumed to be along the bond
direction between nearest neighboring manganese ions.
In other words, the displacement of the oxygen ion perpendicular to 
the Mn-Mn bond, i.e., the buckling mode, is usually ignored.
As shown later, even in this simplified treatment, several
interesting results have been obtained for the spin, charge, and
orbital ordering in manganites.

Rewriting Eqs.~(\ref{eq:q2}), (\ref{eq:q3}), and (\ref{eq:q1})
in terms of the displacement of oxygens from the equilibrium
positions, we express the distortions as
\begin{eqnarray}
\label{Eq:coopQ}
  \begin{array}{l}
  Q_{1{\bf i}}=
  (\Delta_{\bf xi}+\Delta_{\bf yi}+\Delta_{\bf zi})/\sqrt{3},\\
  Q_{2{\bf i}}=
  (\Delta_{\bf xi}-\Delta_{\bf yi})/\sqrt{2},\\
  Q_{3{\bf i}}=
  (2\Delta_{\bf zi}-\Delta_{\bf xi}-\Delta_{\bf yi})/\sqrt{6},
  \end{array}
\end{eqnarray}
where $\Delta_{\bf ai}$ is given by
\begin{equation}
  \Delta_{\bf ai}=u_{\bf i}^{\bf a}-u_{\bf i-a}^{\bf a},
\end{equation}
with $u_{\bf i}^{\bf a}$ being the displacement of oxygen ion at site
${\bf i}$ from the equilibrium position along the ${\bf a}$-axis.
In the $cooperative$ treatment, the $\{u\}$'s are directly optimized
in the numerical calculations \cite{Hotta1999,Allen1999}.
On the other hand, in the $non$-$cooperative$ calculations,
$\{Q\}$'s are treated instead of the $\{u\}$'s.
In the simulations, variables are taken as $\{Q\}$'s or $\{u\}$'s,
depending on the treatments of lattice distortion.

Finally, we briefly comment on the effect of macroscopic distortion.
In the above treatment, we assume the cubic symmetry
for the equilibrium positions of oxygens, but in actuality,
the crystal structure is frequently deviated from the cubic symmetry.
Although we cannot determine the stable crystal structure
in the present treatment, the effect of macroscopic distortions
is included as offset values for the distortions, which are given by
\begin{eqnarray}
  \begin{array}{l}
  Q_{1}^{(0)}=
  (\delta L_{\bf x}+\delta L_{\bf y}+\delta L_{\bf z})/\sqrt{3},\\
  Q_{2}^{(0)}=
  (\delta L_{\bf x}-\delta L_{\bf y})/\sqrt{2},\\
  Q_{3}^{(0)}=
  (2\delta L_{\bf z}-\delta L_{\bf x}-\delta L_{\bf y})/\sqrt{6},
  \end{array}
\end{eqnarray}
where $\delta L_{\bf a}$=$L_{\bf a}-L$,
the non-distorted lattice constants are $L_{\bf a}$,
and $L$=$(L_{\bf x}+L_{\bf y}+L_{\bf z})/3$.
Note that $L_{\bf a}$ is determined from the experimental results.
By adding $Q_{\mu}^{(0)}$ in the right-hand side of $Q_{\mu{\bf i}}$
in Eq.~(\ref{Eq:coopQ}), it is possible to consider the effect
of the deviation from the cubic symmetry

\subsection{Electron hopping}

In previous subsections, we have discussed the local electron state
due to CEF potential and Coulomb interaction.
We have also considered an additional electron-phonon coupling term.
Since the possible local terms have been completed, let us consider
the intersite effect in the next step.
In actual materials, there are several kinds of intersite effects.
Among them, we consider the electron hopping between adjacent
$d$ electron orbitals.
For transition metal oxides, such a hopping process occurs
through the oxygen ion, but in the formalism, it is enough to
consider the direct hopping $d$-electron orbitals.
Effect of oxygen will be discussed later.

\begin{table*}
\begin{center}
    \begin{tabular}{|c|c|c|}
    \hline    $\gamma$ & $\gamma'$ & $E_{\gamma,\gamma'}$ \\ \hline
    $x^2-y^2$  & $x^2-y^2$ &
    $(3/4)(\ell^2-m^2)^2(dd\sigma)$+$[\ell^2+m^2-(\ell^2-m^2)^2](dd\pi)$
    \\ \hline
    $3z^2-r^2$ & $3z^2-r^2$ &
    $[n^2-(\ell^2+m^2)/2]^2(dd\sigma)$+$3n^2(\ell^2+m^2)(dd\pi)$
    \\ \hline
    $x^2-y^2$  & $3z^2-r^2$ &
    $(\sqrt{3}/2)(\ell^2-m^2)[n^2-(\ell^2+m^2)/2](dd\sigma)$
    +$\sqrt{3}n^2(m^2-\ell^2)(dd\pi)$ \\ \hline
    $xy$ & $xy$ &
    $3\ell^2m^2(dd\sigma)$+$(\ell^2+m^2-4\ell^2m^2)(dd\pi)$
    \\ \hline
    $xy$ & $yz$ &
    $3\ell m^2 n (dd\sigma)$+$\ell n(1-4m^2)(dd\pi)$ \\ \hline
    $xy$ & $zx$ &
    $3\ell^2 m n (dd\sigma)$+$mn(1-4\ell^2)(dd\pi)$ \\ \hline
    \end{tabular}
\end{center}
\caption{
Expressions for hopping amplitude between $d$ orbitals.
As for details, see Ref.~\cite{Slater1954}.
We use a direction cosine as $(\ell, m, n)$ for the direction
of hopping from $\gamma$ to $\gamma'$ orbitals.
We show the contributions from the hoppings through
$\sigma$- and $\pi$-bonds.}
\end{table*}

Fortunately, the hopping amplitudes have been already evaluated
and tabulated for any combination of electron orbitals.
We can simply consult with the table of Slater-Koster integral
\cite{Slater1954}, which is the general scheme for the overlap
integral between adjacent electron orbitals.
The kinetic term for $e_{\rm g}$ or $t_{\rm 2g}$ electrons is
expressed in a common form as
\begin{equation}
  H_{\rm kin}^{\Gamma} =
  \sum_{{\bf i},{\bf a},\sigma} \sum_{\gamma,\gamma'}
  t^{\bf a}_{\gamma \gamma'} d_{{\bf i} \gamma \sigma}^{\dag}
  d_{{\bf i+a} \gamma' \sigma},
\end{equation}
where $\Gamma$ is the irreducible representation,
${\bf a}$ is the vector connecting nearest-neighbor sites,
and $t^{\bf a}_{\gamma \gamma'}$ is the nearest-neighbor
hopping amplitude between $\gamma$- and $\gamma'$-orbitals
along the ${\bf a}$-direction.

For the cubic lattice composed of transition metal ion,
we consider the three axis directions, ${\bf x}$=(1,0,0),
${\bf y}$=(0,1,0), and ${\bf z}$=(0,0,1).
Then, $t_{\gamma\gamma'}^{\bf a}$ for $e_{\rm g}$ orbital
is given in a 2$\times$2 matrix form as
\begin{equation}
 \label{eg-hop-x}
  t^{\bf x}=t_1
  \left(
  \begin{array}{cc}
        3/4 & -\sqrt{3}/4 \\
        -\sqrt{3}/4 & 1/4 \\
  \end{array}
  \right),
\end{equation}
for $x$-direction,
\begin{equation}
 \label{eg-hop-y}
  t^{\bf y}=t_1
  \left(
   \begin{array}{cc}
        3/4 & \sqrt{3}/4  \\
        \sqrt{3}/4 & 1/4 \\
   \end{array}
  \right),
\end{equation}
for $y$-direction, and
\begin{equation}
 \label{eg-hop-z}
  t^{\bf z}=t_1
  \left(
   \begin{array}{cc}
        0 & 0  \\
        0 & 1 \\
   \end{array}
  \right),
\end{equation}
for $z$-direction, where $t_1$=$(dd\sigma)$ and $(dd\sigma)$
is one of Slater integrals \cite{Slater1954}.
It should be noted that the signs in the hopping amplitudes between
different orbitals are different between $x$- and $y$-directions.

On the other hand, for $t_{\rm 2g}$ orbitals, we obtain
the hopping amplitudes in a 3$\times$3 matrix form as
\begin{equation}
  t^{\bf x}=t_2
  \left(
  \begin{array}{ccc}
        1 & 0 & 0 \\
        0 & 0 & 0 \\
        0 & 0 & 1 \\
  \end{array}
  \right),
\end{equation}
for $x$-direction,
\begin{equation}
  t^{\bf y}=t_2
  \left(
  \begin{array}{ccc}
        1 & 0 & 0 \\
        0 & 1 & 0 \\
        0 & 0 & 0 \\
  \end{array}
  \right),
\end{equation}
for $y$-direction, and
\begin{equation}
  t^{\bf z}=t_2
  \left(
  \begin{array}{ccc}
        0 & 0 & 0 \\
        0 & 1 & 0 \\
        0 & 0 & 1 \\
  \end{array}
  \right),
\end{equation}
for $z$-direction, where $t_2$=$(dd\pi)$.

Now let us consider explicitly the effect of oxygen orbitals.
Since oxygen ion is placed in the middle of transition metal ions
in the cubic perovskite, the main hopping process should occur
via oxygen $2p$ orbitals.
Thus, the $d$-electron hopping can be expressed by $(pd\sigma)$
or $(pd\pi)$, which is 
the overlap integral between $d$ and $p$ orbitals, divided by
the energy difference between $d$ and $p$ orbitals.
It is possible to calculate
the $d$-electron hopping via oxygen $2p$ orbitals,
by consulting again the Slater-Koster table for the overlap
integral between $d$ and $p$ orbitals.
However, due to the symmetry argument, we easily understand that
the form of hopping amplitude is invariant, after the redefinition
of $t_1$=$-(dp\sigma)^2/(\varepsilon_p-\varepsilon_d)$
and $t_2$=$-(dp\pi)^2/(\varepsilon_p-\varepsilon_d)$
for $e_{\rm g}$ and $t_{\rm 2g}$ electron cases, respectively,
where $\varepsilon_d$ and $\varepsilon_p$ are the energy levels
for $d$ and $p$ electrons, respectively.

We have considered the nearest neighbor hopping,
but in actuality, it is necessary to consider
higher neighbors in order to reproduce the Fermi surface
observed in the experiments such as de Haas-van Alphen
measurements.
However, there is no essential difficulty for the consideration
of higher neighbors by consulting the Slater-Koster table.

\subsection{Intersite interaction term}

In the previous subsections, we have considered the local
term and kinetic motion of $d$ electrons.
Of course, due to the combination of these terms,
intersite interaction terms effectively appear.
In particular, in the strong-coupling limit,
orbital-dependent superexchange terms can be obtained,
leading to the complex magnetic structure with orbital ordering.
Such effects are considered automatically, as long as we consider
the problem within the electronic model.

However, in the model for the high-spin state of $d^4$ electron
configuration, it is necessary to explicitly
add an extra term between localized $t_{\rm 2g}$ spin.
As explained above, due to the Hund's rule coupling between
$e_{\rm g}$ and $t_{\rm 2g}$ interaction, we can easily understand
that $e_{\rm g}$ electrons can move smoothly without any energy loss,
if spins of $e_{\rm g}$ and $t_{\rm 2g}$ are in the same direction.
Namely, in order to gain the kinetic energy of $e_{\rm g}$,
$t_{\rm 2g}$ spins array ferromagnetically.
This is a simple explanation for the appearance of
ferromagnetism in the orbital degenerate system,
in particular, in manganites.
However, there should exist AF intersite coupling between
neighboring $t_{\rm 2g}$ spins due to the superexchange
interaction, given by
\begin{equation}
  H_{\rm inter-site}^{\rm AF}=J_{\rm AF}
  \sum_{\langle {\bf i},{\bf j} \rangle}
  {\bf S}_{\bf i} \cdot {\bf S}_{\bf j},
\end{equation}
where $\langle {\bf i},{\bf j} \rangle$ denotes the pair of
nearest neighbor sites and $J_{\rm AF}$ is the AF
coupling between neighboring $t_{\rm 2g}$ spins.
This term should be added to the model for manganites.
As we will explain later, the competition between FM
tendency to gain kinetic energy and AF energy to
gain magnetic energy is a possible source of complex
magnetic structure in manganites.

In addition to the effective coupling among localized spins,
sometimes we consider another intersite effect originating from
long-range Coulomb interaction, even if it is screened in
actual materials.
In order to include such effect in an effective manner,
we also add
\begin{equation}
  H_{\rm inter-site}^{\rm C}=
  V \sum_{\langle {\bf i},{\bf j} \rangle}
  n_{\bf i} n_{\bf j},
\end{equation}
where $V$ denotes the nearest-neighbor Coulomb interaction.
Since $V$ has a tendency to stabilize the charge ordering,
there occurs competition between striped spin order
and bipartite charge ordering.
This is another source of complex spin-charge-orbital
ordering.

\subsection{Summary}

We have completed the preparation of all components
for the model Hamiltonian.
As a short summary of this section, we show three types of model
Hamiltonians due to the appropriate combination of several terms.

For the system with active $e_{\rm g}$ or $t_{\rm 2g}$ orbital
degree of freedom, we can consider the orbital degenerate
model, expressed in a common form as
\begin{equation}
  \label{Model}
   H_{\Gamma} = H_{\rm kin}^{\Gamma}
   + H_{\rm el-el}^{\Gamma} + H_{\rm el-ph}^{\Gamma}
   + H^{\rm C}_{\rm inter-site},
\end{equation}
where $\Gamma$ denotes $e_{\rm g}$ or $t_{\rm 2g}$.
Note that the inter-site Coulomb interaction term is explicitly
added here, but depending on the nature of the problem,
this term may be ignored.

For the case of $d^4$(H) in which both $e_{\rm g}$ and $t_{\rm 2g}$
orbitals are included, we can define the following model:
\begin{equation}
   H_{\rm DE} = H_{\rm kin}^{e_{\rm g}} + H_{\rm el-el}^{\rm DE}
     + H_{\rm el-ph}^{e_{\rm g}} + H^{\rm AF}_{\rm inter-site}
     + H^{\rm C}_{\rm inter-site}.
\end{equation}
This expression is believed to define an appropriate starting
model for manganites, but unfortunately, it is quite difficult to
solve such a Hamiltonian.
In order to investigate further the properties of manganites,
further simplifications are needed.
This point will be discussed in detail in the next section.

%
%
\section{Orbital physics in manganites}

In the complicated phase diagram for manganites,
there appear so many magnetic phases.
A key issue to understand such richness is the competition between
itinerant and localized tendencies contained in manganites.
As mentioned in the model construction, 
$e_{\rm g}$ electrons can gain the kinetic energy when the background
$t_{\rm 2g}$ spins array ferromagnetically, leading to a metallic
FM phase in manganites.
On the other hand, in order to gain magnetic energy between localized
$t_{\rm 2g}$ spins, there occurs AF phase with
insulating tendency.
In one word, the competition between FM metallic and AF insulating
phases is the origin of complex phase diagram of manganites.

As we will review very briefly in the next subsection,
the metallic tendency has been discussed in the concept of double exchange
for a long time, and the essential point has been considered to
be well understood.
However, the tendency toward insulating phase has not been satisfactorily
understood, mainly due to the complexity of multi-degrees of freedom
such as spin, charge, and orbital.
In particular, ``orbital ordering'' is the remarkable feature,
characteristic to manganites with active $e_{\rm g}$ orbital.
In this section, spin, charge, and orbital structure for the typical
hole doping in the phase diagram of manganites is focused by stressing
the importance of orbital ordering.

\subsection{Concept of double-exchange}

Since the historical review of theoretical and experimental works
on manganites have been found in some articles and textbooks,
we simply refer literatures such as
Refs.~\cite{Dagotto2001,Hotta2004a,Tokura2000b,Dagotto2002}.
However, it is instructive to mention here the meaning of
``double exchange'' (DE), which is important basic concept for
manganites, by following the previous review article \cite{Dagotto2001}.

\begin{figure}[t]
\begin{center}
\includegraphics[width=0.5\textwidth]{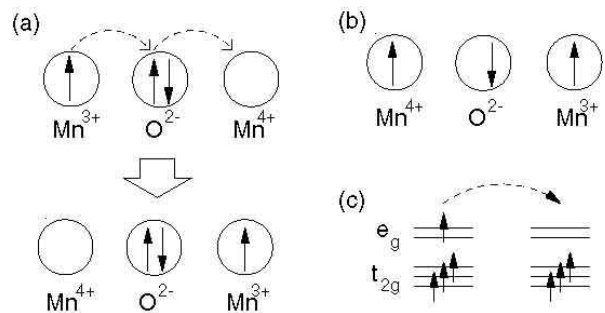}
\end{center}
\caption{%
(a) Sketch of the double exchange mechanism
which involves two Mn ions and one O ion.
(b) The intermediate state of the process (a).
(c) The mobility of $e_{\rm g}$-electrons improves if the localized
spins are polarized.
}
\label{fig3-1}
\end{figure}

In the earth stage of the research on manganites,
it was the task to clarify the qualitative aspects of
the experimentally discovered relation between
transport and magnetic properties, namely the increase in conductivity
upon the polarization of the spins.
The concept of ``double exchange'' was proposed by Zener \cite{Zener1951}
as a way to allow for charge to move in manganites by the generation
of a spin polarized state.
The DE process has been historically explained in two somewhat
different ways.
Originally, Zener considered the explicit movement of electrons,
as shown in Fig.~\ref{fig3-1}(a).
This process is schematically written as
$\rm Mn^{3+}_{1\uparrow}$$\rm O_{2\uparrow,3\downarrow}$$\rm Mn^{4+}$
$\rightarrow$
$\rm Mn^{4+}$$\rm O_{1\uparrow,3\downarrow}$$\rm Mn^{3+}_{2\uparrow}$
\cite{Cieplak},
where 1, 2, and 3 label electrons that belong either to the oxygen
between manganese, or to the $e_{\rm g}$-level of the Mn-ions.
In this process, there are two simultaneous motions
involving electron 2 moving from the oxygen to the
right Mn-ion, and electron 1 from the left Mn-ion to the oxygen.
This is the origin of the name of ``double exchange''.

The second way to visualize DE processes was presented in detail by
Anderson and Hasegawa \cite{Anderson-Hasegawa},
and it involves a second-order process in which the two states
described above go from one to the other using an
intermediate state $\rm Mn^{3+}_{1\uparrow}$$\rm O_{3\downarrow}$
$\rm Mn^{3+}_{2\uparrow}$, as shown in Fig.~\ref{fig3-1}(b).
In this context, the effective hopping for the electron to move from
one Mn-site to the next is proportional to the square of the hopping
involving the $p$-oxygen and $d$-manganese orbitals ($t_{\rm pd}$).
Following Anderson and Hasegawa, let us consider a two-site problem,
in which one itinerant electron with hopping amplitude $t$ between
sites 1 and 2 is coupled with localized spin ${\bf S}$ at each site.
The coupling is assumed to be ferromagnetic and
the magnitude is defined as $J$.
For $t$=0, the local ground state is labelled by $S$+$1/2$ with
the energy of $-JS$ per site, while the excited state is specified as
$S$$-$$1/2$ with the energy of $J(S+1)$ per site.
For $t$$\ne$0 with $t$$\ll$$J$, as observed in manganese ions,
the effective hopping amplitude $t_{\rm eff}$ is
in proportion to the overlap integral between
$|S+1/2,S,S_0 \rangle$ and $|S,S+1/2,S_0\rangle$,
where $|S_1,S_2,S_0 \rangle$ indicates the state
with spin $S_i$ at site $i$ and total spin $S_0$.
Note that $S_0$ is composed of two localized spins
and one itinerant spin.
The overlap integral is evaluated as
$\langle S+1/2,S,S_0 |S,S+1/2,S_0 \rangle$
=$2(S+1)W(S,1/2,S_0,S;S+1/2,S+1/2)$,
where $W$ is the so-called Racah coefficient
for the combination of three spins.
By using some relations for the Racah coefficient \cite{Biedenharn},
we obtain $|t_{\rm eff}/t|$=$(S_0+1/2)/(2S+1)$.

For $S$=1/2, we can intuitively understand the meaning of
the reduction factor.
When the two localized spins are parallel,
we obtain $S_0$=3/2 due to the large $J$
and the reduction factor is unity.
This is understood by the fact that
the local triplet formed by the large $J$
is not destroyed in the course of electron hopping motion.
However, when the two localized spins are anti-parallel,
$S_0$=1/2 and the reduction factor is 1/2.
In this case, it is necessary to reconstruct
the local triplet state after the hopping of electron,
leading to the effective reduction of the electron hopping.
For the case of large $S$, the localized spins are considered classical.
When we define an angle $\theta$ between nearest-neighbor ones,
the above overlap integral is easily evaluated by rotating
the itinerant electron basis so as to be parallel to the localized
spin direction.
Then, we obtain $t_{\rm eff}$=$t\cos (\theta/2)$,
as shown by Anderson and Hasegawa.
If $\theta$=0 the hopping is the largest, while if $\theta$=$\pi$,
corresponding to an AF background, then the hopping cancels.

Note that the oxygen linking the Mn-ions is crucial to understand the
origin of the word ``double'' in this process.
Nevertheless, the majority of the theoretical work carried out in the
context of manganites simply forgets the
presence of the oxygen and uses a manganese-only Hamiltonian.
It is interesting to observe that FM states appear in this
context even without the oxygen.
It is clear that the electrons simply need a polarized background to
improve their kinetic energy, as shown in Fig.~\ref{fig3-1}(c),
in similar ways as the Nagaoka phase is
generated in the one-band Hubbard model at large $U/t$. 
This tendency to optimize the kinetic energy is at work in a variety
of models and the term double-exchange appears unnecessary.
However, in spite of this fact, it has become customary to refer to
virtually any FM phase found in manganese models as 
``DE induced'' or ``DE generated'', forgetting the historical origin
of the term.
In this review, a similar convention will be followed, namely the
credit for the appearance of FM phases will be given to the DE
mechanism, although a more general and simple kinetic-energy
optimization is certainly at work.
This is also the reason why we have used the abbreviation ``DE''
in the model for manganites.

Early theoretical work on manganites carried out by Goodenough
\cite{Goodenough1955}
explained many of the features observed in the neutron scattering
experiments on $\rm La_{1-x}Ca_xMnO_3$
by Wollan and Koehler \cite{Wollan1955}, notably the appearance
of the A-type AF phase at x=0 and the CE-type phase at x=0.5.
The approach of Goodenough was based on the notion of
``semicovalent exchange''.
Analyzing the various possibilities for the orbital directions and
generalizing to the case where Mn$^{4+}$ ions are also present,
Goodenough arrived to the A- and CE-type phases of manganites
very early in the theoretical study of these compounds.
In this line of reasoning, note that the Coulomb interactions are
important to generate Hund-like rules and the oxygen is also important 
to produce the covalent bonds.
The lattice distortions are also quite relevant in deciding
which of the many possible states minimizes the energy.
However, it is interesting to observe that in more recent
theoretical work described below in this review, both the A- and
CE-type phases can be generated without the explicit appearance of
oxygens in the models and also without including long-range Coulomb
terms.

\subsection{Topological aspects of orbital ordering}

In Sec.~2, we have already set
the model Hamiltonian for manganites $H_{\rm DE}$.
Before proceeding to the exhibition of the theoretical results
on this model, we explain the essential point of
manganites from a purely theoretical viewpoint.
We believe that it is an important issue
to establish a simple principle
penetrating the complicated phenomena.

\subsubsection{A simplified model}

In order to extract the essential feature of manganites,
let us define a minimal model, since $H_{\rm DE}$ is still
a complex model.
We note that there exist two important ingredients which should
be kept even in the minimal model.
One is the existence of orbital degree of freedom and another is
a competition between FM metallic and AF insulating tendencies.
In order to minimize the model by keeping these two issues,
first we simply ignore the interaction terms,
$H_{\rm el-el}^{e_{\rm g}}$, $H_{\rm el-ph}^{e_{\rm g}}$,
and $H^{\rm C}_{\rm inter-site}$.
Second, we take an infinite limit of the Hund's rule coupling,
$J_{\rm H}$, between $e_{\rm g}$ electron and $t_{\rm 2g}$ spins.
Then, the direction of $e_{\rm g}$-electron spin perfectly follows
that of $t_{\rm 2g}$ spin.
We can suppress the spin index, if we define the spinless
operator at each site in which the spin direction is fixed
as that of $t_{\rm 2g}$ spin at each site.
Namely, the model is virtually expressed by using
spinless operators with orbital degree of freedom.
Then, we obtain a simplified double-exchange model as
\begin{equation}
 \label{minimalH}
 H = -\sum_{{\bf i,a}\gamma,\gamma'} D_{\bf i,i+a}
   t_{\gamma,\gamma'}^{\bf a}
   c_{{\bf i}\gamma}^{\dag}c_{{\bf i+a}\gamma'}
   + J_{\rm AF} \sum_{\langle {\bf i,j} \rangle}
   S_{\bf i}^z S_{\bf j}^z,
\end{equation}
where $c_{{\bf i}\gamma}$ is the annihilation operator for spinless
$d$ electron in the $\gamma$ orbital at site ${\bf i}$,
the hopping amplitudes are given in
Eqs.~(\ref{eg-hop-x}), (\ref{eg-hop-y}), and
(\ref{eg-hop-z}), where $t_1$=$-t_0$
with $t_0$=$(pd\sigma)^2/(\varepsilon_p-\varepsilon_d)$,
and $D_{\bf i,j}$ is the so-called double-exchange factor,
given by
\begin{eqnarray}
  D_{\bf i,j} = \cos (\theta_{\bf i}/2)\cos (\theta_{\bf j}/2)
  + \sin (\theta_{\bf i}/2)\sin (\theta_{\bf j}/2)
  e^{-i(\phi_{\bf i}-\phi_{\bf j})}.
\end{eqnarray}
Here $\theta_{\bf i}$ and $\phi_{\bf i}$ denote
the polar and azimuthal angles of $t_{\rm 2g}$ spin
at site ${\bf i}$, respectively.
This factor expresses the change of hopping amplitude due to the
difference in angles between $t_{\rm 2g}$-spins
at sites ${\bf i}$ and ${\bf j}$.
Note that the effective hopping in this case is a complex number
(Berry phase), contrary to the real number widely used in a large
number of previous investigations.
As for the effect of the Berry phase in the case of the
one-orbital DE model, readers should refer Ref.~\cite{Muller-Hartmann}.

Furthermore, when we assume the Ising $t_{\rm 2g}$ spins,
the double-exchange factor $D_{\bf i,j}$ denotes 0 or 1
depending on the spin configuration.
Namely, $D_{\bf i,j}$=1 for FM $t_{\rm 2g}$ spin pair
at sites ${\bf i}$ and ${\bf j}$,
while $D_{\bf i,j}$=0 for AF pair.
One may feel that the model seems to be oversimplified,
but as will see later, we can grasp an essential point of
complex phases of manganites,
since the competition between FM metallic
and AF insulating natures is correctly
included in this model.

\subsubsection{Band-insulating state}

First let us consider two limiting situations,
$J_{\rm AF}$=0 and $J_{\rm AF} \gg t_0$.
For simplicity, a two-dimensional (2D) square lattice is taken here.
For $J_{\rm AF}$=0, it is easy to understand the appearance of
FM metallic phase, since the Hamiltonian includes only the
kinetic term of electrons in this limit.
On the other hand, for $J_{\rm AF} \gg t_0$, AF insulating phase
should appear due to the magnetic energy gain.
Then, what happens for intermediate value of $J_{\rm AF}$?
Naively thinking, it is possible to consider the mixture of
FM and AF regions, in order to gain both kinetic and magnetic
energies.
For instance, we can consider the C-type AF phase,
in which one-dimensional (1D) FM chains are antiferromagnetically
coupled with each other.
However, there is no special reason to fix the shape of the FM region
as straight 1D chain.
It may be possible to have zigzag shape for the FM region.

\begin{figure}[t]
\begin{center}
\includegraphics[width=0.7\textwidth]{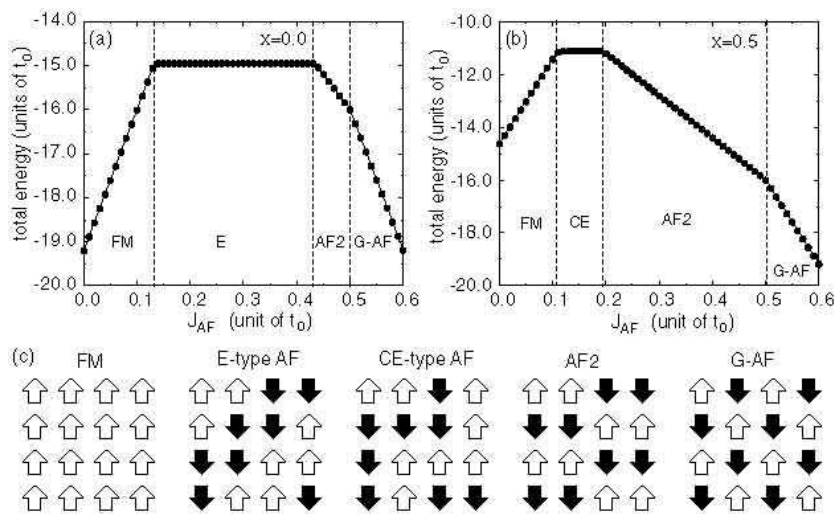}
\end{center}
\caption{%
Ground-state energy vs. $J_{\rm AF}$ for (a) x=0.0 
and (b) x=0.5 of a simplified model.
The unit cell is taken as a $4 \times 4$ lattice.
(c) Schematic views for spin structure of
FM, E-type AF, CE-type AF, AF2, and G-AF phases.
}
\label{fig4}
\end{figure}

In order to determine the optimal shape of the FM region,
we perform simple simulations for magnetic structure
in the 2D lattice \cite{Hotta2000}.
Since the model does not include explicit many-body
interaction among $e_{\rm g}$ electrons,
we can solve the problem on the periodic lattice
composed of an appropriate unit cell such as 
$4 \times 4$ cluster.
Namely, we prepare all possible patterns for $t_{\rm 2g}$
spin configuration and evaluate the ground state
energy on each magnetic structure by changing the value
of $J_{\rm AF}$.
Note that the calculations have been done on the momentum
space by introducing the Bloch phase factor at the boundary.

The results for x=0.0 and 0.5 are shown in
Figs.~\ref{fig4}(a) and \ref{fig4}(b),
where x denotes hole doping from the case of quarter filling,
i.e., one electron per site
with two orbitals.
Note that x=0.0 corresponds to the case in which all sites
are occupied by trivalent manganese ions, while x=0.5
indicates the situation in which half of the sites are
occupied by Mn$^{4+}$.
For x=0.0, there are four regions in Fig.~\ref{fig4}(a).
As mentioned above, we obtain 2D FM phase at $J_{\rm AF}$=0,
while for large $J_{\rm AF}$, the G-AF phase appears, as expected.
In the intermediate region, we observe the striped spin phase,
characterized by zigzag FM path, as shown in Fig.~\ref{fig4}(c).
This structure is just the E-type AF phase.
Note that the magnetic energy is cancelled in the E-type phase,
since the numbers of FM and AF bonds are exactly equal.
We also find a narrow window for another AF phase,
called ``AF2'', between the E-type and G-type AF phases.
In the present context, the appearance of this phase is not
important.

For x=0.5, as shown in Fig.~\ref{fig4}(b),
the situation looks similar with the case of x=0.0,
but among four magnetic phases,
the spin structure in the region labeled by ``CE'' differs from
that of the E-type phase at x=0.0.
Namely, the period of the zigzag is different from
that for x=0.0, as shown in Fig.~\ref{fig4}(c).
We emphasize here that the spin structure in the intermediate
coupling is composed of a bundle of spin FM chains, each
with the zigzag geometry, and with AF interchain coupling.
This is just the CE-type AF phase.
Note that the magnetic energy is also cancelled,
since the numbers of FM and AF bonds are exactly equal
in the CE-type phase.

Now we consider the reason why such complicated structure appears.
For the time being, let us discuss what happens if the zigzag
geometry of CE- or E-type is assumed,
and how it compares with a straight line.
A straightforward way is to calculate the energy band for
the $e_{\rm g}$ electron system on the zigzag 1D path,
since $e_{\rm g}$ electrons can move only in the FM region
in the simplified model due to the double exchange factor.
First we consider the C-type AF phase characterized by straight
1D path, even though it is not the ground state.
As shown in Sec.~2, the hopping amplitudes of $e_{\rm g}$ electrons
depend on the direction, but as easily checked by a diagonalization
of $2 \times 2$ matrix, due to the cubic symmetry,
the energy band does not depend on the chain direction.
Then, by choosing the hopping direction along the $z$-axis,
we easily obtain $E_{k}$=0 and $-2t_0\cos k$,
since there is non-zero hopping amplitude only between
$3z^2-r^2$ orbitals along this direction.

\begin{figure}[t]
\begin{center}
\includegraphics[width=0.5\textwidth]{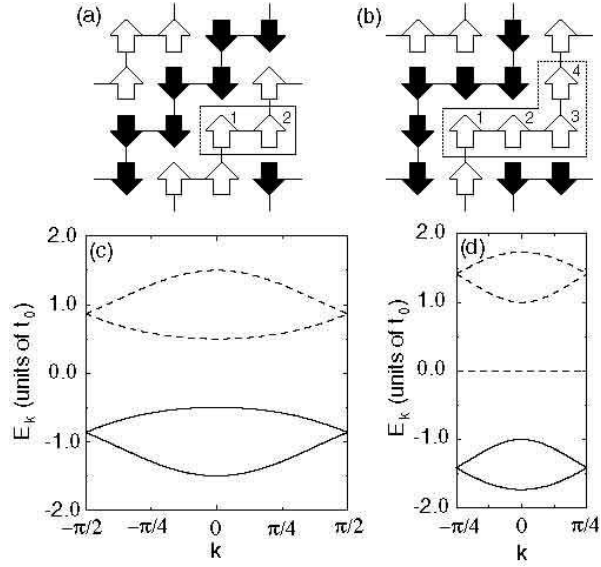}
\end{center}
\caption{%
Definition of the unit cell for (a) E-type and (b)
CE-type 1D zigzag chain.
Momentum is defined along the zigzag chain.
Energy band structure for (c) E-type and (d) CE-type
structure.
Solid and broken curves denote occupied and unoccupied bands,
respectively.
Note that the flat band in (d) has four-fold degeneracy.
}
\label{fig5}
\end{figure}

To solve the present one-body problem in the zigzag 1D case,
unit cells are defined as shown in
Figs.~\ref{fig5}(a) and \ref{fig5}(b),
in which the hopping amplitudes change with a period of
two or four lattice spacings for E- and CE-type structure,
respectively, since the hopping direction changes as
$\{ \cdots, x, y, x, y, \cdots \}$
and $\{ \cdots, x, x, y, y, \cdots \}$ along the zigzag chain,
with $t_{\mu \nu}^{\bf x}$=$-t_{\mu \nu}^{\bf y}$ for $\mu \ne \nu$
according to the values of the hopping amplitudes discussed before.
This difference in sign, i.e., the phase change, is essential for this
problem.
To make this point clear, it is useful to transform the
spinless $e_{\rm g}$-electron operators by using a unitary matrix as
\begin{equation}
  \left(
    \begin{array}{l}
      \alpha_{\bf i} \\
      \beta_{\bf i}
    \end{array}
  \right)
  = {1 \over \sqrt{2}} \left(
   \begin{array}{cc}
   1 &  i \\
   1 & -i 
    \end{array}
  \right)
  \left(
    \begin{array}{l}
      c_{{\bf i}{\rm a}} \\
      c_{{\bf i}{\rm b}}
    \end{array}
  \right).
\end{equation}
After simple algebra, $H_{\rm kin}$ is rewritten as
\begin{eqnarray}
  H_{\rm kin} = -t_0/2
  \sum_{\bf i,a} (\alpha_{\bf i}^{\dag}\alpha_{\bf i+a}
  +\beta_{\bf i}^{\dag} \beta_{\bf i+a}
  +e^{i \phi_{\bf a}}\alpha_{\bf i}^{\dag} \beta_{\bf i+a}
  +e^{-i \phi_{\bf a}}\beta_{\bf i}^{\dag} \alpha_{\bf i+a}),
\end{eqnarray}
where the phase $\phi_{\bf a}$ depends only on the hopping direction,
and it is given by $\phi_{\bf x}$=$-\phi$,  $\phi_{\bf y}$=$\phi$,
and $\phi_{\bf z}$=$\pi$, with $\phi$=$\pi/3$.
Note that the $e_{\rm g}$-electron picks up a phase change
when it moves between different neighboring orbitals.
In this expression, the effect of the change of the local phase
is correctly included in the Hamiltonian.

Here we solve the problem in the E-type structure.
Details of the solutions of the CE-type structure have found
in the previous review \cite{Dagotto2001}.
To introduce the momentum $k$ along the zigzag chain,
the Bloch's phase $e^{ \pm i k}$ is added to the hopping
term between adjacent sites.
Then, the problem is reduced to finding the eigenvalues
of a $4 \times 4$ matrix, given by
\begin{equation}
  {\hat h} \left(
    \begin{array}{c}
        \psi_{\alpha 1} \\ 
        \psi_{\beta 1} \\ 
        \psi_{\alpha 2} \\ 
        \psi_{\beta 2}
    \end{array}
  \right)
  =\varepsilon_{k}
  \left(
    \begin{array}{c}
        \psi_{\alpha 1} \\ 
        \psi_{\beta 1} \\ 
        \psi_{\alpha 2} \\ 
        \psi_{\beta 2}
    \end{array}
  \right),
\end{equation}
where $\psi_{\alpha j}$ and $\psi_{\beta j}$ are the basis function
for $\alpha$- and $\beta$-electrons at the $j$-site of the unit cell,
respectively, and the Hamiltonian matrix ${\hat h}$ is given by
\begin{equation}
{\hat h}=
  -t_0 \left(
    \begin{array}{cccc}
        0 & 0 & \cos k & \cos (k-\phi) \\
        0 & 0 & \cos (k+\phi) & \cos k \\
        \cos k & \cos (k+\phi) & 0 & 0 \\
        \cos (k-\phi) & \cos k & 0 & 0
  \end{array}
  \right).
\end{equation}
To make the Hamiltonian in a block diagonalized form,
we introduce two kinds of bonding and antibonding states as
$\Phi^{\pm}_1$=
($\psi_{\alpha 1} \pm \psi_{\beta 1} \pm
\psi_{\alpha 2}+\psi_{\beta 2}$)/2
and
$\Phi^{\pm}_2$=
($\psi_{\alpha 1} \mp \psi_{\beta 1} \pm
\psi_{\alpha 2}-\psi_{\beta 2}$)/2.
For the states 1 and 2, we obtain two eigen equations as
\begin{equation}
  -(t_0/2)
  \left(
    \begin{array}{cc}
      3 \cos k  &  \sqrt{3} \sin k \\
      \sqrt{3} \sin k & -\cos k \\
    \end{array}
  \right)
  \left(
    \begin{array}{c}
      \Phi_1^{+} \\
      \Phi_1^{-}
    \end{array}
  \right)
  =E_{k}^{(1)}
  \left(
    \begin{array}{c}
      \Phi_1^{+} \\
      \Phi_1^{-} 
    \end{array}
  \right),
\end{equation}
and
\begin{equation}
  -(t_0/2)
  \left(
    \begin{array}{cc}
      \cos k  &  -\sqrt{3} \sin k \\
      -\sqrt{3} \sin k & -3\cos k \\
    \end{array}
  \right)
  \left(
    \begin{array}{c}
      \Phi_2^{+} \\
      \Phi_2^{-}
    \end{array}
  \right)
  =E_{k}^{(2)}
  \left(
    \begin{array}{c}
      \Phi_2^{+} \\
      \Phi_2^{-} 
    \end{array}
  \right),
\end{equation}
respectively.
We can easily diagonalize each $2 \times 2$ matrix and obtain
\begin{equation}
 E_k^{(1)} = (t_0/2)(-\cos k \pm \sqrt{\cos ^2 k + 3})
\end{equation}
and
\begin{equation}
 E_k^{(2)} = (t_0/2)( \cos k \pm \sqrt{\cos ^2 k + 3}).
\end{equation}

For the case of CE-type AF phase, it is necessary to solve
eigen value problem in the $8 \times 8$ matrix.
Readers interested in the detail of the calculations
can refer the review article by Dagotto, Hotta, and Moreo
\cite{Dagotto2001}.
Here we show only the results.
Eight eigen energies have been obtained as
\begin{equation}
  E_k = 0, ~\pm t_0\sqrt{2+\cos(2k)}, ~
  \pm t_0\sqrt{2-\cos(2k)},
\end{equation}
where the flat band $\varepsilon_{\bf k}$=0 has four-fold degeneracy.

The band structures for E-type and CE-type zigzag path
are shown in Figs.~\ref{fig5}(c) and \ref{fig5}(d).
Note that the solid curves denote the occupied bands.
The most remarkable feature is that the system is {\it band-insulating},
with a bandgap of value $t_0$ for x=0.0 \cite{Hotta2003a,Hotta2003b}
and x=0.5 \cite{Hotta2000}.
Remark that the band-insulator state at x=0.5 was independently
obtained in Refs.~\cite{vandenBrink1999,Solovyev1999,Solovyev2001}.
This band insulating state, without any explicit potential
among the electrons moving along the zigzag chains,
is caused by the phase difference between
$t^{\bf x}_{\mu \nu}$ and $t^{\bf y}_{\mu \nu}$.
Since $t_0$ is at least of the order of 1000K, this band-insulating
state is considered to be very robust.

Intuitively, the band-insulating state of the zigzag AF structure
originates in the presence of
a standing-wave state due to the interference between
two traveling waves running along the $x$- and $y$-directions.
In this interference picture, the nodes of the wavefunction can exist
on the ``corner" of the zigzag structure, and the probability
amplitude becomes larger in the ``straight" segment of the path.
Thus, even a weak potential can produce the charge and orbital
ordering based on this band-insulating phase.
In fact, if some potential is included into such an insulating phase,
the system maintains its ``insulating'' properties,
and a spatial modulation in the charge density appears.
For x=0.5, since the charge density should be increased in the
sites 2 and 4 in Fig.~\ref{fig5}(b), it is easy to understand that
checker-board type charge ordering occurs, when some potential is
further included in the model.

Concerning the orbital shape, we should point out that
the $e_{\rm g}$-electron orbital is maximally
polarized along the transfer direction
in order to gain the kinetic energy.
This effect may be called an ``orbital double-exchange''
in the sense that the orbitals align along the axis direction 
to make the transfer of the electron smooth, similarly
as the FM alignment of $t_{2g}$ spins in the usual DE mechanism.
Namely, on the straight-line part in the $x$- and $y$-direction, 
the orbital is polarized as $3x^2-r^2$
and $3y^2-r^2$, respectively.

On the other hand, for the case of x=0, there is
no straight-line part in the E-type zigzag structure.
In this case, rather the cooperative Jahn-Teller effect is
essential to determined the orbital shape, since each site
is Jahn-Teller active.
We cannot determine the orbital ordering pattern within
the present simple discussion.
It is necessary to consider a more realistic Hamiltonian.
The actual orbital ordering will be discussed later.

\subsubsection{Topological number}

In the previous subsection, we have emphasized that the shape
of zigzag path plays an important role for the determination
of the CE- and E-type AF phases.
Now let us consider the quantity to specify the zigzag shape.
For the purpose, we include the coupling between $e_{\rm g}$
electrons and JT phonons.
Effect of Coulomb interaction will be discussed later.
The model is given by
\begin{eqnarray}
  \label{H-JT}
  H &=& -\sum_{{\bf ia}\gamma \gamma'}
  D_{\bf i,i+a} t^{\bf a}_{\gamma \gamma'} 
  c_{{\bf i} \gamma}^{\dag}c_{{\bf i+a} \gamma'}
  + J_{\rm AF} \sum_{\langle {\bf i,j} \rangle}
  {\bf S}_{\bf i} \cdot {\bf S}_{\bf j} \nonumber \\
  &+& E_{\rm JT} \sum_{\bf i}
  [2(q_{2{\bf i}} \tau_{x{\bf i}}
  + q_{3{\bf i}} \tau_{z{\bf i}}) 
  +q_{2{\bf i}}^2 +q_{3{\bf i}}^2],
\end{eqnarray}
where $E_{\rm JT}$ is the static Jahn-Teller energy,
$\tau_{x\bf i}$=
$c_{{\bf i}a}^{\dag}c_{{\bf i}b}$+$c_{{\bf i}b}^{\dag}c_{{\bf i}a}$,
and 
$\tau_{z\bf i}$=
$c_{{\bf i}a}^{\dag}c_{{\bf i}a}$$-$$c_{{\bf i}b}^{\dag}c_{{\bf i}b}$.
By using the phase $\xi_{\bf i}$ defined in Eq.~(\ref{eq:polar}),
which is the angle to specify the Jahn-Teller distortion,
it is convenient to transform
$c_{{\bf i}{\rm a}}$ and $c_{{\bf i}{\rm b}}$ into the
``phase-dressed" operators, 
${\tilde c}_{{\bf i}{\rm a}}$ and ${\tilde c}_{{\bf i}{\rm b}}$, as
\begin{equation}
 \label{trans1}
  \left(
    \begin{array}{l}
      {\tilde c}_{{\bf i}a} \\
      {\tilde c}_{{\bf i}b}
    \end{array}
  \right)
  ={\hat R}(\xi_{\bf i})
  \left(
    \begin{array}{l}
      c_{{\bf i}a} \\
      c_{{\bf i}b}
    \end{array}
  \right).
\end{equation}
where the unitary matrix ${\hat R}(\xi_{\bf i})$ is given by
\begin{equation}
  \label{trans2}
  {\hat R}(\xi_{\bf i}) = e^{i \xi_{\bf i}/2} 
  \left(
    \begin{array}{cc}
      \cos [\xi_{\bf i}/2] & \sin [\xi_{\bf i}/2] \\
      -\sin [\xi_{\bf i}/2] & \cos [\xi_{\bf i}/2] \\
    \end{array}
  \right).
\end{equation}
Note that if $\xi_{\bf i}$ is increased by $2\pi$,
the SU(2) matrix itself changes its sign.
This is the same phenomenon found in spin wavefunction,
since in general, spinor is isomorphic to the wavefunction
of a two-level system.
In 1950's, Longuet-Higgins et al. have pointed out that
the electron wavefunction of Jahn-Teller molecule
changes its sign for the $2\pi$-rotation in the parameter space
in the adiabatic approximation \cite{Longuet-Higgins}.
Note that the total wavefunction, given by the product of
electron and phonon wavefunctions, is always single-valued,
since the phonon part also changes its sign for the $2\pi$-rotation.
The spinor-like wavefunction for the electron part appears
due to the adiabatic approximation for the JT system.
It has been also mentioned that the change of sign is regarded
as the effect of the Berry phase \cite{Ham}.

In order to keep the transformation unchanged upon a $2\pi$-rotation
in $\xi_{\bf i}$, a phase factor $e^{i\xi_{\bf i}/2}$ is needed.
This is also regarded as the effect from the phonon wavefunction.
In the expression for the ground state of the single JT molecule,
namely, the single-site problem discussed before,
this phase factor has not been considered explicitly,
since the electron does not hop around from site to site and
the phases do not correlate with each other.
It was enough to pay attention to the fact that
the electron wavefunction at a single site is double-valued.
However, in the JT crystal in which $e_{\rm g}$ electrons move
in the periodic array of the JT centers,
the addition of this phase factor is useful to take into
account the effect of the Berry phase
arising from the circular motion of $e_{\rm g}$-electrons
around the JT center \cite{Koizumi1998a,Koizumi1998b}.
It could be possible to carry out the calculation
without including explicitly this phase factor,
but in that case,
it is necessary to pay due attention to the inclusion of the
effect of the Berry phase.
The qualitative importance of this effect will be explained later.

Note also that the phase $\xi_{\bf i}$ determines the electron orbital
set at each site.
In the previous section, the single-site problem was discussed
and the ground-state at site ${\bf i}$ was found to be
\begin{equation}
  |``{\rm b}" \rangle =
  [-\sin (\xi_{\bf i}/2) d_{{\bf i}{\rm a}\sigma}^{\dag}
  + \cos (\xi_{\bf i}/2) d_{{\bf i}{\rm b}\sigma}^{\dag}] |0 \rangle,
\end{equation}
which is referred to as the ``b"-orbital, namely
the combination with the lowest-energy at a given site.
The excited-state or ``a"-orbital is simply obtained by requesting it to be 
orthogonal to ``b" as
\begin{equation}
  |``{\rm a}" \rangle =
  [\cos (\xi_{\bf i}/2) d_{{\bf i}{\rm a}\sigma}^{\dag}
  + \sin (\xi_{\bf i}/2) d_{{\bf i}{\rm b}\sigma}^{\dag}] |0 \rangle.
\end{equation}
For instance, at $\xi_{\bf i}$=$2\pi/3$, ``a'' and ``b'' denote the 
$d_{y^2-z^2}$- and $d_{3x^2-r^2}$-orbitals, respectively.
It should be noted here that $d_{3x^2-r^2}$ and $d_{3y^2-r^2}$ 
never appear as the local orbital set.
Sometimes those were treated as an orthogonal orbital
set to reproduce the experimental results, but such a treatment is
an approximation, since the orbital ordering is $not$ due to the
simple alternation of two arbitrary kinds of orbitals.

Using the above described transformations, the model Eq.~(\ref{H-JT})
is rewritten after some algebra as 
\begin{eqnarray}
  {\tilde H} &=& -\sum_{{\bf ia}\gamma \gamma'}
  D_{\bf i,i+a} 
  {\tilde t}^{\bf a}_{\gamma \gamma'}({\bf i},{\bf i+a})
  {\tilde c}_{{\bf i} \gamma}^{\dag}{\tilde c}_{{\bf i+a} \gamma'}
  + J_{\rm AF} \sum_{\langle {\bf i,j} \rangle}
  {\bf S}_{\bf i} \cdot {\bf S}_{\bf j} \nonumber \\
  &+& E_{\rm JT} \sum_{\bf i}
  [2q_{\bf i}({\tilde n}_{{\bf i}{\rm a}}-{\tilde n}_{{\bf i}{\rm b}})
  + q_{{\bf i}}^2],
\end{eqnarray}
where
${\tilde n}_{{\bf i} \gamma}$=
${\tilde c}_{{\bf i} \gamma}^{\dag}{\tilde c}_{{\bf i} \gamma}$
and the hopping amplitude is changed as
\begin{eqnarray}
  {\tilde t}^{\bf a}_{\gamma \gamma'}({\bf i},{\bf j})
  ={\hat R}(\xi_{\bf i})_{\gamma \eta} t^{\bf a}_{\eta \eta'} 
   {\hat R}^{-1}(\xi_{\bf j})_{\eta'\gamma'}.
\end{eqnarray}

In order to characterize the shape of the zigzag path,
it is useful to formulate the change of the phase.
For the purpose, we use the concept of ``the Berry-phase connection''
and define ``the winding number'' by following Ref.~\cite{Hotta1998}.
The phase-dressed operator, ${\tilde c}_{{\bf i} \gamma}$, naturally
introduces a connection form, the Berry-phase connection, as
\cite{Wilczek}
\begin{eqnarray}
 A d{\bf r}
&\equiv&
\left(
\begin{array}{cc}
i \langle 0|
\tilde{c}_{{\bf r}{\rm a}} \nabla \tilde{c}_{{\bf r}{\rm a}}^{\dagger}
|0 \rangle \cdot d{\bf r}
&
i \langle 0|
\tilde{c}_{{\bf r}{\rm a}} \nabla \tilde{c}_{{\bf r}{\rm b}}^{\dagger}
|0 \rangle \cdot d{\bf r}
\\
i \langle 0|
\tilde{c}_{{\bf r}{\rm b}} \nabla \tilde{c}_{{\bf r}{\rm a}}^{\dagger}
|0 \rangle \cdot d{\bf r}
&
i \langle 0|
\tilde{c}_{{\bf r}{\rm b}} \nabla \tilde{c}_{{\bf r}{\rm b}}^{\dagger}
|0 \rangle \cdot d{\bf r}
\end{array}
\right)
\nonumber
\\
&=& { 1 \over 2}
\left( 
\begin{array}{cc}
\nabla \xi \cdot d{\bf r}   & -i\nabla \xi \cdot d{\bf r} \\
i\nabla \xi \cdot d{\bf r} & \nabla \xi \cdot d{\bf r} 
\end{array}
\right),
\end{eqnarray}
where $|0\rangle$ denotes the vacuum state.

Now we consider a 2D sheet of the JT crystal.
In two space dimensions, there is a topologically conserved current
for an arbitrary vector field ${\bf v}$ as
\begin{eqnarray}
 \label{eq:current}
 j^{\mu}={1 \over {2\pi}} \sum_{\nu, \lambda}
 \epsilon^{\mu \nu \lambda}\partial_{\nu} v_{\lambda},
\end{eqnarray}
where ${\bf j}$ is the current and
$\epsilon^{\mu \nu \lambda}$ is the antisymmetric Levi-Civita tensor.
This current obeys the continuity equation, expressed as 
\begin{eqnarray}
  \sum_{\mu}\partial_{\mu} j^{\mu}=0.
\end{eqnarray}
Therefore, if $j^{1}$ and $j^{2}$ are zero
at the boundary of a closed surface $S$, the quantity $Q$ (Chern number),
defined by
\begin{eqnarray}
  Q \equiv \int_{S} d^2 {\bf r}\ j^{0}
\end{eqnarray}
is conserved, where indices 1 and 2 indicate the space-components
and $0$ indicates the time-component of a vector, respectively.

For time-independent solutions with which we are now concerned,
we have $\partial_{0}j^{0}=0$.
Thus, $Q$ is conserved for an arbitrary surface $S$.
Substituting ${\bf a}={\rm Tr}(A)=\nabla \xi $ for ${\bf v}$
into Eq.~(\ref{eq:current}), we obtain the topologically 
conserved quantity, or ``the winding number'' as
\begin{eqnarray}
  w&=&\int_{S} d^2 {\bf r} \ j^{0}=
  {1 \over {2\pi}} \int_{S} d^2 {\bf r}
  (\partial_1 a_2-\partial_2 a_1) \nonumber
  \\
  &=&{1 \over {2 \pi}} \oint_{C} d{\bf r}\cdot {\bf a}
  ={1 \over {2 \pi}} \oint_{C} d{\bf r}\cdot \nabla \xi
  =m,
\end{eqnarray}
where $m$ is an integer representing the number for
the twisting-around of the JT distortions
along a path $C$ enclosing $S$.
Because of the conserved nature, we will used the winding number
$w$ to label a state hereafter.

In the system with zigzag AF structure,
$C$ is considered to be a closed loop for the 1D path in the periodic
boundary condition.
In this case, the winding number $W$ may be decomposed into two terms
as $w$=$w_{\rm g}$+$w_{\rm t}$.
The former, $w_{\rm g}$, is the geometric term, which becomes $0$ ($1$) 
corresponding to the periodic (anti-periodic) boundary condition 
in the $e_{\rm g}$-electron wavefunction.
The discussion on the kinetic energy leads us to conclude that 
the state with $w_{\rm g}$=$0$ has lower energy than that with 
$w_{\rm g}$=$1$ for x $\ge 1/2$,
in agreement with the two-site analysis \cite{Takada1999}.
Thus, $w_{\rm g}$ is taken as zero hereafter.

\begin{figure}[t]
\begin{center}
\includegraphics[width=0.4\textwidth]{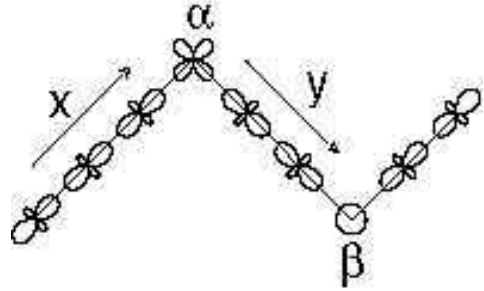}
\end{center}
\caption{%
A typical building block for a zigzag 1D FM path
for an $e_{\rm g}$ electron with orbital ordering.
}
\label{fig6}
\end{figure}

In order to understand that only the number of corners
in the zigzag path, $N_{\rm v}$,
determines the topological term $w_{\rm t}$,
let us consider the transfer of a single $e_{\rm g}$ electron
along the path shown in Fig.~\ref{fig6}.
As mentioned above, on the straight-line part in the $x$-($y$-)direction,
the phase is fixed at $\theta_{\bf x}$=$2\pi/3$
($\theta_{\bf y}$=$4\pi/3$), because the $e_{\rm g}$-electron orbital is
polarized along the transfer direction.
Thus, $w_{\rm t}$ does not change on the straight-line part of the path.
However, when the electron passes the vertex $\alpha$ ($\beta$),
the phase changes from $\theta_{\bf x}$ to $\theta_{\bf y}$
($\theta_{\bf y}$ to $\theta_{\bf x}$),
indicating that the electron picks up a phase change of $2\pi/3$ ($4\pi/3$).
Since these two vertices appear in pairs, $w_{\rm t}$(=$w$) is evaluated
as $w_{\rm t}$=$(N_{\rm v}/2)(2\pi/3 \!+\! 4\pi/3)/(2\pi)$=$N_{\rm v}/2$.
The phases at the vertices are assigned as an average of the phases 
sandwiching those vertices, $\theta_{\alpha}$=$\pi$ and
$\theta_{\beta}$=$0$, to keep $w_{\rm g}$ invariant.
Then, the phases are determined at all the sites,
once $\theta_{\bf x}$, $\theta_{\bf y}$, $\theta_{\alpha}$, 
and $\theta_{\beta}$ are known.

Finally, we note that the problem in the zigzag one-dimensional chain
provided us with a typical example to better understand
the importance of the additional
factor $e^{i\xi_{\bf i}/2}$ in front of the $2 \times 2$ SU(2) unitary
matrix to generate the phase dressed operator at each site.
As clearly shown above, the ``a'' and ``b'' orbitals should be chosen
as ``a''=$y^2-z^2$ and ``b''=$3x^2-r^2$ at site 2, and ``a''=$z^2-x^2$
and ``b''=$3y^2-r^2$ at site 4, respectively. 
Namely, $\xi_2$=$2\pi/3$ and $\xi_4$=$4\pi/3$. 
The reason for these choices of $\xi_{\bf i}$ is easily understood
due to the fact that the orbital tends to polarize along the hopping
direction to maximize the overlap. 
Thus, to make the Hamiltonian simple, it is useful to fix the orbitals
at sites 2 and 4 as $\xi_2$=$2\pi/3$ and $\xi_4$=$4\pi/3$.
Here, the phase factor $e^{i\xi_{\bf i}/2}$ in the basis function is
essential to reproduce exactly the same solution as obtained in the
discussion above.
As already mentioned, in a single-site problem, this phase factor can
be neglected, since it provides only an additional phase to the whole
wave function.
However, if the $e_{\rm g}$-electron starts moving from site to site, 
the accumulation of the phase difference between adjacent sites
does not lead just to an additional phase factor to the whole wave
function.
In fact, if this additional phase is accidentally neglected,
the band structure will shift in momentum space as
$k$$\rightarrow$$k+\pi$,
indicating that the minimum of the lowest-energy band is not located 
at $k$=0, but at $k$=$\pi$, as already pointed out by 
Koizumi et al. \cite{Koizumi1998b}.
Of course, this can be removed by the redefinition of $k$ by including 
``the crystal momentum'', but it is not necessary to redefine $k$,
if the local phase factors are correctly included in the problem.

\subsubsection{Stability of zigzag structure}

Concerning the stability of the zigzag AF phase,
from the results in Fig.~\ref{fig5},
we can understand the following two points:
(i) The zigzag structure of E- and CE-type
shows the lowest energy compared
with other zigzag paths with the same periodicity
and compared with the straight 1D path.
(ii) The energy of the zigzag AF phase becomes lower than
that of the FM or other AF phases in the parameter region
of $J_{\rm AF}$ around $J_{\rm AF}$$\approx$0.1$t_0$.
However, in the calculation, we have just assumed the
periodicity of four lattice spacing, but it is unclear whether
such period actually produces the $global$ ground state or not.
As emphasized above, it is true that the zigzag 1D FM chain
has a large band-gap, but this fact does not guarantee
that this band-insulating phase is the lowest-energy state.
In other words, we cannot exclude the possibility that
the zigzag structure with another periodicity becomes
the global ground state.

Unfortunately, it is quite difficult to carry out the direct
comparison among the energies for all possible states,
since there are infinite
possibilities for the combinations of hopping directions.
Instead, to mimic the periodic change of the phase $\phi_{\bf a}$ in
the hopping process, let us imagine a virtual situation in which a JT
distortion occurs in the 1D $e_{\rm g}$-electron system, 
by following Koizumi at al. \cite{Koizumi1998a}.
To focus on the effect of the local phase, it is assumed that the
amplitude of the JT distortion $q_i$ is independent of the site index,
i.e., $q_i=q$, and only the phase $\xi_i$ is changed periodically.
For simplicity, the phase is uniformly twisted with the period of $M$
lattice spacings, namely, $\xi_j$=$j$$\times$$(2\pi)/M$
for 1$\le$$j$$\le$$M$.
Since the periodic change of the hopping direction is mimicked by the
phase change of the JT distortion, $t_{\mu \nu}^{\bf a}$ is simply
taken as the unit matrix $t_0 \delta_{\mu \nu}$ to avoid the
double-counting of the effect of the phase change.
If the potential amplitude is written as $v$=$2qE_{\rm JT}$, the
Hamiltonian for the present situation is given by
\begin{eqnarray}
  H &=& -t_0 \sum_{\langle i,j \rangle}
  (c_{i{\rm a}}^{\dag} c_{j{\rm a}}
  + c_{i{\rm b}}^{\dag} c_{j{\rm b}} + {\rm h.c.}) \nonumber \\
  &+& v \sum_i [\sin \xi_i (c_{i{\rm a}}^{\dag} c_{i{\rm b}}
  + c_{i{\rm b}}^{\dag} c_{i{\rm a}})
  + \cos \xi_i (c_{i{\rm a}}^{\dag} c_{i{\rm a}}
  - c_{i{\rm b}}^{\dag} c_{i{\rm b}})],
\end{eqnarray}
where the spinless $e_{\rm g}$-electron operator is used,
since the 1D FM chain is considered here.
The potential term for the JT distortion is ignored,
since it provides only a constant energy shift in this case.
By using the transformation Eqs.~(\ref{trans1}) and (\ref{trans2}),
the Hamiltonian is rewritten as 
\begin{eqnarray}
  H &=& -t_0 \sum_{\langle i,j \rangle} [ e^{i(\xi_i-\xi_j)/2}
  \{ \cos \frac{\xi_i-\xi_j}{2}
  ({\tilde c}_{i{\rm a}}^{\dag} {\tilde c}_{j{\rm a}}+
  {\tilde c}_{i{\rm b}}^{\dag} {\tilde c}_{j{\rm b}}) \nonumber \\
  &+& \sin \frac{\xi_i-\xi_j}{2}
  ({\tilde c}_{i{\rm a}}^{\dag} {\tilde c}_{j{\rm b}}-
  {\tilde c}_{i{\rm b}}^{\dag} {\tilde c}_{j{\rm a}}) \}
  + {\rm h.c.} ]
  +v \sum_i ({\tilde c}_{i{\rm a}}^{\dag} {\tilde c}_{i{\rm a}}
  -{\tilde c}_{i{\rm b}}^{\dag} {\tilde c}_{i{\rm b}}).
\end{eqnarray}
The Hamiltonian in momentum space is obtained by the Fourier transform as
\begin{eqnarray}
  H &=& \sum_{k} [\varepsilon_{k}
   \cos (\pi/M) ({\tilde c}_{k{\rm a}}^{\dag} {\tilde c}_{k{\rm a}} 
  +{\tilde c}_{k{\rm b}}^{\dag} {\tilde c}_{k{\rm b}})
  +i s_{k} \sin (\pi/M)
  ({\tilde c}_{k{\rm a}}^{\dag} {\tilde c}_{k{\rm b}}-
  {\tilde c}_{k{\rm b}}^{\dag} {\tilde c}_{k{\rm a}})] \nonumber \\
  &+& v \sum_{k} ({\tilde c}_{k{\rm a}}^{\dag} {\tilde c}_{k{\rm a}}
  -{\tilde c}_{k{\rm b}}^{\dag} {\tilde c}_{k{\rm b}}),
\end{eqnarray}
where $\varepsilon_k$=$-2t_0 \cos k$, $s_k$=$-2t_0 \sin k$,
and the periodic boundary condition is imposed. 
Note that in this expression, $k$ is the generalized quasimomentum,
redefined as $k-\pi/M \rightarrow k$, to incorporate the additional
phase $\pi/M$ which appears to arise from a fictitious magnetic field
(see Ref.~\cite{Koizumi1998b}).
The eigenenergies are easily obtained by diagonalization as
\begin{eqnarray}
  E_{k}^{\pm} &=& \varepsilon_k \cos (\pi/M)
  \pm \sqrt{v^2 + s_k^2 \sin ^2 (\pi/M)} \nonumber \\
  &=& (1/2) [
  \varepsilon_{k+\pi/M}+\varepsilon_{k-\pi/M}
  \pm \sqrt{4v^2 +(\varepsilon_{k+\pi/M}-\varepsilon_{k-\pi/M})^2}~].
\end{eqnarray}
Since this is just the coupling of two bands,
$\varepsilon_{k+\pi/M}$ and $\varepsilon_{k-\pi/M}$,
it is easily understood that the energy gain due to the opening of
the bandgap is the best for the filling of $n$=$2/M$,
where $n$=1$-$x with doping x.
In other words, when the periodicity $M$ is equal to $2/n$,
the energy becomes the lowest among the states considered here
with several possible periods.
Although this is just a proof in an idealized special situation,
it is believed that it captures the essence of the problem.

Here the effect of the local phase factor $e^{i \xi_{\bf i}/2}$
should be again noted.
If this factor is dropped, the phase $\pi/M$ due to the fictitious
magnetic field disappears and the eigenenergies are given by the
coupling of $\varepsilon_{k+\pi+\pi/M}$ and $\varepsilon_{k+\pi-\pi/M}$,
which has been also checked by the computational calculation.
This ``$\pi$" shift in momentum space appears at the boundary, 
modifying the periodic boundary condition to anti-periodic,
even if there is no intention to use anti-periodic boundary condition.
Of course, this is avoidable when the momentum $k$ is redefined as
$k+\pi \rightarrow k$, as pointed out in Ref.~\cite{Koizumi1998b}.
However, it is natural that the results for periodic boundary condition
are obtained in the calculation using periodic boundary condition.
Thus, also from this technical viewpoint, it is recommended that the
phase factor $e^{i \xi_{\bf i}/2}$ is added for the local rotation in
the orbital space.

\subsubsection{Summary}

In summary, at x=0.5, the CE-type AF phase can be stabilized 
even without the Coulomb and/or the JT phononic interactions,
only with large Hund and finite $J_{\rm AF}$ couplings.
We have also pointed out the appearance of E-type phase
due to the same mechanism.
Of course, as we will see in the next subsection,
Coulombic and JT phononic interactions are needed to reproduce
the charge and orbital ordering.
However, as already mentioned in the above discussion, because of
the special geometry of the one-dimensional zigzag FM chain,
for instance, at x=0.5, it is easy to imagine that
the checkerboard type charge-ordering and $(3x^2-r^2/3y^2-r^2)$
orbital-ordering pattern will be stabilized.
Furthermore, the charge confinement in the straight segment,
i.e., sites 2 and 4 in Fig.~\ref{fig5}(b),
will naturally lead to charge stacking along the $z$-axis,
with stability caused by the special geometry of the zigzag
structure.
Thus, the complex spin-charge-orbital structure for half-doped
manganites can be understood intuitively simply from the viewpoint of
its band-insulating nature.

\subsection{Spin, charge, and orbital ordering}

Now we review the theoretical results on spin, charge, and orbital ordering
in undoped and doped manganites on the basis of realistic models.
The Hamiltonian mainly used here is two-orbital double exchange model
strongly coupled with Jahn-Teller phonons, explicitly given by
\begin{equation}
 H=H_{\rm kin}^{e_{\rm g}}+H_{\rm Hund}+H_{\rm inter-site}^{\rm AF}
  + H_{\rm el-ph}^{e_{\rm g}}.
\end{equation}
In the infinite limit for $J_{\rm H}$,
we can further simplify the model into the following form.
\begin{eqnarray}
  H_{\rm JT} &=& -\sum_{{\bf ia}\gamma \gamma'}
  D_{\bf i,i+a} t^{\bf a}_{\gamma \gamma'}
  c_{{\bf i} \gamma}^{\dag}c_{{\bf i+a} \gamma'}
  + J_{\rm AF} \sum_{\langle {\bf i,j} \rangle}
  {\bf S}_{\bf i} \cdot {\bf S}_{\bf j} \nonumber \\
  &+& E_{\rm JT} \sum_{\bf i}
  [2(q_{1{\bf i}} n_{\bf i}+
  q_{2{\bf i}} \tau_{x{\bf i}}
  + q_{3{\bf i}} \tau_{z{\bf i}}) 
  +\beta q_{1{\bf i}}^2 + q_{2{\bf i}}^2 +q_{3{\bf i}}^2],
\end{eqnarray}
where $n_{\bf i}$=
$\sum_{\gamma} c_{{\bf i} \gamma}^{\dag}c_{{\bf i} \gamma}$
and $\beta$=$k_{\rm br}/k_{\rm JT}$, the ratio
of spring constants of breathing and Jahn-Teller phonons.
Concerning the value of $\beta$,
from experimental results and band-calculation data for
the energy of breathing and Jahn-Teller modes \cite{Iliev},
it is estimated as $\beta \approx 2$ for manganites.
It is convenient to introduce non-dimensional electron-phonon
coupling constant $\lambda$ as
\begin{equation}
 \lambda=\sqrt{2E_{\rm JT}/t_0}=g/\sqrt{k_{\rm JT}t_0}.
\end{equation}
Here we simply drop the Coulomb interaction terms, but
the reason will be discussed in the next subsection.

This model is analyzed by using both the numerical techniques
(Monte Carlo simulation and relaxation method) and mean-field
approximation.
Note that in the numerical simulations,
depending on the non-cooperative and cooperative treatments,
the variables are angles $\theta_{\bf i}$ and $\phi_{\bf i}$
which specifies $t_{\rm 2g}$ spin directions
and coordinates $\{ q \}$ and oxygen positions $\{ u \}$,
respectively.
In any case, all variables are classical and thus,
there is no essential problems to perform almost exactly
the simulation, within the limit of the power of computers.
Recently, there has been an important progress in the
simulation for the electron systems coupled with
classical variables
\cite{Alonso2001c,Motome1999,Motome2000}.
In particular, Motome and Furukawa have developed
the efficient simulation technique
for the acceleration of the calculation and the increase of
the precision \cite{Motome1999,Motome2000}.

\subsubsection{Effect of Coulomb interaction}

Let us discuss briefly the effect of the Coulomb interaction.
Since we consider the strong Hund's rule interaction between
$e_{\rm g}$ electron and $t_{\rm 2g}$ localized spins,
$e_{\rm g}$ electron spins tend to array in a site
and thus, the effect of intra-orbital Coulomb interaction
is automatically suppressed.
However, inter-orbital Coulomb interaction  still works
between electrons with the same spin.
Here we explain the reason why we ignore the on-site Coulomb
interaction.
The effect of inter-site Coulomb interaction is discussed
in the stabilization mechanism of charge stacking in the x=0.5
CE-type structure.

In the spinless model, the inter-orbital Coulomb interaction term is
written by
\begin{equation}
  H_{\rm el-el} = U'\sum_{\bf i} n_{{\bf i}{\rm a}}n_{{\bf i}{\rm b}},
\end{equation}
where $n_{{\bf i}{\gamma}}$=
$c_{{\bf i} \gamma}^{\dag}c_{{\bf i} \gamma}$
and the present $U'$ means $U'-J$ in the standard notation for
the on-site Coulomb interaction.
We also consider the inter-site Coulomb interaction term, given by
\begin{equation}
  H_{\rm inter-site} = V \sum_{\langle {\bf i},{\bf j} \rangle}
  n_{\bf i}n_{\bf j},
\end{equation}
where
$n_{\bf i}$=$\sum_{\gamma}c^{\dag}_{{\bf i}\gamma}c_{{\bf i}\gamma}$.
In order to understand the ignorance of on-site Coulomb
interaction term, it is quite instructive to consider
the mean-field approximation.
As for the detail of the formulation,
readers can refer the original paper \cite{Hotta2000b}
and the previous review \cite{Dagotto2001}.
Here we show the result of the mean-field Hamiltonian.
\begin{eqnarray}
  H_{\rm MF} &=& -\sum_{{\bf ia}\gamma \gamma'}
  D_{\bf i.i+a} t^{\bf a}_{\gamma \gamma'} 
  c_{{\bf i} \gamma}^{\dag} c_{{\bf i+a} \gamma'}
  + J_{\rm AF} \sum_{\langle {\bf i,j} \rangle}
  {\bf S}_{\bf i} \cdot {\bf S}_{\bf j} \nonumber \\ 
  &+& {\tilde E_{\rm JT}} \sum_{\bf i} 
  [-2(\langle \tau_{x{\bf i}} \rangle \tau_{x{\bf i}}
  +\langle \tau_{z{\bf i}} \rangle \tau_{z{\bf i}})
  +\langle \tau_{x{\bf i}} \rangle^2
  + \langle \tau_{z{\bf i}} \rangle^2]    \nonumber \\ 
  &+& \sum_{\bf i} [({\tilde U'}/2) \langle n_{{\bf i}} \rangle
  + V \sum_{\bf a} \langle n_{{\bf i+a}} \rangle]
  (n_{{\bf i}}- \langle n_{{\bf i}} \rangle/2),
\end{eqnarray}
where $\langle \cdots \rangle$ denotes the average value.
The renormalized JT energy is given by
\begin{equation}
  {\tilde E_{\rm JT}}=E_{\rm JT}+U'/4,
\end{equation}
and the renormalized inter-orbital Coulomb interaction is 
expressed as 
\begin{equation}
  {\tilde U'}=U'- 4E_{\rm br},
\end{equation}
where $E_{\rm br}$=$g^2/(2k_{\rm br})$.
Physically, the former relation indicates that the JT energy is
effectively enhanced by $U'$.
Namely, the strong on-site Coulombic correlation plays the $same$ role
as that of the JT phonon, at least at the mean-field level,
indicating that it is not necessary to include $U'$ explicitly
in the models, as emphasized in Ref.~\cite{Hotta2000b}.
See also Ref.~\cite{Benedetti1999}.
The latter equation for ${\tilde U'}$ means that the one-site 
inter-orbital Coulomb interaction is effectively reduced by the
breathing-mode phonon, since the optical-mode phonon provides an
effective attraction between electrons.
The expected positive value of ${\tilde U'}$ indicates that 
$e_{\rm g}$ electrons dislike double occupancy at the site, since the 
energy loss is proportional to the average local electron number in
the mean-field argument.
Thus, to exploit the gain due to the static JT energy and avoid 
the loss due to the on-site repulsion, 
an $e_{\rm g}$ electron will singly occupy a given site.

\subsubsection{x=0}

First let us consider the mother material LaMnO$_3$ with one $e_{\rm g}$
electron per site.
This material has the insulating AF phase with A-type AF spin order,
in which $t_{\rm 2g}$ spins are ferromagnetic in the $a$-$b$ plane and
AF along the $c$-axis.
For the purpose to understand the appearance of A-AF phase,
it is enough to consider a $2 \times 2 \times 2$ cube as a minimal
cluster for undoped manganites.
Results in a larger size cluster will be discussed later.

\begin{figure}[t]
\begin{center}
\includegraphics[width=0.7\textwidth]{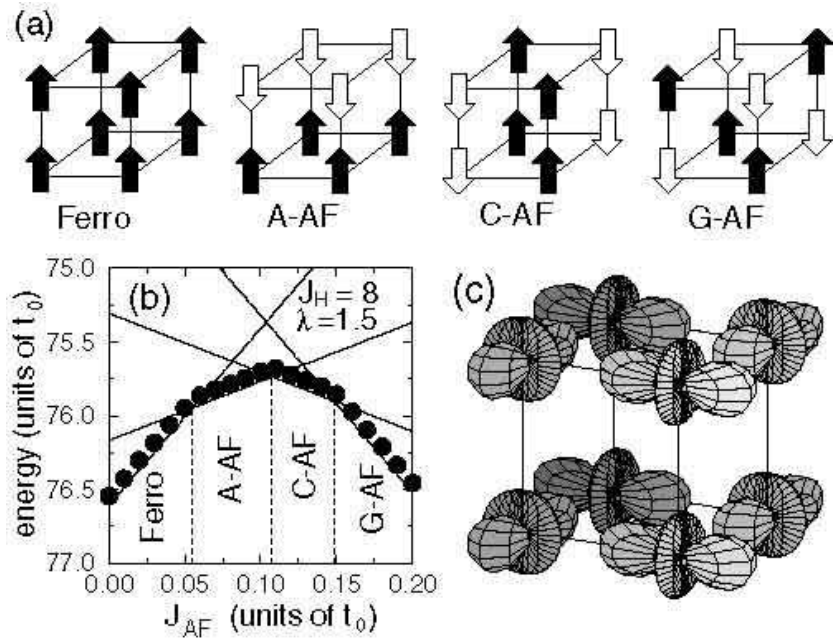}
\end{center}
\caption{%
(a) The four spin arrangements for Ferro, A-AF, C-AF, and G-AF.
(b) Total energy vs $J_{\rm AF}$ on a 2$^3$ cluster at low temperature
with $J_{\rm H}$=8$t_0$ and $\lambda$=1.5.
The results were obtained using Monte Carlo and relaxational
techniques, with excellent agreement among them.
(c) Orbital order corresponding to the A-type AF state.
For more details, the reader should consult Ref.~\cite{Hotta1999}.
}
\label{fig7}
\end{figure}

Recent investigations by Hotta et al. \cite{Hotta1999} have shown that,
in the context of the model with Jahn-Teller phonons, the important
ingredient to understand the A-type AF phase is $J_{\rm AF}$,
namely by increasing this coupling from 0.05 to larger values,
a transition from a FM to an A-type AF exists
The relevance of Jahn-Teller couplings at x=0.0
has also been remarked in Ref.~\cite{Capone2000}.
This can be visualized easily in Fig.~\ref{fig7},
where the energy vs. $J_{\rm AF}$ at fixed intermediate $\lambda$ and
$J_{\rm H}$ is shown.
Four regimes are identified: FM, A-AF, C-AF, and G-AF states that
are sketched also in that figure.
The reason is simple: As $J_{\rm AF}$ grows, the tendency toward spin
AF must grow, since this coupling favors such an order.
If $J_{\rm AF}$ is very large, then it is clear that a G-AF state must
be the one that lowers the energy, in agreement with the Monte Carlo
simulations.
If $J_{\rm AF}$ is small or zero, there is no reason why spin AF will
be favorable at intermediate $\lambda$ and the density under consideration,
and then the state is ferromagnetic to improve the electronic mobility.
It should be no surprise that at intermediate $J_{\rm AF}$, the
dominant state is intermediate between the two extremes, with A-type
and C-type antiferromagnetism becoming stable in intermediate regions
of parameter space.

It is interesting to note that similar results regarding the relevance 
of $J_{\rm AF}$ to stabilize the A-type order have been found by
Koshibae et al. \cite{Koshibae1997} in a model with Coulomb interactions.
An analogous conclusion was found by
Solovyev, Hamada, and Terakura \cite{Solovyev1996a,Solovyev1996b}
and Ishihara et al. \cite{Ishihara1997}.
Betouras and Fujimoto \cite{Betouras}, using bosonization techniques for the 
one-dimensional one-orbital model,
also emphasized the importance of $J_{\rm AF}$,
similarly as did by Yi, Yu, and Lee based on Monte Carlo studies
in two dimensions of the same model \cite{Yi2000}.
The overall conclusion is that there are clear analogies between
the strong Coulomb and strong Jahn-Teller coupling approaches.
Actually in the mean-field approximation,
it was shown by Hotta, Malvezzi, and Dagotto \cite{Hotta2000b}
that the influence of the Coulombic terms can be hidden
in simple redefinitions of the electron-phonon couplings
(see also Ref.~\cite{Benedetti1999}).
In our opinion, both approaches (Jahn-Teller and Coulomb) have strong
similarities and it is not surprising that basically the same physics
is obtained in both cases.
Actually, Fig.~2 of Maezono, Ishihara, and Nagaosa \cite{Maezono1998b}
showing the energy vs. $J_{\rm AF}$ in mean-field calculations of
the Coulombic Hamiltonian without phonons is very similar to our
Fig.~\ref{fig7}(b), aside from overall scales.
On the other hand, Mizokawa and Fujimori \cite{Mizokawa1995,Mizokawa1996}
states that the A-type AF is stabilized only when the Jahn-Teller distortion
is included, namely, the FM phase is stabilized in the purely Coulomb model,
based on the unrestricted Hartree-Fock calculation for the $d$-$p$ model.

The issue of what kind of orbital order is concomitant with A-type AF
order is an important matter.
This has been discussed at length by Hotta et al. \cite{Hotta1999},
and the final conclusion, after the introduction
of perturbations caused by the experimentally known difference in
lattice spacings between the three axes, is that the order shown in
Fig.~\ref{fig7}(c) minimizes the energy. 
This state has indeed been identified in resonant X-ray scattering
experiments \cite{Murakami1998a}, and it is quite remarkable that
such a complex pattern of spin and orbital degrees of freedom
indeed emerges from mean-field and computational studies.
Studies by van den Brink et al. \cite{Brink1999}
using purely Coulombic models arrived at similar conclusions. 
The orbital ordering has been also captured from the viewpoint
of orbital density wave state by Koizumi et al.
\cite{Koizumi1998a,Koizumi1998b}.
The similar discussion has been done recently by Efremov and Khomskii
\cite{Efremov}.

Why does the orbital order occur here?
In order to respond to this question,
it is quite instructive to consider the situation
perturbatively in the electron hopping.
A hopping matrix only connecting the same orbitals,
with hopping parameter $t$, is assumed for simplicity. 
The energy difference between $e_{\rm g}$ orbitals at a given site is 
$E_{\rm JT}$, which is a monotonous function of $\lambda$.
For simplicity, in the notation let us refer to orbital uniform
(staggered) as orbital ``FM'' (``AF'').
Case (a) corresponds to spin FM and orbital AF: In this
case when an electron moves from orbital a on the left to the same
orbital on the right, which is the only possible hopping by assumption, 
an energy of order $E_{\rm JT}$ is lost, but kinetic energy is gained.
As in any second order perturbative calculation the energy gain is then
proportional to $t^2/E_{\rm JT}$.
In case (b), both spin and orbital FM, the electrons do not move and
the energy gain is zero (again, the nondiagonal hoppings are assumed
negligible just for simplicity).
In case (c), the spin are AF but the orbitals are FM.
This is like a one orbital model and the gain in energy is
proportional to $t^2/(2J_{\rm H})$.
Finally, in case (d) with AF in spin and orbital, both Hund and
orbital splitting energies are lost in the intermediate state, and the
overall gain becomes proportional to $t^2/(2J_{\rm H} + E_{\rm JT})$.
As a consequence, if the Hund coupling is larger than $E_{\rm JT}$,
then case (a) is the best, as it occurs at intermediate $E_{\rm JT}$
values.
Then, the presence of orbital order can be easily understood from
a perturbative estimation,
quite similarly as done by Kugel and Khomskii
in their pioneering work on orbital order \cite{Kugel-Khomskii}.

\begin{figure}[t]
\begin{center}
\includegraphics[width=1.0\textwidth]{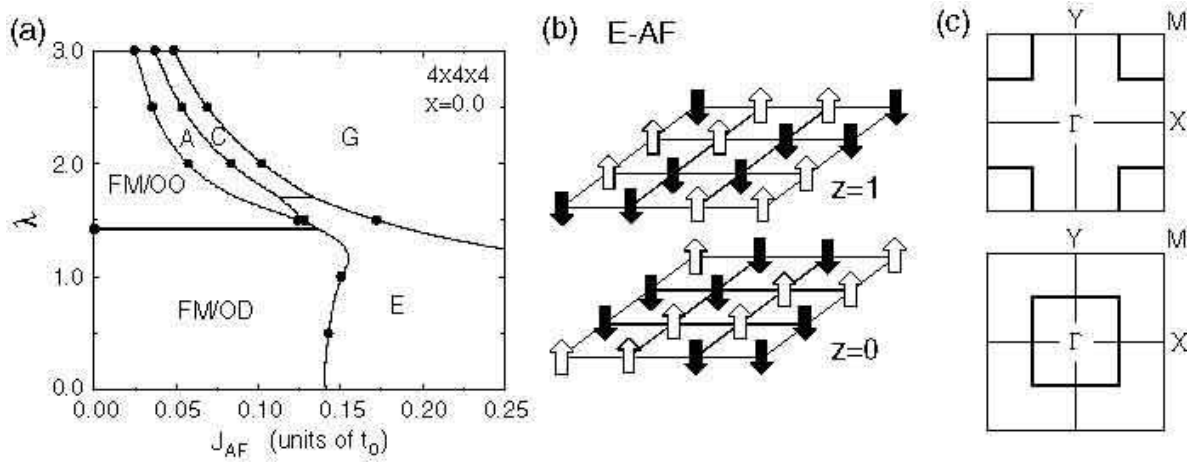}
\end{center}
\caption{%
(a) Ground-state phase diagram for undoped manganites by using the
4$\times$4$\times$4 lattice.
Solid curves denote the mean-field results, while solid circles
indicate the result for optimization.
(b) Spin structure for E-AF phase in three-dimensional environment.
(c) Fermi-surface lines of 2D $e_{\rm g}$ electron system
at x=0.0.
}
\label{fig8}
\end{figure}

Here readers may have a question in their mind.
Where is the E-type phase emphasized in the previous subsection?
In order to respond to this question, it is necessary to
treat a larger-size cluster.
In Fig.~\ref{fig8}(a), we show the phase diagram of undoped manganites
with the direct comparison between the mean-field and numerical results
in the 4$\times$4$\times$4 lattice \cite{Hotta2003b}.
Solid curves are depicted from the cooperative mean-field approximation,
while solid circles denote the result for optimization both
for $t_{\rm 2g}$ spin directions and oxygen positions.
The good agreements clearly indicate that the present mean-field
procedure works quite well for undoped manganites.

Now the phase diagram includes six phases,
but there is a clear separation around at $\lambda \approx 1.5$.
For $\lambda > 1.5$, there occurs a chain of transitions
in the order of FM, A-AF, C-AF, and G-AF phases
with increasing $J_{\rm AF}$,
already obtained in 2$\times$2$\times$2 calculations.
Note that the boundary curve always indicates the first order transition.
The present result shows that size effects are small in undoped
strongly-coupled manganites, which is intuitively reasonable.
The spin arrangement for each phase is shown in Fig.~\ref{fig7}(a).
Note that the FM phase is concomitant with orbital ordering (OO),
which will be discussed later, and this FM/OO phase is considered
to be insulating.

On the other hand, in the weak or intermediate coupling region
for $\lambda < 1.5$, there is a transition between
the FM orbital-disordered (OD) phase and E-type AF phase.
The spin arrangement for E-AF phase is shown in Fig.~\ref{fig8}(b).
Along the zigzag chains, $t_{\rm 2g}$ spins order ferromagnetically,
but they are antiparallel perpendicular to the zigzag direction.
This is just the new AF phase, suggested by recent experiments
on HoMnO$_3$.
As suggested in Fig.~\ref{fig8}(b),
the spin directions are reversed from plane to plane.
Note that the orbital structure in the E-AF phase is the same
as that of the A-AF phase, namely, the staggered pattern of
($3x^2$$-$$r^2$)- and ($3y^2$$-$$r^2$)-like orbitals.
It is easily understood that the orbital ordering is closely
related to the cooperative Jahn-Teller distortion
in undoped manganites and such a cooperative effect
should be very strong irrespective of the $t_{\rm 2g}$
spin configuration.

Note that near $\lambda$$\sim$1.5,
which is a realistic value for manganites,
the A-AF phase is adjacent to the E-type state.
This region could correspond to the actual situation observed
in experiments for RMnO$_3$:
When the ionic radius of the R-site decreases,
a N\'eel temperature $T_{\rm N}$ of the A-AF phase decreases as well,
and eventually the E-AF phase is stabilized for R=Ho.
Another interesting point of the phase diagram is that the E-type spin
arrangement is the ground-state for a wide range of $J_{\rm AF}$,
even at $\lambda$=0, indicating that the coupling with JT phonons
is not a necessary condition for its stabilization.
As pointed out in the previous subsection, the E-type phase is stable
due to the zigzag geometry of the FM chains that induce a band-insulator.
Namely, E-AF phase is always insulating irrespective of the value of
$\lambda$.

Concerning the appearance of the E-AF phase,
Kimura et al. have explained it on the basis of
a frustrated spin system with
FM nearest-neighbor and AF next-nearest-neighbor interactions
within the MnO$_2$ plane \cite{Kimura2003}.
They have found that the staggered orbital order
associated with the GdFeO$_3$-type distortion induced
the anisotropic next-nearest-neighbor interaction,
leading to unique sinusoidal and up-up-down-down AF order,
i.e., E-type phase, in undoped manganites.
In a conceptual level, the spin model is considered to be
obtained in the strong coupling limit of the
$e_{\rm g}$-orbital degenerate double-exchange model.
Thus, the band-insulating picture
for the appearance of E-type phase in the present scenario
is complementary to the result of Kimura et al.,
in the sense that the weak-coupling state
is continuously connected to that in the
strong-coupling limit.

In addition to the explanation of
the A-AF of LaMnO$_3$ and E-AF of HoMnO$_3$,
Kimura et al. have also examined systematically the magnetic and
orbital structures in a series of RMnO$_3$ as a function of
$r_{\rm R}$, the radius of rare-earth ion R.
They have pointed out that the effect on the crystal structure
by decreasing $r_{\rm R}$ appears as the enhancement of
the GdFeO$_3$-type distortion, indicating the shortening of
oxygen-oxygen distance,
Then, the superexchange interaction between next-nearest-neighbor sites
is enhanced due to the shortened oxygen-oxygen path,
leading to the frustrated spin model with the competition between
FM nearest-neighbor and AF next-nearest-neighbor interactions
By analyzing the frustrated spin model on the staggered
orbital-ordered background, Kimura et al. have explained the
phase diagram of RMnO$_3$.
It is considered that the phase diagram can be also understood
from the band-insulating picture,
but for comparison with actual materials,
it is necessary to include the effect of the GdFeO$_3$-type distortions,
which has not been considered in the present model.

Near the transition region between A- and E-type AF phases,
Salafranca and Brey have mentioned the importance of
the competition between the nearest neighbor AF superexchange
interaction and the
double exchange induced long-range FM interaction \cite{Salafranca}.
They concluded that such competition results in the appearance
of incommensurate phases.
These phases consist of a periodic array of domain walls.

As discussed above, in the strong-coupling region, FM/OO insulating
state appears, but when $\lambda$ decreases, OO disappears and instead,
an OD phase is observed.
This is considered as a metallic phase, as deduced from the result
of the density of states.
Note that this metallic OD/FM phase is next to the insulating E-AF phase
for $\lambda < 1.5$, which is a new and important result
in the study of undoped manganites \cite{Hotta-Nova}.
Namely, the competition between FM metallic and insulating phases
is at the heart of the CMR phenomena, and then, 
by tuning experimentally the lattice parameters in RMnO$_3$
it may be possible to observe the magnetic-field induced metal-insulator
transition even in undoped manganites.

Let us consider the reason why the metallic phase can exist
even at half-filling.
To clarify this point, it is quite useful to depict the Fermi-surface
lines. As shown in Fig.~\ref{fig8}(c),
the nesting vector is $(\pi,0)$ or $(0,\pi)$, {\it not} $(\pi,\pi)$.
These nesting vectors are $not$ compatible with the staggered orbital
ordering pattern that is stabilized increasing $\lambda$. 
This is one of the remarkable features of the multiorbital
$e_{\rm g}$-electron system, which is not specific to two dimensionality.
In fact, in the results for the three-dimensional (3D) case,
we also observe the signal of the metal-insulator transition
at a finite value of $\lambda$.
In this case, the orbital ordering pattern becomes very complicated,
but the pattern repeats periodically on lattice larger than
2$\times$2$\times$2.
In the 3D case, an intrinsic incompatibility between the Fermi surface
and the orbital ordering pattern is also found.
Even without invoking the numerical results discussed before,
the qualitative arguments related with the nesting effects
in $H_{\rm kin}$ incompatible with staggered orbital ordering
strongly suggests the presence of a metallic phase
in two and three dimensions at small $\lambda$.

\begin{figure}[t]
\begin{center}
\includegraphics[width=0.8\textwidth]{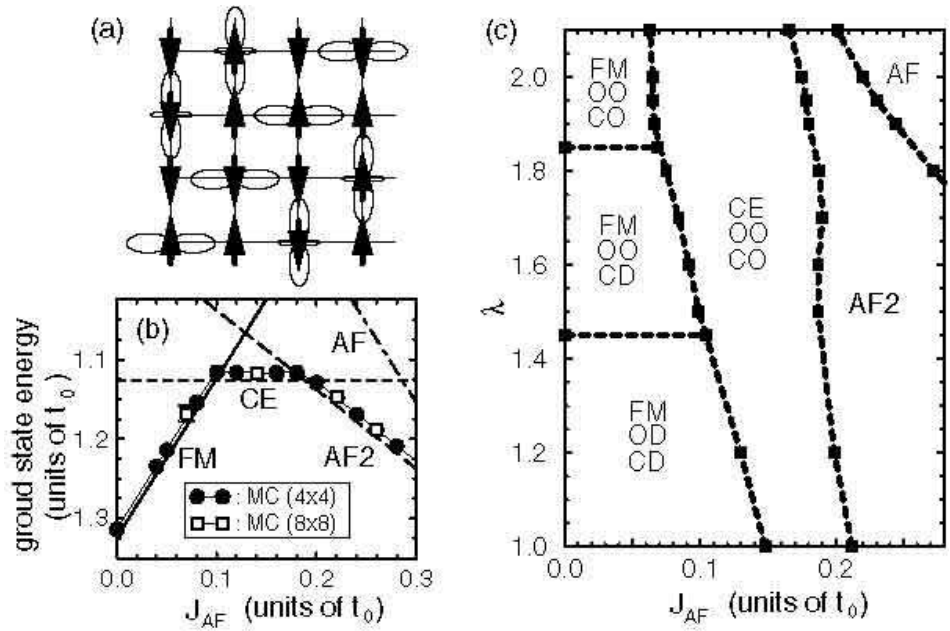}
\end{center}
\caption{(a) Schematic view of CE-type structure at x=0.5
for 2D case.
(b) Monte Carlo energy per site vs $J_{\rm AF}$ at density x=0.5,
$\lambda$=1.5, low temperature $T$=1/100, and $J_{\rm H}$=$\infty$,
using the two-orbital model in two dimensions with 
non-cooperative Jahn-Teller phonons.
As for AF2, see Fig.~\ref{fig5}(c).
(c) Phase diagram in the plane $\lambda$-$J_{\rm AF}$ at x=0.5,
obtained numerically using up to 8$\times$8 clusters.
All transitions are of first-order.
The notation is the standard one (CD = charge
disorder, CO = charge order, OO = orbital order, OD = orbital
disorder). Results are reproduced from Ref.~\cite{Yunoki2000}
where more details can be found.
}
\label{fig9}
\end{figure}

\subsubsection{x=0.5}

Now let us move to another important doping x=0.5.
For half-doped perovskite manganites, the so-called CE-type AF phase
has been established as the ground state in the 1950's. 
This phase is composed of zigzag FM arrays of $t_{\rm 2g}$-spins,
which are coupled antiferromagnetically
perpendicular to the zigzag direction.
Furthermore, the checkerboard-type charge ordering in the $x$-$y$
plane, the charge stacking along the $z$-axis, and
($3x^2-r^2$/$3y^2-r^2$) orbital ordering are associated with this phase.
A schematic view of CE-type structure with charge and orbital
ordering is shown in Fig.~\ref{fig9}(a) for 2D case.
In 3D, this patters repeats along the $z$-axis by keeping charge and
orbital structure, but changing spin directions.

Although there is little doubt that the famous CE-state of Goodenough
is indeed the ground state of x=0.5
intermediate and low bandwidth manganites, only very recently such a
state has received theoretical confirmation using unbiased techniques, 
at least within some models.
In the early approach of Goodenough it was $assumed$ that the charge
was distributed in a checkerboard pattern, upon which spin and orbital
order was found. But it would be desirable to obtain the CE-state
based entirely upon a more fundamental theoretical analysis, as
the true state of minimum energy of a well-defined and realistic
Hamiltonian.
If such a calculation can be done, as a bonus one would find out which 
states compete with the CE-state in parameter space, an
issue very important in view of the mixed-phase tendencies of
Mn-oxides, which cannot be handled within the approach of Goodenough.

One may naively believe that it is as easy as introducing a huge
nearest-neighbor Coulomb repulsion $V$ to stabilize a charge-ordered
state at x=0.5, upon which the reasoning of Goodenough can be applied.
However, there are at least two problems with this approach
\cite{Hotta2001com}.
First, such a large $V$ quite likely will destabilize the
FM charge-disordered state and others supposed to be
competing with the CE-state. It may be possible to explain the
CE-state with this approach, but not others also observed at
x=0.5 in large bandwidth Mn-oxides.
Second, a large $V$ would produce a checkerboard pattern in the
$three$ directions.
However, experimentally it has been known for a long time \cite{Wollan1955}
that the charge $stacks$ along the $z$-axis, namely the
same checkerboard pattern is repeated along $z$-axis,
rather than being shifted by one lattice spacing from plane to plane.
A dominant Coulomb interaction $V$ cannot be the whole story
for x=0.5 low-bandwidth manganese oxides.

The nontrivial task of finding a CE-state without the use of
a huge nearest-neighbors repulsion has been recently performed by
Yunoki, Hotta, and Dagotto \cite{Yunoki2000},
using the two-orbital model with strong electron Jahn-Teller phonon
coupling.
The calculation proceeded using an unbiased Monte Carlo simulation,
and as an output of the study, the CE-state indeed emerged as the
ground-state in some region of coupling space.
Typical results are shown in Figs.~\ref{fig9}(b) and \ref{fig9}(c).
In part (b) the energy at very low temperature is shown as a function
of $J_{\rm AF}$ at fixed density x=0.5, $J_{\rm H}$=$\infty$ for
simplicity, and with a robust electron-phonon coupling $\lambda$=1.5
using the two orbital model $H_{\rm JT}$
At small $J_{\rm AF}$, a FM phase was found to be
stabilized, according to the Monte Carlo simulation.
Actually, at $J_{\rm AF}$=0.0 it has not been possible to stabilize a
partially AF-state at x=0.5, namely the states are always
ferromagnetic at least within the wide range of $\lambda$'s
investigated (but they can have charge and orbital order).
On the other hand, as $J_{\rm AF}$ grows, a tendency to form AF links
develops, as it happens at x=0.0.
At large $J_{\rm AF}$ eventually the system transitions to
states that are mostly antiferromagnetic, such as the so-called
``AF(2)'' state of Fig.~\ref{fig9}(b) (with an up-up-down-down spin pattern
repeated along one axis, and AF coupling along the other axis),
or directly a fully AF-state in both directions.

However, the intermediate values of $J_{\rm AF}$ are the most
interesting ones. In this case the energy of the two-dimensional clusters 
become flat as a function of $J_{\rm AF}$
suggesting that the state has the same
number of FM and AF links, a property that the CE-state indeed has.
By measuring charge-correlations it was found that a checkerboard
pattern is formed particularly at intermediate and large $\lambda$'s,
as in the CE-state.
Finally, after measuring the spin and orbital correlations, it was
confirmed that indeed the complex pattern of the CE-state was fully
stabilized in the simulation. This occurs in a robust portion of the
$\lambda$-$J_{\rm AF}$ plane, as shown in Fig.~\ref{fig9}(c).
The use of $J_{\rm AF}$ as the natural parameter to vary in order
to understand the CE-state is justified based on Fig.~\ref{fig9}(c),
since the region of stability of the CE-phase is elongated
along the $\lambda$-axis,
meaning that its existence is not so much dependent on that coupling
but much more on $J_{\rm AF}$ itself.
It appears that some explicit tendency in the Hamiltonian toward the
formation of AF links is necessary to form the CE-state.
If this tendency is absent, a FM state if formed, while if it is too
strong an AF-state appears.
The x=0.5 CE-state, similar to the A-type AF at x=0.0,
needs an intermediate value of $J_{\rm AF}$ for stabilization.
The stability window is finite and in this respect there is no need
to carry out a fine tuning of parameters to find the CE phase.
However, it is clear that there is a balance of AF and FM
tendencies in the CE-phase that makes the state somewhat fragile.

Note that the transitions among the many states obtained when varying
$J_{\rm AF}$ are all of $first$ order, namely they correspond to
crossings of levels at zero temperature.
The first-order character of these transitions is a crucial ingredient
of the recent scenario proposed by Moreo et al. \cite{Moreo2000}
involving mixed-phase tendencies with coexisting clusters with equal
density.
Recently, first-order transitions have also been reported in the
one-orbital model at x=0.5 by Alonso et al.
\cite{Alonso2001a,Alonso2001b},
as well as tendencies toward phase separation.

\begin{figure}[t]
\begin{center}
\includegraphics[width=0.8\textwidth]{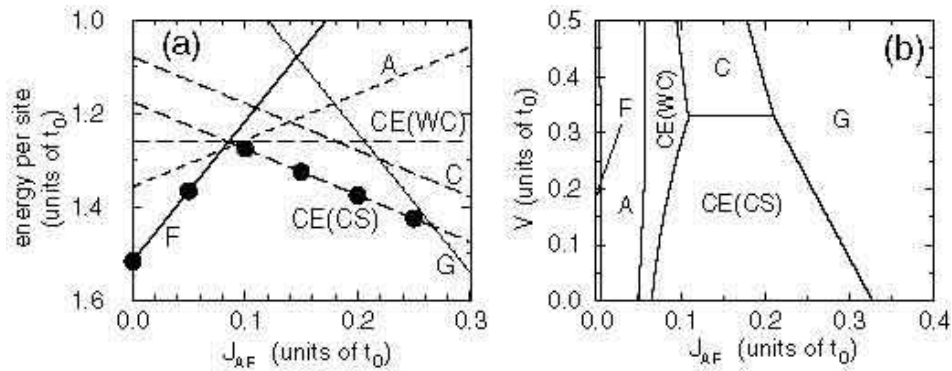}
\end{center}
\caption{(a) Energy per site as a function of $J_{\rm AF}$ for
$\lambda$=1.6 and $J_{\rm H}$=$\infty$ for $H_{\rm JT}$
using a 4$\times$4$\times$4 lattice.
The curves denote the mean-field results and the solid symbols
indicate the energy obtained by the relaxation method.
Thick solid, thick broken, thin broken, thick dashed, thin dashed,
thin broken, and thin solid lines denotes FM, A-type, CE-type
with WC structure, charge-stacked CE-type, C-type, and G-type states,
respectively.
Note that the charge-stacked CE-state is observed in experiments.
(b) Phase diagram in the $(J_{\rm AF},V)$ plane
for 4$\times$4$\times$4 lattice.
Note that the charge-stacked structure along the $z$-axis can be
observed only in the CE-type AF phase.
Results are reproduced from Ref.~\cite{Hotta2000b}
where more details can be found.}
\label{fig10}
\end{figure}

Let us address now the issue of charge-stacking (CS) along the $z$-axis.
For this purpose simulations using 3D clusters were carried out
\cite{Hotta2000b}.
The result for the energy vs. $J_{\rm AF}$ is shown
in Fig.~\ref{fig10}(a),
with $J_{\rm H}$=$\infty$ and $\lambda$=1.6 fixed.
The CE-state with charge-stacking has been found to be the ground state 
on a wide $J_{\rm AF}$ window.
The reason that this state has lower energy than the so-called
``Wigner-crystal'' (WC) version of the CE-state, namely with the
charge spread as much as possible, is once again the influence of 
$J_{\rm AF}$. With a charge stacked arrangement, the links along the
$z$-axis can all be simultaneously antiferromagnetic, thereby minimizing
the energy. In the WC-state this is not possible.

It should be noted that this charge stacked CE-state is not
immediately destroyed when the weak nearest-neighbor repulsion $V$
is introduced to the model, as shown in Fig.~\ref{fig10}(b),
obtained in the mean-field calculations by
Hotta, Malvezzi, and Dagotto \cite{Hotta2000b}.
If $V$ is further increased for a realistic value of $J_{\rm AF}$,
the ground state eventually changes from the charge stacked
CE-phase to the WC version of the CE-state or the C-type AF phase 
with WC charge ordering.
As explained above, the stability of the charge stacked phase to 
the WC version of the CE-state is due to the magnetic energy difference.
However, the competition between the charge-stacked CE-state
and the C-type AF phase with the WC structure is not simply understood 
by the effect of $J_{\rm AF}$, since those two kinds of AF phases
have the same magnetic energy.
In this case, the stabilization of the charge stacking originates
from the difference in the geometry of the one-dimensional FM path,
namely a zigzag-path for the CE-phase and a straight-line path
for the C-type AF state.
As discussed above, the energy for $e_{\rm g}$
electrons in the zigzag path is lower than that in the straight-line
path, and this energy difference causes the stabilization of the
charge stacking.
In short, the stability of the charge-stacked structure
at the expense of $V$ is supported by ``the geometric energy''
as well as the magnetic energy.
Note that each energy gain is just a fraction of $t_0$.
Thus, in the absence of other mechanisms to understand the
charge-stacking, another consequence of this analysis is that $V$
actually must be substantially $smaller$ than naively expected,
otherwise such a charge pattern would not be stable.
In fact, estimations given by Yunoki, Hotta, and Dagotto \cite{Yunoki2000}
suggest that the manganites must have a large dielectric function at
short distances (see Ref.~\cite{Arima1995}) to prevent the melting
of the charge-stacked state.

Note also that the mean-field approximations by Hotta, Malvezzi, and
Dagotto \cite{Hotta2000b} have shown that on-site Coulomb interactions $U$
and $U'$ can also generate a two-dimensional CE-state,
in agreement with the calculations by van den Brink et al.
\cite{vandenBrink1999}.
Then, we believe that strong Jahn-Teller and
Coulomb couplings tend to give similar results.
This belief finds partial confirmation in the mean-field
approximations of Hotta, Malvezzi, and Dagotto \cite{Hotta2000b},
where the similarities between a strong $\lambda$ and $(U,U')$ were
investigated.
Even doing the calculation with Coulombic interactions, the influence
of $J_{\rm AF}$ is still crucial to inducing charge-stacking.
The importance of this parameter has also been remarked by
Mathieu, Svedlindh and Nordblad \cite{Mathieu2000}
based on experimental results.

Many other authors carried out important work in the context of the
CE-state at x=0.5.
For example, with the help of Hartree-Fock calculations, Mizokawa and
Fujimori \cite{Mizokawa1997}
reported the stabilization of the CE-state at x=0.5
only if Jahn-Teller distortions were incorporated into a model with
Coulomb interactions.
This state was found to be in competition with a uniform FM state, as
well as with an A-type AF-state with uniform orbital order.
In this respect the results are very similar to those found by Yunoki,
Hotta and Dagotto \cite{Yunoki2000} using Monte Carlo simulations.
In addition, using a large nearest-neighbor repulsion and the
one-orbital model, charge ordering and a spin structure compatible
with the zigzag chains of the CE state was found by Lee and Min
at x=0.5 \cite{Lee1997}.
Jackeli, Perkins, and Plakida also obtained charge-ordering
at x=0.5 using mean-field approximations and a large $V$
\cite{Jackeli}.
Charge-stacking was not investigated by those authors.
The CE-state in Pr$_{0.5}$Ca$_{0.5}$MnO$_3$ was also obtained
by Anisimov et al. using LSDA+U techniques \cite{Anisimov}.

\subsubsection{x$>$0.5}

In the previous subsection, the discussion focused on the CE-type AF 
phase at x=0.5.
Naively, it may be expected that similar arguments can be extended to
the regime x$>$1/2, since in the phase diagram for
${\rm La_xCa_{1-x}MnO_3}$,
the AF phase has been found at low temperatures in the region
0.50$<$x$<$0.88.
Then, let us try to consider the band-insulating phase for density
x=2/3 based on the minimal model Eq.~(\ref{minimalH})
without both the Jahn-Teller phononic and Coulombic interactions,
since this doping is quite important for the appearance of
the bi-stripe structure (see Refs.~\cite{Mori1998a,Mori1998b}).

\begin{figure}[t]
\begin{center}
\includegraphics[width=0.8\textwidth]{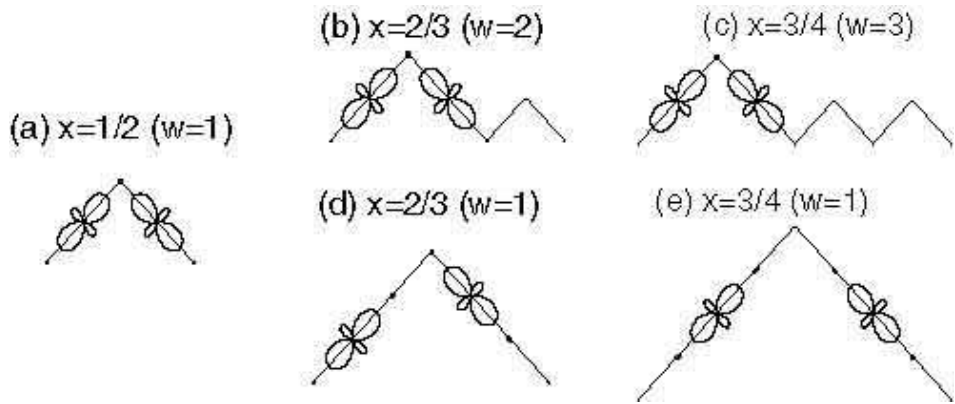}
\end{center}
\caption{
(a) Path with $w$=$1$ at x=$1/2$.
Charge and orbital densities are
calculated in the MFA for $E_{\rm JT}$=$2t$. At each site, the orbital
shape is shown with its size in proportion to the orbital density.
(b) The BS-structure path with $w$=$2$ at x=$2/3$. 
(c) The BS-structure path with $w$=$3$ at x=$3/4$. 
(d) The WC-structure path with $w$=$1$ at x=$2/3$. 
(e) The WC-structure path with $w$=$1$ at x=$3/4$.}
\label{fig11}
\end{figure}

After several calculations for x=2/3, as reported by Hotta et al.
\cite{Hotta2000},
the lowest-energy state was found to be characterized by the
straight path, not the zigzag one, leading to the C-type AF phase
which was also discussed in previous subsection.
For a visual representation of the C-type state,
see Fig.~4 of Ref.~\cite{Kajimoto1999}.
At first glance, the zigzag structure similar to that for x=0.5
could be the ground-state for the same reason
as it occurs in the case of x=0.5. 
However, while it is true that the state with such a zigzag structure
is a band-insulator, the energy gain due to the opening of the bandgap
is not always the dominant effect.
In fact, even in the case of x=0.5, the energy of the bottom of the
band for the straight path is $-2t_0$, while for the zigzag path,
it is $-\sqrt{3}t_0$. For x=1/2, the energy gain due to the gap
opening overcomes the energy difference at the bottom of the band,
leading to the band-insulating ground-state. 
However, for x=2/3 even if a band-gap opens the energy of the zigzag
structure cannot be lower than that of the metallic straight-line
phase. Intuitively, this point can be understood as follows: 
An electron can move smoothly along the one-dimensional path
if it is straight. However, if the path is zigzag, ``reflection" of
the wavefunction occurs at the corner, and then a smooth movement of
one electron is no longer possible. Thus, for small numbers of
carriers, it is natural that the ground-state is characterized by the
straight path to optimize the kinetic energy of the $e_{\rm g}$
electrons.

However, in neutron scattering experiments a spin pattern similar
to the CE-type AF phase has been suggested by Radaelli et al.
\cite{Radaeli1999}.
In order to stabilize the zigzag AF phase to reproduce those
experiments it is necessary to include the Jahn-Teller distortion effectively. 
As discussed by Hotta et al. \cite{Hotta2000},
a variety of zigzag paths could
be stabilized when the Jahn-Teller phonons are included.
In such a case, the classification of zigzag paths is an important
issue to understand the competing ``bi-stripe" vs. ``Wigner-crystal"
structures.
The former has been proposed by Mori et al. \cite{Mori1998a,Mori1998b},
while the latter was claimed to be stable by Radaelli et al.
\cite{Radaeli1999}.
As shown in the previous subsection,
the shape of the zigzag structure is
characterized by the winding number $w$ associated with the 
Berry-phase connection of an $e_{\rm g}$-electron parallel-transported 
through Jahn-Teller centers, along zigzag one-dimensional paths.
As shown above,
the total winding number is equal to half of the number of corners
included in the zigzag unit path.
Namely, the winding number $w$ is a good label to specify the shape of
the zigzag one-dimensional FM path.

\begin{figure}[t]
\begin{center}
\includegraphics[width=1.0\textwidth]{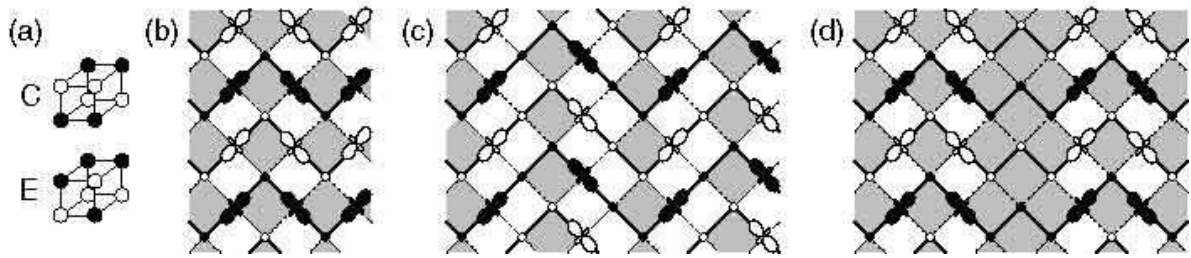}
\end{center}
\caption{
(a) C- and E-type unit cells \cite{Wollan1955}.
(b) The spin structure in the $a$-$b$ plane at x=1/2.
Open and solid symbols denote the spin up and down, respectively.
The thick line indicates the zigzag FM path.
The open and shaded squares denote the C- and E-type unit cells.
At x=1/2, C-type unit cell occupies half of the two-dimensional
plane, clearly indicating the ``CE'' type phase.
(c) The spin structure at x=2/3 for Wigner-crystal type
phase. Note that 66\% of the two-dimensional lattice is occupied
by C-type unit cell. Thus, it is called ``C$_{2/3}$E$_{1/3}$''-type AF phase.
(d) The spin structure at x=2/3 for bi-stripe type
phase. Note that 33\% of the two-dimensional lattice is occupied
by C-type unit cell. Thus, it is called ``C$_{1/3}$E$_{2/3}$''-type AF phase.
}
\label{fig12}
\end{figure}

After several attempts to include effectively the Jahn-Teller phonons,
it was found that the bi-stripe phase and the Wigner crystal phase
universally appear for $w$=$\rm x/(1-x)$ and $w$=1, respectively.
Note here that the winding number for the bi-stripe structure has a
remarkable dependence on x, reflecting  the fact that the distance
between adjacent bi-stripes changes with x.
This x-dependence of the modulation vector of the lattice distortion
has been observed in electron microscopy experiments
\cite{Mori1998a,Mori1998b}.
The corresponding zigzag paths with the charge and orbital ordering 
are shown in Fig.~\ref{fig11}.
In the bi-stripe structure, the charge is
confined in the short straight segment as in the case of the CE-type
structure at x=0.5.
On the other hand, in the Wigner-crystal structure, the straight
segment includes two sites, indicating that the charge prefers to
occupy either of these sites.
Then, to minimize the Jahn-Teller energy and/or the Coulomb repulsion,
the $e_{\rm g}$ electrons are distributed with equal spacing. 
The corresponding spin structure is shown in Fig.~\ref{fig12}.
A difference in the zigzag geometry can produce a significant
different in the spin structure.
The definitions for the C- and E-type AF structures \cite{Wollan1955}
are shown in Fig.~\ref{fig12}(a) for convenience.
At x=1/2, as clearly shown in Fig.~\ref{fig12}(b),
half of the plane is filled by the C-type,
while another half is covered by the E-type, clearly illustrating the
meaning of ``CE" in the spin structure of half-doped manganites.
On the other hand, as shown in Figs.~\ref{fig12}(c) and \ref{fig12}(d),
the bi-stripe and Wigner crystal structure have $\rm C_{1-x}E_{x}$-type
and $\rm C_{x}E_{1-x}$-type AF spin arrangements, respectively.
Such zigzag-based AF structure has been discussed experimentally
in single layered manganites $\rm Nd_{1-x}Sr_{1+x}MnO_4$
by Kimura et al. \cite{Kimura2002}.

\begin{figure}[t]
\begin{center}
\includegraphics[width=0.8\textwidth]{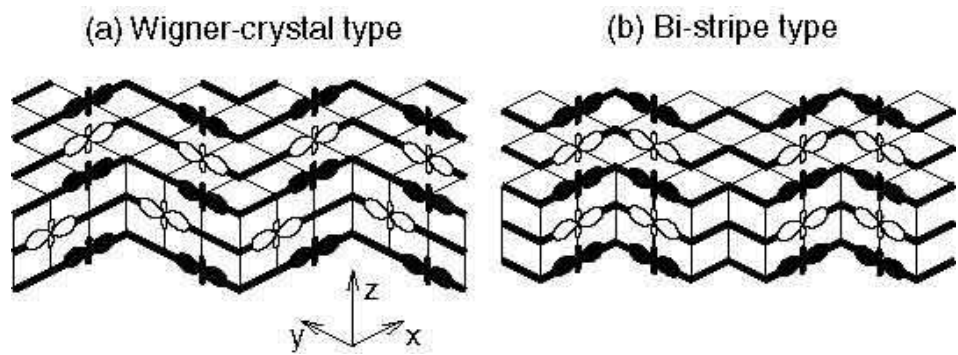}
\end{center}
\caption{
Schematic views for spin, charge, and orbital ordering for
(a) Wigner-crystal structure and
and (b) bi-stripe type structures at x=2/3 in 3D environment.
The open and solid symbols indicate
the spin up and down, respectively. The FM one-dimensional path is
denoted by the thick line.
The empty sites denote Mn$^{4+}$ ions, while the robes
indicate the Mn$^{3+}$ ions in which $3x^2-r^2$ or $3y^2-r^2$ orbitals
are occupied.
}
\label{fig13}
\end{figure}

The charge structure along the $z$-axis for x=2/3 has been discussed by
Hotta et al. \cite{Hotta2000c}.
As schematically shown in Figs.~\ref{fig13}(a) and (b),
a remarkable feature can be observed.
Due to the confinement of charge in the short straight segment for the
bi-stripe phase, the charge stacking is suggested from our topological
argument.
On the other hand, in the Wigner-crystal type structure, charge is not
stacked, but it is shifted by one lattice constant to avoid the
Coulomb repulsion. 
Thus, if the charge stacking is also observed in the experiment for
x=2/3, our topological scenario suggests the bi-stripe phase as the
ground-state in the low temperature region.
To establish the final ``winner" in the competition between the
bi-stripe and Wigner-crystal structure at x=2/3, more precise
experiments, as well as quantitative calculations, will be
further needed.

\subsubsection{x$<$0.5}

Regarding densities smaller than 0.5, the states at x=1/8, 1/4 and
3/8 have received considerable attention.
See Refs.~\cite{Mizokawa2000,Korotin2000,Hotta2000d}.
These investigations are still in a ``fluid'' state, and
the experiments are not quite decisive yet, and for this reason,
this issue will not be discussed in much detail here.
However, without a doubt, it is very important to clarify the
structure of charge-ordered states that may be in competition
with the FM states in the range in which the latter
is stable in some compounds.
``Stripes'' may emerge from this picture, as recently remarked in
experiments \cite{Adams2000,Dai2000,Kubota2000,Vasiliu-Doloc1999}
and calculations \cite{Hotta2001}, and surely the
identification of charge/orbital arrangements at x$<$0.5 will be an
important area of investigations in the very near future.

\begin{figure}[t]
\begin{center}
\includegraphics[width=0.7\textwidth]{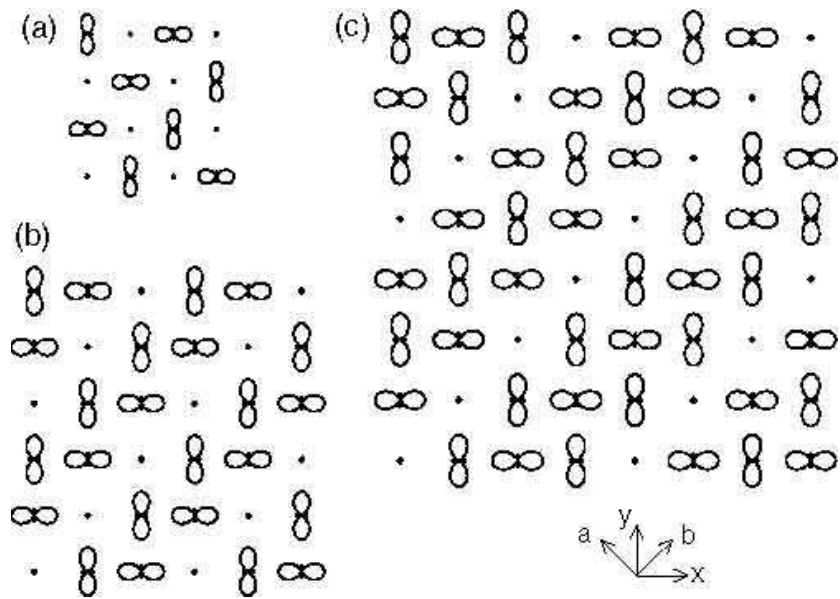}
\end{center}
\caption{%
Numerical results for orbital densities in the FM phase
for (a)x=1/2, (b)1/3, and (c)1/4 \cite{Hotta2001}.
The charge density in the lower-energy orbital is shown,
and the size of the orbital is exactly in proportion to
this density.
}
\label{fig14}
\end{figure}

Here a typical result for this stripe-like charge ordering is shown in
Fig.~\ref{fig14}, in which the lower-energy orbital at each site is
depicted,
and its size is in proportion to the electron density occupying that
orbital.
This pattern is theoretically obtained by the relaxation technique for
the optimization of oxygen positions, namely including the cooperative
Jahn-Teller effect. 
At least in the strong electron-phonon coupling region, the stripe
charge ordering along the $diagonal$ direction in the $xy$ plane
becomes the global ground-state.
Note, however, that many meta-stable states can appear very close to
this ground state. 
Thus, the shape of the stripe is considered to fluctuate both in space 
and time, and in experiments it may occur that only some fragments of
this stripe can be detected. 
It should also be emphasized that the orbital ordering occurs concomitant
with this stripe charge ordering.  
In the electron-rich region, the same antiferro orbital-order exists
as that corresponding to x=0.0.
On the other hand, the pattern around the diagonal array of
electron-poor sites is quite similar to the building block of the
charge/orbital structure at x=0.5.

In Fig.~\ref{fig14}, it is found that
the same charge and orbital structure stacks along the $b$-axis.
Namely, it is possible to cover the whole two-dimensional plane by 
some periodic charge-orbital array along the $a$-axis.
If this periodic array is taken as the closed loop $C$,
the winding numbers are $w$=1, 2, and 3,
for x=1/2, 1/3, and 1/4, respectively.
Note that in this case $w$ is independent of the path along the $a$-axis.
The results imply a general relation $w$=$(1-x)/x$
for the charge-orbital stripe in the FM phase, reflecting the fact that
the distance between the diagonal arrays of holes changes with x.
Our topological argument predicts stable charge-orbital stripes at special
doping such as x=$1/(1+w)$, with $w$ an integer.

This orbital ordering can be also interpreted as providing 
a ``$\pi$"-shift in the orbital sector,
by analogy with the dynamical stripes found in cuprates \cite{Buhler2000},
although in copper oxides
the charge/spin stripes mainly appear along the $x$- or $y$-directions.
The study of the similarities and differences between stripes in manganites
and cuprates is one of the most interesting open problems in the study of
transition metal oxides, and considerable work is expected in the near future.

\begin{figure}[t]
\begin{center}
\includegraphics[width=0.4\textwidth]{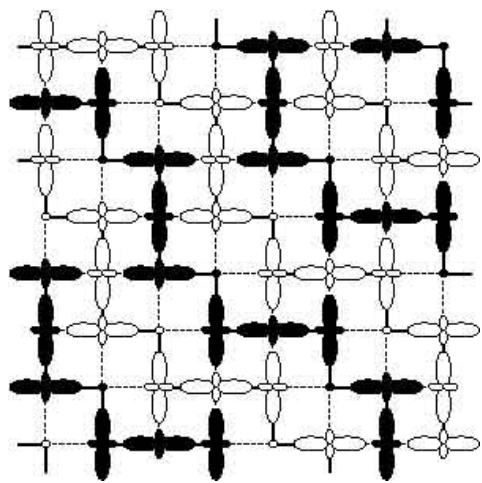}
\end{center}
\caption{%
Schematic figure of the possible spin-charge-orbital structure at x=1/4
in the zigzag AF phase at low temperature and large electron-phonon
coupling \cite{Hotta2001}.
This figure was obtained using numerical techniques, and cooperative
phonons, for $J_{\rm H}$=$\infty$ and $J_{\rm AF}$=$0.1t_0$.
For the non-cooperative phonons, basically the same pattern
can be obtained.
}
\label{fig15}
\end{figure}

Finally, a new zigzag AF spin configuration for x$<$0.5 is here briefly
discussed \cite{Hotta2001}.
In Fig.~\ref{fig15}, a schematic view of this novel spin-charge-orbital
structure on the 8$\times$8 lattice at x=1/4 is shown,
deduced using the numerical relaxation technique applied to
cooperative Jahn-Teller phonons in the strong-coupling region.
This structure appears to be the global ground state, but many excited states
with different spin and charge structures are also found with small
excitation energy, suggesting that the AF spin structure for x$<$0.5
in the layered manganites is easily disordered due to this
``quasi-degeneracy'' in the ground state.
This result may be related to the ``spin-glass'' nature of the single layer
manganites reported in experiments \cite{Moritomo1995}.

It should be noted that the charge-orbital structure is essentially the
same as that in the two-dimensional FM phase,
as shown in Fig.~\ref{fig14}.
This suggests the following scenario for the layered manganites:
When the temperature is decreased from the higher temperature region,
first charge ordering occurs due to the cooperative Jahn-Teller
distortions in the FM (or paramagnetic) region.
If the temperature is further decreased, the zigzag AF spin arrangement
is stabilized, adjusting itself to the orbital structure.
Thus, the separation between the charge ordering temperature $T_{\rm CO}$
and the N\'eel temperature $T_{\rm N}$ occurs naturally in this context.
This is not surprising, since $T_{\rm CO}$ is due to the electron-lattice 
coupling, while $T_{\rm N}$ originates in the coupling $J_{\rm AF}$.
However, if the electron-phonon coupling is weak, then $T_{\rm CO}$ becomes
very low. In this case, the transition to the zigzag AF phase may occur
prior to the charge ordering. 
As discussed above, the $e_{\rm g}$ electron hopping is confined to
one dimensional structures in the zigzag AF environment.
Thus, in this situation, even a weak coupling electron-phonon coupling can
produce the charge-orbital ordering, as easily understood from the Peierls
instability argument.
Namely, just at the transition to the zigzag AF phase, the charge-orbital
ordering occurs simultaneously, indicating that $T_{\rm CO}$=$T_{\rm N}$.
Note also that in the zigzag AF phase, there is no essential difference
in the charge-orbital structures for the non-cooperative and cooperative
phonons, due to the one-dimensionality of those zigzag chains.

\subsection{Summary}

In this section, we have reviewed the theoretical results on
spin, charge, and orbital ordering in manganites.
We believe that the complicated ordering in manganites is caused
by (i) competition between FM metallic and AF insulating phases
and (ii) active $e_{\rm g}$ orbital degree of freedom.
The existence of the FM metallic phase in the issue (i) has been
understood by the double-exchange concept,
while the variety of AF insulating originating from the point (ii)
has not been considered satisfactorily
in the standard double-exchange mechanism.
Here we stress that the existence of active orbital
does not simply indicate the increase of internal degrees of freedom
in addition to spin and charge.
We should remark an important effect of the orbital shape,
leading to the geometrical pattern in the spin configuration.
This point has been emphasized in this section in the context of
topological aspect of orbital ordering.

We have not mentioned another important characteristic issue
of manganites, i.e., phase separation tendency, which is a
driving force of colossal magneto-resistance phenomenon
in manganites.
The strong tendency of the phase separation is easily
understood in the complex phase diagram including
several kinds of first order transition.
Readers should refer the previous review and textbook
\cite{Dagotto2001,Dagotto2002},
in which the phase-separation tendency and related physics
have been explained in detail.

%
%
\section{Orbital physics in other $d$-electron materials}

In the previous section, we have concentrated on the orbital physics
of manganites. However, we can also observe orbital ordering phenomena
in other transition metal oxides.
Here we introduce possible orbital ordering in nickelates and
ruthenates as typical materials of $e_{\rm g}$ and $t_{\rm 2g}$
electron systems, respectively, in the sense that $H_{e_{\rm g}}$
and $H_{t_{\rm 2g}}$ can be applied.
Finally, we also discuss a potential role of orbital ordering
in geometrically frustrated electron systems
with orbital degeneracy.

%
\subsection{$e_{\rm g}$ electron systems}

The existence and origin of ``striped'' structures continues attracting
considerable attention in the research field of
transition metal oxides \cite{Tranquada1995,Mook1998}.
In a system with dominant electron-electron repulsion,
the Wigner-crystal state should be stabilized, but in real materials,
more complicated non-uniform charge structures have been found.
In Nd-based lightly-doped cuprates, neutron scattering experiments
revealed incommensurate spin structures
\cite{Cheong1991,Mason1992,Thurston1992},
where AF spin stripes are periodically separated by domain walls of holes.
In $\rm La_{2-x}Sr_xCuO_4$, dynamical stripes are believed to exist
along vertical or horizontal directions (Cu-O bond direction).
Note that for x$<$0.055, the spin-glass phase exhibits a diagonal spin
modulation \cite{Yamada1998,Matsuda2003}.
In nickelates, the charge-ordered stripes are along the diagonal
direction \cite{Tranquada1994,Sachan1995,Yoshizawa2000,Kajimoto2003}.
In manganites, as mentioned in the previous section,
evidence for striped charge-ordering also along the diagonal
direction has been reported in the AF phase for x$>$1/2
\cite{Mori1998a,Mori1998b},
while short-range diagonal stripe correlations have been found
in the FM phase at x$<$1/2 \cite{Adams2000,Dai2000}.

In general, stripes can be classified into metallic or insulating.
In $\rm La_{2-{\it x}}Sr_{\it x}CuO_4$, the dynamical stripes
exhibit metallic properties, but they are easily pinned
by lattice effects and impurities.
In $\rm La_{1.6-{\it x}}Nd_{0.4}Sr_{\it x}CuO_4$, stripes along
the bond-direction are pinned by lattice distortions
\cite{Tranquada1995}, but they are still metallic.
Intuitively, vertical or horizontal
stripes could be associated with the formation of ``rivers of holes'',
to prevent individual charges from fighting against
the AF background \cite{Emery,Buhler2000,Malvezzi}.
Such stripes should be metallic, even if they are pinned,
since they are induced by the optimization of hole motion
between nearest-neighbor Cu-sites via oxygens.

However, in the diagonal stripes observed in manganites and nickelates,
charges are basically localized,
indicating that such insulating stripes are $not$ determined
just by the optimization of the hole motion.
In the FM state of manganites, the hole movement is already
optimal and, naively, charges should not form stripes.
Obviously, an additional effective local potential must be
acting to confine electrons into stripes.
If such a potential originates in lattice distortions, it
is expected to occur along the bond direction
to avoid energy loss due to the conflict
between neighboring lattice distortions sharing the same oxygens.
Then, static stripes stabilized by lattice distortions tends to
occur along the $diagonal$ direction, as shown Fig.~\ref{fig15},
which are stabilized by Jahn-Teller distortions \cite{Hotta2001}.

In simple terms, vertical or horizontal stripes in cuprates
can be understood by the competition between Coulomb interaction
and hole motion, while diagonal stripes are better explained as a
consequence of a robust electron-lattice coupling.
However, a difficulty has been found
for theoretical studies of stripe formation in doped nickelates,
since both Coulomb interaction and
electron-lattice coupling appear to be important.
Since the Ni$^{2+}$ ion has two electrons in the $e_{\rm g}$ orbitals,
on-site Coulomb interactions certainly play a crucial role to form
spins $S$=1.
When holes are doped, one electron is removed and another remains
in the $e_{\rm g}$ orbitals, indicating that the hole-doped site
should become JT active. Then,
in hole-doped nickelates $both$ Coulombic and phononic interactions
could be of relevance, a fact not considered in previous theoretical
investigations.

\begin{figure}[t]
\begin{center}
\includegraphics[width=0.7\textwidth]{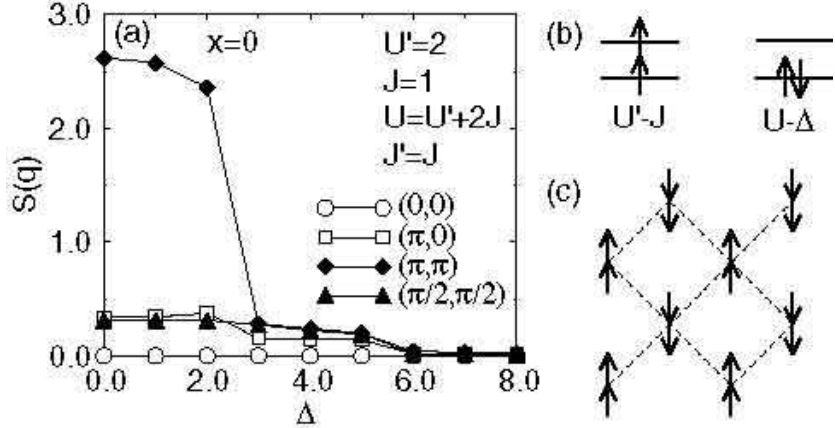}
\end{center}
\caption{
(a) Spin correlation $S(\bm{q})$ vs. $\Delta$ for x=0.
(b) Two kinds of local $e_{\rm g}$-electron arrangements for x=0.
(c) AF spin pattern theoretically determined for $\Delta$$<$3.
}
\label{fig16}
\end{figure}

In the following, we will review the recent results by
Hotta and Dagotto \cite{Hotta2004Ni}.
The model for nickelates is the $e_{\rm g}$ orbital
Hubbard Hamiltonian Eq.~(\ref{Model}),
but another important ingredient is added here.
Namely, the electron-lattice term is divided into
couplings for the apical and in-plane oxygen motions.
In layered nickelates, all NiO$_6$ octahedra are significantly
elongated along the $c$-axis, splitting the $e_{\rm g}$ orbitals.
This splitting from apical oxygens should be included explicitly
from the start as the level splitting between a- and b-orbitals.
Then, the model is defined as
\begin{equation}
 H=H^{e_{\rm g}}_{\rm kin}+H^{e_{\rm g}}_{\rm el-el}
 +\Delta \sum_{\bf i}(n_{{\bf i}{\rm a}}-n_{{\bf i}{\rm b}})/2,
\end{equation}
where $\Delta$ is the level splitting.
Later, the in-plane motion of oxygens should be studied
by adding $H_{\rm ep-ph}^{e_{\rm g}}$.
Note that the energy unit is also $t_0$ in this subsection.

\begin{figure}[t]
\begin{center}
\includegraphics[width=0.7\textwidth]{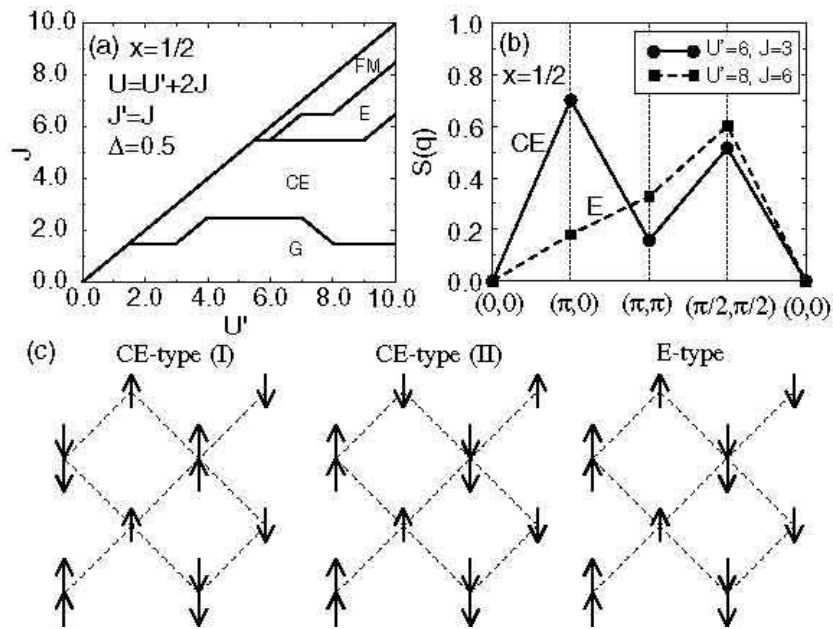}
\end{center}
\caption{(a) Ground-state phase diagram at x=1/2 without
electron-phonon coupling.
(b) $S(\bm{q})$ for the CE- and E-type phases,
at the couplings indicated.
(c) Spin and charge patterns for the CE- and E-type phases.
These are schematic views, since local charge-densities in practice
are not exactly 1 and 2.
}
\label{fig17}
\end{figure}

First, consider the undoped case.
The calculation is done for an 8-site tilted cluster, equivalent
in complexity to a 16-site lattice for the single-band Hubbard model.
Since at all sites the two orbitals are occupied
due to the Hund's rule coupling, the JT distortions are not active and
it is possible to grasp the essential ground-state properties using $H$.
In Fig.~\ref{fig16}(a), the Fourier transform of spin correlations
is shown vs. $\Delta$, where
$S(\bm{q})$=$(1/N)$$\sum_{\bf i,j}e^{i\bm{q}\cdot({\bf i}-{\bf j})}
\langle S^{z}_{{\bf i}} S^{z}_{{\bf j}} \rangle$, with
$S^{z}_{{\bf i}}$=
$\sum_{\gamma}(d_{{\bf i} \gamma\uparrow}^{\dag}d_{{\bf i}\gamma\uparrow}$
$-$$d_{{\bf i} \gamma\downarrow}^{\dag}d_{{\bf i}\gamma\downarrow})/2$.
As expected, a robust $(\pi, \pi)$ peak can be observed for
$\Delta$$<$3, suggesting that the AF phase is stabilized by
super-exchange interactions.
The rapid decrease of $S(\pi, \pi)$ for $\Delta$$>$3 is understood by
comparing the energies for local triplet and singlet states,
as shown in Fig.~\ref{fig16}(b).
The ground-state properties change at $U'$$-$$J$=$U$$-$$\Delta$,
leading to $\Delta$=$3J$ for the transition.
The spin structure at x=0 is schematically shown
in Fig.~\ref{fig16}(c).

Let us turn our attention to the case x=1/2.
The 8-site tilted lattice is again used for the analysis,
and the phase diagram Fig.~\ref{fig17}(a) is obtained for $\Delta$=0.5.
Since $\Delta$ of nickelates is half of that of cuprates
from the lattice constants for $\rm CuO_6$ and $\rm NiO_6$ octahedra,
it is reasonable to select $\Delta$=0.5 in the unit of $t_0$.
Increasing $J$, an interesting transformation from AF
to FM phases is found. This is natural, since at large $J$
the system has a formal similarity with manganite models,
where kinetic-energy gains lead to ferromagnetism, while at 
small $J$ the magnetic energy dominates.
However, between the G-type AF for $J$$\approx$0
and FM phase for $J$$\approx$$U'$, unexpected states appear which are
mixtures of FM and AF phases, due to the competition between
kinetic and magnetic energies.
Typical spin correlations $S(\bm{q})$ are shown in Fig.~\ref{fig17}(b).
Note that peaks at $\bm{q}$=$(\pi, 0)$ and $(\pi/2,\pi/2)$
indicate ``C'' and ``E'' type spin-structures, respectively.
Double peaks at $\bm{q}$=$(\pi, 0)$ and $(\pi/2,\pi/2)$
denote the CE-type structure, frequently observed in half-doped
manganites \cite{Murakami1998b}.
In half-doped nickelates, the CE-phase is expressed as
a mixture of type (I) and (II) in Fig.~\ref{fig17}(c),
depending on the positions of the $S$=1 and $S$=1/2 sites,
although the ``zigzag'' FM chain structure is common for both types.
The E-type phase is also depicted in Fig.~\ref{fig17}(c).
Note that the charge correlation always exhibits a peak
at $\bm{q}$=$(\pi, \pi)$ (not shown here),
indicating the checkerboard-type charge ordering.

\begin{figure}[t]
\begin{center}
\includegraphics[width=0.7\textwidth]{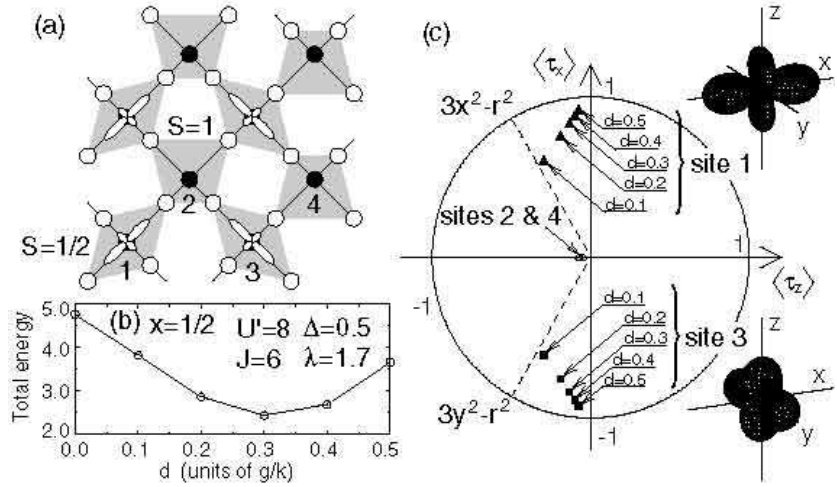}
\end{center}
\caption{
(a) Numerically obtained cooperative distortion pattern for an
8-site lattice at x=1/2 including $H_{\rm el-ph}^{e_{\rm g}}$.
Black and open circles indicate Ni and O ions, respectively.
Open symbols indicate $e_{\rm g}$ orbitals in the optimized state.
(b) Total ground-state energy vs. $d$ for x=1/2.
(c) Orbital densities $\langle \tau_{z{\bf i}} \rangle$
and $\langle \tau_{x{\bf i}} \rangle$ for sites 1--4.
See (d) for the site labels.
Optimized orbitals at $d$=0.3 for sites 1 and 3 are also shown.
}
\label{fig18}
\end{figure}

In experimental results, a peak at $(\pi/2,\pi/2)$ in $S(\bm{q})$
has been reported, suggesting an AF pair of
$S$=1 spins $across$ the singly-occupied sites with holes.
Moreover, the checkerboard-type charge ordering
has been experimentally observed
\cite{Tranquada1994,Sachan1995,Yoshizawa2000,Kajimoto2003}.
Thus, the spin-charge patterns of CE(II)- and E-type
are consistent with the experimental results.
Our phase diagram has a robust region with
a peak at $(\pi/2,\pi/2)$, both for CE- and E-type phases,
although the CE-phase exhibits an extra peak at $(\pi, 0)$.
Whether the E- or CE-phases are present in nickelates can be studied 
experimentally in the future by searching for this $(\pi, 0)$ peak.
Note that if diffuse scattering experiments detect the AF correlation
$along$ the hole stripe, as has been found at x=1/3 \cite{Boothroyd},
the CE(II)-type may be the only possibility.
Summarizing, the spin-charge structure in x=1/2 experiments can be
understood within the Hamiltonian $H$ by assuming a relatively large $J$.

Consider now the effect of in-plane oxygen motion.
Note that apical oxygen motions have already been
included as an $e_{\rm g}$-level splitting.
The extra electron-phonon coupling term is
$H_{\rm el-ph}^{e_{\rm g}}$, Eq.~(\ref{eph-eg}).
Since all oxygens are shared by adjacent NiO$_6$ octahedra,
the distortions are $not$ $independent$.
To consider such cooperative effect, in principle,
the O-ion displacements should be optimized.
However, in practice it is not
feasible to perform both the Lanczos diagonalization and
the optimization of all oxygen positions for 6- and 8-site clusters.
In the actual calculations, $Q_{1{\bf i}}$, $Q_{2{\bf i}}$, and
$Q_{3{\bf i}}$ are expressed by a single parameter $d$, for the shift
of the O-ion coordinate. Note that the unit of $d$ is $g/k$, typically
0.1$\sim$0.3\AA.
Then, the total energy is evaluated as a function of $d$
to find the minimum energy state.
Repeating these calculations for several distortion patterns,
it is possible to deduce the optimal state.

After several trials, the optimal distortion
at x=1/2 is shown in Fig.~\ref{fig18}(a).
The diagonalization has been performed at several values of
$d$ on the 8-site distorted lattice and the minimum in the total
energy is found at $d$=0.3, as shown in Fig.~\ref{fig18}(b).
As mentioned above, even without $H_{\rm el-ph}^{e_{\rm g}}$,
the checkerboard-type charge ordering has been obtained,
but the peak at $\bm{q}$=$(\pi, \pi)$ significantly grows
due to the effect of lattice distortions.
Note that the distortion pattern in Fig.~\ref{fig18}(a) is
essentially the same as that for half-doped manganites.
This is quite natural, since JT active and inactive ions exist
bipartitely also for half-doped nickelates.
Then, due to this JT-type distortion,
orbital ordering for half-doped nickelates is predicted,
as schematically shown in Fig.~\ref{fig18}(a).
The shapes of orbitals are determined from the orbital densities,
$\langle \tau_{z{\bf i}} \rangle$ and
$\langle \tau_{x{\bf i}} \rangle$,
as shown in Fig.~\ref{fig18}(c).
The well-known alternate pattern of $3x^2$$-$$r^2$ and
$3y^2$$-$$r^2$ orbitals
in half-doped manganites is denoted by dashed lines.
Increasing $d$, the shape of orbitals
deviates from $3x^2$$-$$r^2$ and $3y^2$$-$$r^2$,
but it is still characterized
by the orbitals elongating along the $x$- and $y$-directions.
See insets of Fig.~\ref{fig18}(c).
It would be very interesting to search for orbital ordering
in half-doped nickelates,
using the resonant X-ray scattering technique.

\begin{figure}[t]
\begin{center}
\includegraphics[width=0.7\textwidth]{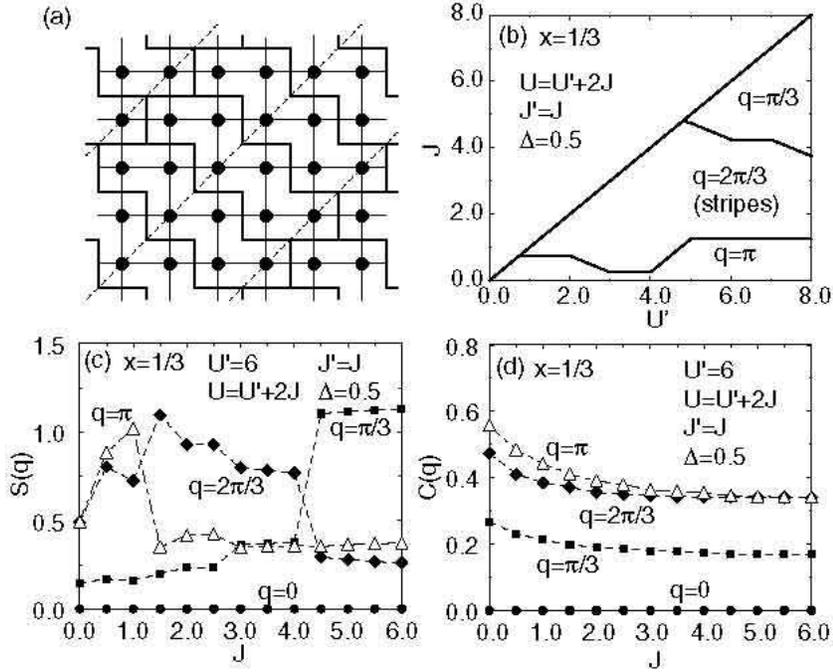}
\end{center}
\caption{(a) Zigzag 6-sites cluster covering the 2D lattice.
Black circles denote Ni ions, and dashed lines indicate hole positions.
(b) Phase diagram at x=1/3 without $H_{\rm el-ph}^{e_{\rm g}}$.
Each phase is characterized by the momentum that shows a peak in $S(q)$.
(c) $S(q)$ and (d) $C(q)$ vs. $J$ for $U'$=6 and $\Delta$=0.5.
}
\label{fig19}
\end{figure}

Now let us move to the case x=1/3.
If the actual expected stripe structure at x=1/3 is faithfully
considered
\cite{Tranquada1994,Sachan1995,Yoshizawa2000,Kajimoto2003},
it is necessary to analyze, at least, a 6$\times$6 cluster.
However, such a large-size cluster with orbital degeneracy
cannot be treated exactly due to the exponential growth of the
Hilbert space with cluster size.
Then, a covering of the two-dimensional (2D) lattice using
zigzag 6-sites clusters as shown in Fig.~\ref{fig19}(a)
is considered
by assuming a periodic structure along the diagonal direction.

First we consider the case without $H_{\rm el-ph}^{e_{\rm g}}$.
The phase diagram obtained by analyzing the zigzag 6-site cluster
for $H$ is in Fig.~\ref{fig19}(b).
Typical spin and charge correlations are
shown in Figs.~\ref{fig19}(c) and \ref{fig19}(d), where
$C(\bm{q})$=$(1/N)$$\sum_{\bf i,j}e^{i\bm{q}\cdot({\bf i}-{\bf j})}$
$\langle (n_{\bf i}-\langle n \rangle)$$\cdot$$(n_{\bf j}-
\langle n \rangle) \rangle$,
with $n_{\bf i}$=$\sum_{\gamma}n_{{\bf i}\gamma}$.

Since the momentum $q$ is defined along the zigzag direction
in the unit of $\sqrt{2}/a$, where $a$ is the lattice constant,
the phase labeled by $q$=$2\pi/3$ in Fig.~\ref{fig19}(b) denotes
an $incommensurate$ $AF$ $phase$ with the proper spin stripe structure.
The phase labeled by $q$=$\pi/3$ indicates
a spin spiral state, which will eventually turn to
the FM phase in the thermodynamic limit.
Thus, the spin stripe phase appears between the commensurate
AF and FM-like phases, similar to the case of x=1/2.
However, as seen in Fig.~\ref{fig19}(d),
$C(q)$ in the spin stripe phase
does $not$ show the striped charge structure ($q$=$2\pi/3$).
Rather, bipartite charge ordering characterized by
a peak at $q$=$\pi$ still remains.
Namely, the Hamiltonian without $H_{\rm el-ph}^{e_{\rm g}}$
can explain the spin stripe,
but does not reproduce the striped charge ordering at x=1/3,
indicating the importance of $H_{\rm eph}$.

\begin{figure}[t]
\begin{center}
\includegraphics[width=0.7\linewidth]{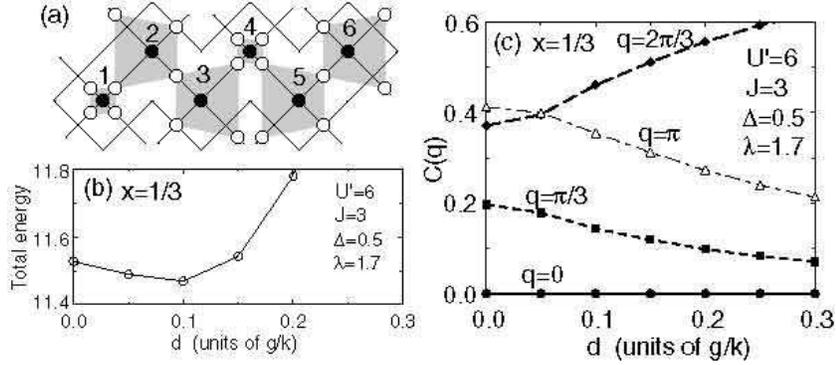}
\end{center}
\caption{
(a) Cooperative distortion pattern for the zigzag 6-sites cluster
at x=1/3.
(b) Total ground-state energy and (g) $C(q)$ vs. $d$ for x=1/3.
}
\label{fig20}
\end{figure}

Consider now the effect of $H_{\rm el-ph}^{e_{\rm g}}$ for x=1/3.
After evaluating total ground-state energies for several kinds of
distortions, the pattern in Fig.~\ref{fig20}(a) has been found to
provide the optimal state at x=1/3.
This type of distortion induces a spatial modulation of
the level splitting as
$-\delta_1/2$=$\delta_2$=$\delta_3$=$-\delta_4$/2=$\delta_5$
=$\delta_6$,
where $\delta_{\bf i}$ is the level splitting caused by the
in-plane oxygen motions.
Note that this breathing-mode modulation is consistent with experimental
results \cite{Tranquada1995b}.
The site numbers are found in Fig.~\ref{fig20}(a).
The minimum energy is found at $d$=0.1, as shown in Fig.~\ref{fig20}(b).
The modulation of level splitting stabilizes the striped charge
ordering characterized by a $q$=$2\pi/3$ peak in $C(q)$,
as clearly shown in Fig.~\ref{fig20}(c).

Note that ($3x^2$$-$$r^2$/$3y^2$$-$$r^2$)-type orbital ordering
does $not$ occur in Fig.~\ref{fig20}(a).
Phenomenologically, such orbital ordering tends to appear
in a hole pair separated by one site,
the unit of the ``bi-stripe'' of manganites \cite{Mori1998a,Mori1998b}.
However, such a bi-stripe-type ordering contradicts the x=1/3 striped 
charge-ordering, and the bi-stripe-type solution was found to be
unstable in these calculations.
One may consider other distortion patterns which satisfy
both ($3x^2$$-$$r^2$/$3y^2$$-$$r^2$)-type orbital and
striped charge-ordering, but in such distortions
no energy minimum was obtained for $d$$>$0.
After several trials, Fig.~\ref{fig20}(a) has provided
the most optimal state.

In summary, possible spin, charge, $and$ orbital structures of
layered nickelates have been discussed
based on the $e_{\rm g}$-orbital degenerate
Hubbard model coupled with lattice distortions.
To understand the nickelate stripes,
both Hund's rule interaction and electron-lattice coupling
appear essentially important.
At x=1/2, ($3x^2$$-$$r^2$/$3y^2$$-$$r^2$)-type
orbital ordering similar to that in half-doped manganites
is predicted. Even FM phases could be stabilized by
chemically altering the carrier's bandwidth.
For x=1/3, a spatial modulation in level splitting
plays an important role for stripe formation.

%
\subsection{$t_{\rm 2g}$ electron systems}

Let us now consider orbital ordering
in a system with active $t_{\rm 2g}$-orbital degree of freedom.
As is well known, $t_{\rm 2g}$-electron systems such as titanates
and vanadates have been studied for a long time.
After the discovery of superconductivity in layered cobalt oxyhydrate
Na$_{0.35}$CoO$_2$$\cdot$1.3H$_2$O \cite{Takada2003},
magnetic properties of cobaltites have been discussed intensively
both from experimental and theoretical sides.
In relation with cobaltites, one-dimensional $t_{\rm 2g}$ electron
model has been studied theoretically for the understanding of
spin and orbital state of $t_{\rm 2g}$ electrons
\cite{Onishi2004b,Onishi2005e}.
When we turn our attention to $4d$ electron systems,
Ru-oxides have been also focused, after the discovery of
triplet superconductivity in the layered ruthenate
Sr$_2$RuO$_4$ \cite{Maeno}.
In addition, the isostructural material Ca$_2$RuO$_4$ has been
studied as a typical stage of spin and orbital ordering
of $t_{\rm 2g}$ electrons.
In this subsection,
we review the orbital ordering phenomenon in Ca$_2$RuO$_4$.

As mentioned in Sec.~2, Ru$^{4+}$ ion include four electrons in
the low-spin state, since the crystalline electric field is
effectively larger than the Hund's rule interaction.
Thus, four electrons occupy $t_{\rm 2g}$ orbitals,
leading to $S$=1 spin.
The G-type AF phase in Ca$_2$RuO$_4$ is characterized as
a standard N\'eel state with spin $S$=1
\cite{Nakatsuji1,Nakatsuji2,Braden}.
The N\'eel temperature $T_{\rm N}$ is 125K.
To understand the N\'eel state observed in experiments,
one may consider the effect of the tetragonal crystal field,
leading to the splitting between xy and \{yz,zx\} orbitals,
where the xy-orbital state is lower in energy than the other levels.
When the xy-orbital is fully occupied, 
a simple superexchange interaction at strong Hund's rule coupling
can stabilize the AF state.
However, X-ray absorption spectroscopy studies have shown that
0.5 holes per site exist in the xy-orbital, while 1.5 holes are
contained in the zx- and yz-orbitals \cite{Mizokawa},
suggesting that the above naive picture
based on crystal field effects seems to be incomplete.
This fact suggests that the orbital degree of freedom may play
a more crucial role in the magnetic ordering in ruthenates than
previously anticipated.

First let us briefly review the result by Hotta and Dagotto
\cite{Hotta2002}.
The Hamiltonian is the $t_{\rm 2g}$ Hubbard model coupled with
Jahn-Teller distortions, already given by $H_{t_{\rm 2g}}$,
Eq.~(\ref{Model}) in Sec.~2.
This model is believed to provide a starting point to study
the electronic properties of ruthenates, but it is difficult
to solve even approximately.
To gain insight into this complex system, an unbiased technique
should be employed first.
Thus, Hotta and Dagotto have analyzed a small 2$\times$2 plaquette
cluster in detail by using the Lanczos algorithm
for the exact diagonalization, and the relaxation technique
to determine the oxygen positions.
In actual calculations, at each step for the relaxation, 
the electronic portion of the Hamiltonian is exactly
diagonalized for a fixed distortion.
Iterations are repeated until the system converges
to the global ground state.

\begin{figure}[t]
\begin{center}
\includegraphics[width=0.4\linewidth]{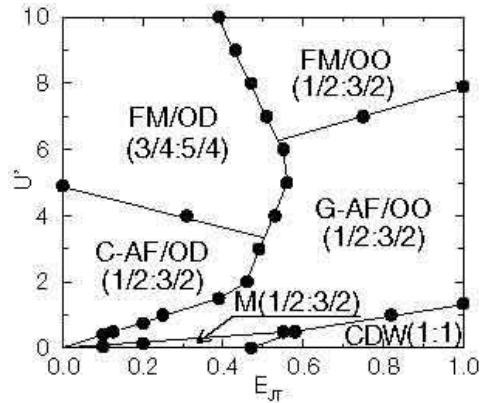}
\end{center}
\caption{%
Ground state phase diagram for the $t_{\rm 2g}$ Hubbard model
coupled with Jahn-Teller phonons for $J$=$3U'/4$.
The notation is explained in the maintext.
}
\label{fig21}
\end{figure}

The ground state phase diagram obtained by Hotta and Dagotto
is shown in Fig.~\ref{fig21}.
There are six phases in total,
which are categorized into two groups.
One group is composed of phases stemming from
the $U'$=0 or $E_{\rm JT}$=0 limits.
The origin of these phases will be addressed later, but first
their main characteristics are briefly discussed.
For $E_{\rm JT}$=0, a C-type AF orbital disordered (OD)
phase appears in the region of small and intermediate $U'$. 
This state is characterized by
$n_{\rm xy}$:$n_{\rm yz}$+$n_{\rm zx}$=1/2:3/2,
where $n_{\gamma}$ is the hole number per site at the $\gamma$-orbital.
Hereafter, a shorthand notation such as ``1/2:3/2'' is used to denote
the hole configuration.
For large $U'$, and still $E_{\rm JT}$=0,
a FM/OD phase characterized by 3/4:5/4 is stable,
which may correspond to Sr$_2$RuO$_4$. 
On the other hand, for $U'$=0 and small $E_{\rm JT}$,
a ``metallic'' (M) phase with small lattice distortion is observed,
while for large $E_{\rm JT}$, a charge-density-wave (CDW) state
characterized by 1:1 was found.
In short, the G-type AF phase observed experimentally \cite{Braden}
does {\it not} appear, neither for $E_{\rm JT}$=0 nor for $U'$=0.

Another group includes two phases which are not connected to either
$E_{\rm JT}$=0 or $U'$=0.
It is only in this group, with {\it both} lattice and Coulomb effects being
relevant, that for intermediate $U'$ the G-type AF and orbital ordered (OO)
phase with 1/2:3/2 found in experiments \cite{Mizokawa} is stabilized.
At larger $U'$, a FM/OO phase occurs with the same hole arrangement.
In the FM phase, since an $S$=1 spin with $S_z$=+1 is formed at each site,
the up-spin number is unity at each orbital,
while the down-spin distribution depends on the orbital.
In the AF state, the configuration of double-occupied orbitals is
the same as in the FM phase, but the single-occupied orbital contains
0.5 up- and 0.5-down spins on average,
since the $S$=1 spin direction fluctuates due to the AF
coupling between neighboring $S$=1 spins.
However, the spin correlations peak at ($\pi$,$\pi$),
indicating the G-AF structure.
Except for the spin direction, the charge and orbital configuration
in the FM/OO phase is the same as in the G-AF/OO state.
An antiferro-orbital ordering pattern including xy, yz, and zx orbitals
has been suggested for these FM and AF phases.

On the other hand, a ferro ``0:2'' xy-orbital ordered state has been
suggested by Anisimov {\it et al.} \cite{Anisimov2002}.
Fang et al. also predicted the ferro-type orbital ordering
\cite{Fang2004}.
It seems to be different from experiments on
the hole distribution in Ref.~\cite{Mizokawa},
but due to the combination of optical conductivity measurement
and LDA+U calculations \cite{Jung}, it has been found that
xy-orbital ferro ordering occurs and the change of hole population
can be explained due to the temperature dependence of
electronic structure.

Recently, Kubota et al. have performed the experiment to
determine the orbital ordering in Ca$_2$RuO$_4$ by using
the resonant X-ray scattering interference technique
at the $K$ edge of Ru \cite{Kubota2005}.
In this new and skillful technique, it is remarkable that
the $d_{xy}$ orbital ordering is observed even at room temperature,
in which the Jahn-Teller distortion is negligible.
Note here that the Jahn-Teller distortion is defined as the ratio
of the apical Ru-O bond length to the equatorial Ru-O bond length
in the RuO$_6$ octahedron.

\begin{figure}[t]
\begin{center}
\includegraphics[width=0.8\linewidth]{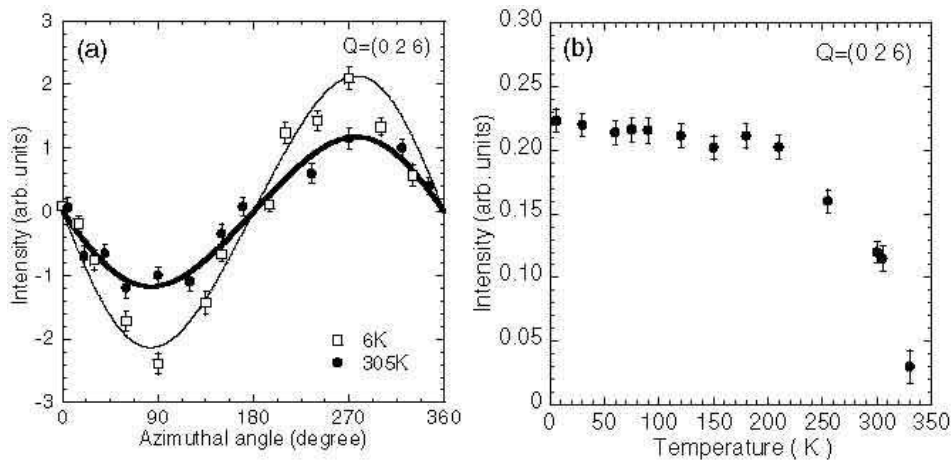}
\end{center}
\caption{%
(a) Azimuthal angle dependence of the interference term for a main edge
peak at 305 and 6 K at $Q$=$(0,2,6)$.
The thick and thin curves denote the analysis results at 305 and 6 K,
respectively.
(b) Temperature dependence of the interference term at $Q$=$(0,2,6)$.
As for details, readers refer Ref.~\cite{Kubota2005}
}
\label{fig22}
\end{figure}

The resonant x-ray scattering (RXS) measurement has been
very powerful method to detect the orbital ordering,
but the conventional RXS measurement is not useful for
the observation of a ferro-type orbital state,
since it is difficult to extract the signal for the ferro-orbital
ordered state at $\Gamma$ point in a momentum space,
which is accompanied with a large amplitude of a fundamental
reflection by Thomson scattering.
However, the RXS interference technique can observe the
ferro-type orbital ordered state, in which the signal
is magnified by the interference with a fundamental signal.

In Figs.~\ref{fig22}, we show typical results at the K edge of Ru,
obtained by Kubota et al.
Figure \ref{fig22}(a) denotes the azimuthal angle dependence of
the interference term for a main edge peak.
We note that the signal exhibits the characteristic oscillation
with the period of 360 degrees.
Moreover, the significant signal can be found even at 305 K,
in which the Jahn-Teller distortion is negligible.
This fact suggests that the interference term is directly
related to the orbital ordering.
In Fig.~\ref{fig22}(b), the temperature dependence of the RXS signal
is shown.
Below 200K, we can observe that the RXS signal for the ferro-type
xy-orbital ordering is almost saturated due to the
occurrence of the G-AF N\'eel state.
Above 200 K, the magnitude of the signal is gradually decreased
and becomes zero at the metal-insulator transition temperature
($\sim$357 K).
Since the apical bond length of RuO$_6$ becomes almost equal to
the averaged equatorial bond length around at 300K \cite{Braden},
it is difficult to consider that the Jahn-Teller distortion is
the origin of the orbital ordering.
Thus, the coupling of $t_{\rm 2g}$ electron
with Jahn-Teller distortion is not the primary term
of the electron-phonon coupling part.
Rather, the tilting and/or buckling modes should be
included seriously in the model Hamiltonian.


Finally, let us briefly mention another result of
a resonant X-ray diffraction study on Ca$_2$RuO$_4$
at the Ru $L_{\rm II}$ and $L_{\rm III}$ edges \cite{Zegkinoglou}.
Zegkinoglou et al. have observed a significant enhancement of
the magnetic scattering intensity at the wave vector
which characterizes the AF ordering.
Then, they have found a phase transition between two paramagnetic
phases around 260 K, in addition to the well-known
AF transition at $T_{\rm N}$=110K.
Due to the analysis of polarization and azimuthal angle dependence
of the diffraction signal, Zegkinoglou et al. have concluded that
the transition at 260K is attributed to the orbital ordering of
Ru $t_{\rm 2g}$ electrons.
This orbital order is characterized by the same propagation
vector as the low-temperature AF phase.
Note, however, that the ferro-orbital component of the ordering pattern
cannot be ruled out, as mentioned by Zegkinoglou et al.

\subsection{Geometrically frustrated systems}

As an important ingredient to understand novel magnetism
of actual strongly correlated electron materials,
thus far we have emphasized a potential role of
orbital degree of freedom,
when electrons partially fill degenerate orbitals.
However, on the lattice with {\it geometrical frustration},
a subtle balance among competing interactions easily
leads to a variety of interesting phenomena
such as unconventional superconductivity and exotic magnetism.
The recent discovery of superconductivity
in layered cobalt oxyhydrate Na$_{0.35}$CoO$_2$$\cdot$1.3H$_2$O
\cite{Takada2003} has certainly triggered intensive
investigations of superconductivity on the triangular lattice.
Concerning the magnetism,
antiferromagnetism on the triangle-based structure
has a long history of investigation \cite{Diep1994}.
In the low-dimensional system, the combined effect of
geometrical frustration and strong quantum fluctuation
is a source of peculiar behavior in low-energy physics,
as typically found in the Heisenberg zigzag chain with spin $S$=1/2.
As the strength of frustration is increased, the ground state is
known to be changed from a critical spin-liquid to a gapped dimer phase
\cite{Majumdar-Ghosh,Tonegawa-zigzag,Okamoto-zigzag,White-zigzag}.
In the dimer phase, neighboring spins form a valence bond to gain
the local magnetic energy, while the correlation among the valence
bonds is weakened to suppress the effect of spin frustration.

Here we have a naive question:
What happens in a system with $both$ active orbital degree
of freedom and geometrical frustration?
It is considered to be an intriguing issue to clarify the influence
of orbital ordering on magnetic properties
in geometrically frustrated systems.
For instance, significant role of $t_{\rm 2g}$-orbital degree of
freedom has been remarked to understand the mechanism of
two phase transitions in spinel vanadium oxides $A$V$_2$O$_4$
($A$=Zn, Mg, and Cd)
\cite{Tsunetsugu-AV2O4,Motome-AV2O4,Lee-ZnV2O4}.
It has been proposed that
orbital ordering brings a spatial modulation in the spin exchange
and spin frustration is consequently relaxed.
Similarly, for MgTi$_2$O$_4$, the formation of a valence-bond crystal
due to orbital ordering has been also suggested
\cite{Matteo-MgTi2O4,Mizokawa2005}.

Since $d$- and $f$-electron orbitals are spatially anisotropic,
there always exist easy and hard directions for electron motion.
Thus, it is reasonable to expect that the effect of geometrical
frustration would be reduced due to orbital ordering,
depending on the lattice structure and the type of orbital,
in order to arrive at the spin structure which minimizes
the influence of frustration.
However, the spin structure on such an orbital-ordered background
may be fragile, since the effect of geometrical frustration never
vanishes, unless the lattice distortion is explicitly taken
into account.
It is a highly non-trivial problem,
whether such an orbital arrangement actually describes
the low-energy physics of geometrically frustrated systems.
In particular, it is important to clarify
how the orbital-arranged background is intrinsically stabilized
through the spin-orbital correlation
even without the electron-lattice coupling.
In this subsection, we review the recent result by
Onishi and Hotta concerning the role of orbital ordering
in the geometrically frustrated lattice
\cite{Onishi2005a,Onishi2005b,Onishi2005c,Onishi2005d}.

\begin{figure}[t]
\begin{center}
\includegraphics[width=0.8\linewidth]{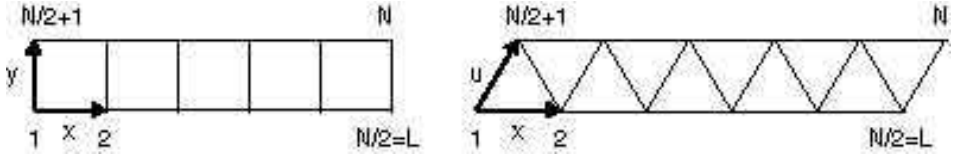}
\end{center}
\caption{
Lattice location and site numbering of $N$-site ladder
and zigzag chain. The length is defined as $L$=$N/2$.
}
\label{fig23}
\end{figure}

Onishi and Hotta have considered an $e_{\rm g}$-orbital model
on the $N$-site ladder or zigzag chain,
including one electron per site with two orbitals,
i.e., quarter filling.
The lattice they have used is shown in Fig.~\ref{fig23}.
Note that the zigzag chain is composed of equilateral triangles.
The $e_{\rm g}$-orbital degenerate Hubbard model is already
given by Eq.~(\ref{Model}), but the electron-phonon term is
not considered.
Namely, the model is written as
\begin{equation}
  H_{e_{\rm g}} = H^{e_{\rm g}}_{\rm kin} + H^{e_{\rm g}}_{\rm el-el}.
\end{equation}
Here the $d$-electron hopping amplitude
$t_{\gamma,\gamma'}^{\bf a}$ for the oblique $u$ direction
is defined by $t_{aa}^{\bf u}$=$t_1/4$,
$t_{ab}^{\bf u}$=$t_{ba}^{\bf u}$=$\sqrt{3}t_1/8$,
$t_{bb}^{\bf u}$=$3t_1/16$.
Note the relation of
$t_{\gamma\gamma'}^{\bf u-x}$=$t_{\gamma\gamma'}^{\bf u}$.
Concerning hopping amplitudes along $x$- and $y$-directions,
see Eqs.~(\ref{eg-hop-x}) and (\ref{eg-hop-y}).
In this subsection, $t_1$ is taken as the energy unit.

In order to analyze the complex model including both
orbital degree of freedom and geometrical frustration,
Onishi and Hotta have employed the finite-system density matrix
renormalization group (DMRG) method, which is appropriate for
the analysis of quasi-one-dimensional systems
with the open boundary condition
\cite{White-DMRG-1,White-DMRG-2,Liang-DMRG}.
Since one site includes two $e_{\rm g}$ orbitals and
the number of bases is 16 per site,
the size of the superblock Hilbert space becomes
very large as $m^2$$\times$$16^2$,
where $m$ is the number of states kept for each block.
To accelerate the calculation and to save memory resources,
Onishi has skillfully reduced the size of the superblock Hilbert space
to $m^2$$\times$$4^2$, by treating each orbital as an effective site.
In the actual calculations, $m$ states up to $m$=200 were kept
in the renormalization process
and the truncation error was estimated to be $10^{-5}$ at most.

\begin{figure}[t]
\begin{center}
\includegraphics[width=0.8\linewidth]{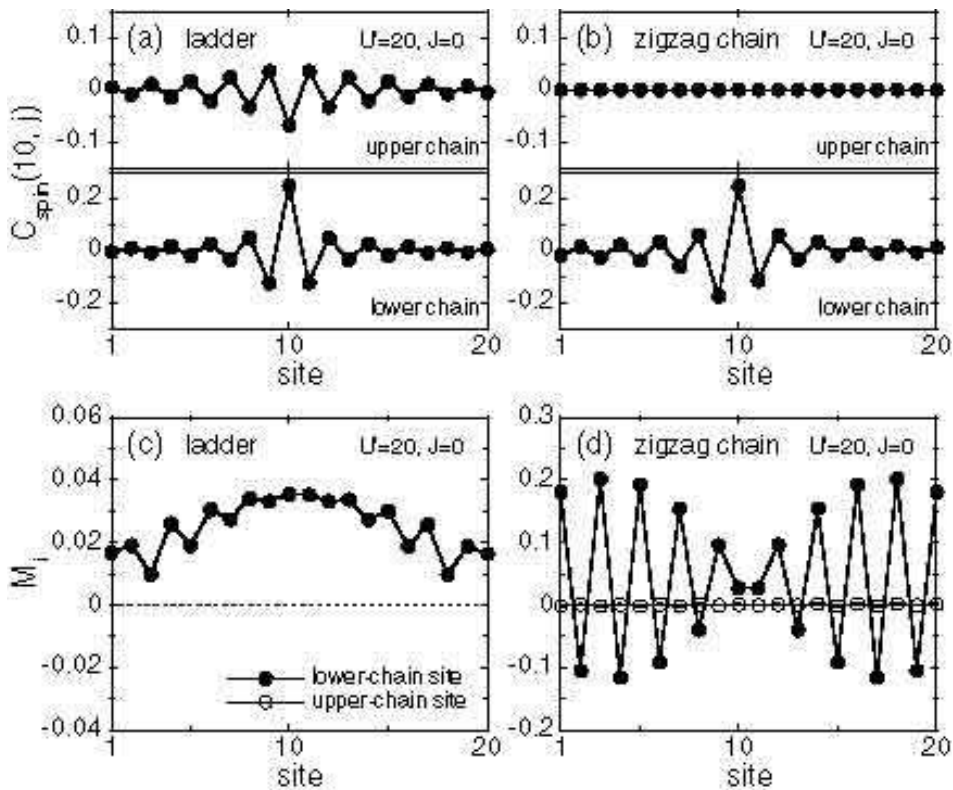}
\end{center}
\caption{
The spin-correlation function measured from the center
of the lower chain for the PM ground state
in (a) the ladder and (b) the zigzag chain.
The local magnetization for the first spin-excited state
in (c) the ladder and (d) the zigzag chain.
}
\label{fig24}
\end{figure}

Now we introduce the results on the spin structure of
the paramagnetic (PM) ground state at $J$=0,
since the zigzag chain is relevant to
a geometrically frustrated antiferromagnet in the spin-singlet PM phase.
Here we refer only the result for $J$=0, but readers
should consult with Ref.~\cite{Onishi2005a}
about the results for $J$$\ne$0.
In Figs.~\ref{fig24}(a) and \ref{fig24}(b),
we show the DMRG results for $N$=40 of the spin-correlation function
$C_{\rm spin}({\bf i},{\bf j})$=
$\langle S_{\bf i}^z S_{\bf j}^z \rangle$
with
$S_{\bf i}^z$=$\sum_{\gamma}(\rho_{{\bf i}\gamma\uparrow}$$-$
$\rho_{{\bf i}\gamma\downarrow})/2$.
Note that a large value of $U'$=20 was used to consider
the strong-coupling region, but the results did not change qualitatively
for smaller values of $U'$.
As shown in Fig.~\ref{fig24}(a), we observe a simple N\'eel structure
in the ladder.
On the other hand, in the zigzag chain, there exists AF correlation
between intra-chain sites in each of lower and upper chains,
while the spin correlation between inter-chain sites is much weak.
Namely, the zigzag chain is considered to be decoupled
to a double chain in terms of the spin structure.

In order to clarify the characteristics of the spin structure
in the excited state, Onishi and Hotta have investigated
the local magnetization
$M_{\bf i}$=$\langle S_{\bf i}^z \rangle$
for the lowest-energy state with $S_{\rm tot}^z$=1,
i.e., the first spin-excited state,
where $S_{\rm tot}^z$ is the $z$ component of the total spin.
In the ladder, the total moment of $S_{\rm tot}^z$=1 is distributed
to the whole system and there is no significant structure,
as shown in Fig.~\ref{fig24}(c).
On the other hand, the situation is drastically changed
in the zigzag chain.
As shown in Fig.~\ref{fig24}(d), the total moment of
$S_{\rm tot}^z$=1 is confined in the lower chain
and it forms a sinusoidal shape with a node,
while nothing is found in the upper chain.
Note that the sinusoidal shape of the local magnetization is
characteristic of the $S$=1/2 AF Heisenberg chain with edges
at low temperatures \cite{Laukamp-edge,Nishino-edge}.
Thus, the double-chain nature in the spin structure remains robust
even for the spin-excited state.

Onishi and Hotta have also discussed the orbital arrangement
to understand the mechanism of the appearance of the spin structures.
For the determination of the orbital arrangement,
orbital correlations are usually measured,
but due care should be paid to the definition.
By analogy with Eqs.~(\ref{trans1}) and (\ref{trans2}) which have
treated the phase of the JT distortions,
phase-dressed operators are introduced as
\begin{equation}
 \left\{
 \begin{array}{l}
 \tilde{d}_{{\bf i}a\sigma}=
 e^{i\theta_{\bf i}/2}[\cos(\theta_{\bf i}/2)d_{{\bf i}a\sigma}+
 \sin(\theta_{\bf i}/2)d_{{\bf i}b\sigma}],
 \\
 \tilde{d}_{{\bf i}b\sigma}=
 e^{i\theta_{\bf i}/2}[-\sin(\theta_{\bf i}/2)d_{{\bf i}a\sigma}+
 \cos(\theta_{\bf i}/2)d_{{\bf i}b\sigma}].
 \end{array}
 \right.
\end{equation}
Then, the optimal set of $\{\theta_{\bf i}\}$ is determined
so as to maximize the orbital-correlation function,
which is defined as
\begin{equation}
 T({\bf q})=
 (1/N^{2})\sum_{{\bf i},{\bf j}}
 \langle \tilde{T}_{\bf i}^{z}\tilde{T}_{\bf j}^{z} \rangle
 e^{i{\bf q}\cdot({\bf i}-{\bf j})},
\end{equation}
with $\tilde{T}_{\bf i}^z$=$\sum_{\sigma}
(\tilde{d}_{{\bf i}a\sigma}^{\dag}\tilde{d}_{{\bf i}a\sigma}$$-$%
$\tilde{d}_{{\bf i}b\sigma}^{\dag}\tilde{d}_{{\bf i}b\sigma})/2$.

\begin{figure}[t]
\begin{center}
\includegraphics[width=0.9\linewidth]{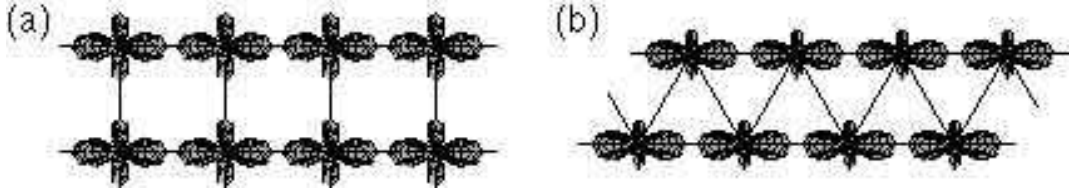}
\end{center}
\caption{
Optimal orbital arrangement in (a) the ladder and
(b) the zigzag chain.
}
\label{fig25}
\end{figure}

As shown in Fig.~\ref{fig25}(a), in the case of the ladder,
Onishi and Hotta have found that a ferro-orbital (FO) ordering,
characterized by $\theta_{\bf i}/\pi$$\sim$1.18,
appears in the ground state.
In the first spin-excited state, the FO structure also appears,
but the angle characterizing the orbital shape is slightly
changed as $\theta_{\bf i}/\pi$$\sim$1.20
to further extend to the leg direction.
On the other hand, in the zigzag chain,
it is observed that $both$ in the ground and
first spin-excited states, $T({\bf q})$ becomes maximum
at ${\bf q}$=0 with $\theta_{\bf i}/\pi$$\sim$$1.32$,
indicating a 3$x^2$$-$$r^2$ orbital at each site,
as shown in Fig.~\ref{fig25}(b).
Note that the orbital arrangement is unchanged
even in the spin-excited state.
Namely, the orbital degree of freedom spontaneously becomes
``dead'' in low-energy states to suppress the effect of
spin frustration.

It is interesting to remark that the spin-exchange interactions
become anisotropic due to the orbital-arranged background.
First let us consider the zigzag chain, which is effectively
described by the Hubbard model composed of 3$x^2$$-$$r^2$ orbital.
It is intuitively understood that the AF exchange interaction
along the $u$ direction $J_1$ should be much weaker than
that along the $x$ direction $J_2$,
since the orbital shape extends along the double chain,
not along the zigzag path, as shown in Fig.~\ref{fig25}(b).
In order to estimate the ratio of $J_1/J_2$,
it is enough to consider the hopping amplitudes between adjacent
optimal ``a''-orbitals, $3x^2$$-$$r^2$ in this case,
which are given by
$\tilde{t}_{aa}^{\bf x}$=1 and $\tilde{t}_{aa}^{\bf u}$=1/64.
Then, taking account of the second-order process in terms of
electron hopping between only 3$x^2$$-$$r^2$ orbitals, we obtain
$J_1/J_2$
=$[2(\tilde{t}_{aa}^{\bf u})^2/U]/[2(\tilde{t}_{aa}^{\bf x})^2/U]$
=$1/64^2$.
This small value of $J_1/J_2$ clearly indicates that
the spin correlation on the zigzag path is
reduced due to the spatial anisotropy of 3$x^2$$-$$r^2$ orbital.
Thus, the zigzag chain is effectively reduced to
a double-chain system of the $S$=1/2 AF Heisenberg chain,
suggesting that the spin gap should be extremely suppressed,
since the spin gap decreases exponentially with the increase of $J_2/J_1$
in the gapped dimer phase in the zigzag spin chain \cite{White-zigzag}.

On the other hand, in the ladder with the ferro-orbital structure
as shown in Fig.~\ref{fig25}(a),
the orbital shape extends to the rung direction
as well as to the leg direction.
When we define $J_{\rm leg}$ and $J_{\rm rung}$
as the AF exchange interactions
along the leg and rung directions, respectively,
we obtain $J_{\rm rung}/J_{\rm leg}$=0.26,
which is much larger than $J_1/J_2$=$1/64^2$
in the case of the zigzag chain.
The spin correlation on the rung is considered to remain finite,
leading to the simple N\'eel structure.
Thus, the spin excitation in the ladder is expected to be gapful
similar to the spin ladder \cite{Barnes-ladder,Greven-ladder}.

We have reviewed both ground- and excited-state properties
of the $e_{\rm g}$-orbital degenerate Hubbard model
on the ladder and the zigzag chain.
It has been found that the zigzag chain is reduced to
a decoupled double-chain spin system
due to the selection of a specific orbital.
It is considered as a general feature of geometrically frustrated
multi-orbital systems that the orbital selection spontaneously
occurs so as to suppress the effect of spin frustration.

Finally, let us briefly comment on the effect of level splitting
between $x^2$$-$$y^2$ and $3z^2$$-$$r^2$ orbitals,
which has not been considered in the present Hamiltonian.
In particular, when $3z^2$$-$$r^2$ orbital is the lower
level which is well separated from $x^2$$-$$y^2$ orbital,
the hopping amplitude does not depend on the direction
and the effect of spin frustration revives for the system
with isotropic AF interactions.
In such a region with strong spin frustration, a finite energy gap
between ground and first-excited states can be clearly observed.
Naively thinking, it may be called a spin gap,
but we should note that the orbital arrangement is significantly
influenced by the spin excitation.
In general, the energy gap between ground and first-excited states
in multi-orbital systems should be called a {\it spin-orbital} gap.
As for details, readers consult with Ref.~\cite{Onishi2006b}

%
%
\section{Model Hamiltonian for $f$-electron systems}

Thus far, we have reviewed the theoretical results
on orbital ordering phenomena of $d$-electron systems.
As typical examples, we have picked up manganites, nickelates,
and ruthenates.
However, there exists another spin-charge-orbital complex system
such as $f$-electron compounds.
In the latter half of this article, we review orbital ordering
phenomena of $f$-electron systems.
Before proceeding to the description of the theoretical results,
again it is necessary to set the model Hamiltonian for $f$-electron
systems.
In order to construct such a microscopic Hamiltonian,
we must include simultaneously the itinerant nature of $f$ electrons
as well as the effects of strong electron correlation, CEF,
and spin-orbit interaction.
Among them, the existence of strong spin-orbit interaction
is essential difference from $d$-electron systems.
The inclusion of the spin-orbit interaction is a key issue,
when we construct the model Hamiltonian for $f$ electron materials.
Since this is a complicated problem, it is instructive to start the
discussion with a more basic level.
Namely, we first review in detail the single ion problem
focusing on the properties of local $f$-electron states
in comparison with those obtained
in the $LS$ and $j$-$j$ coupling schemes.
Then, we move on to the explanation of the microscopic $f$-electron model
on the basis of the $j$-$j$ coupling scheme.

%
\subsection{$LS$ vs. $j$-$j$ coupling schemes}

In the standard textbook, it is frequently stated that
for rare-earth ion systems, the $LS$ coupling scheme works well,
while for actinides, in particular, heavy actinides,
the $j$-$j$ coupling scheme becomes better.
However, do we simply accept such a statement?
Depending on the level of the problem in the condensed matter physics,
the validity of the approximation should be changed,
but such a point has not been explained in the textbook.
It is important to clarify which picture is appropriate
for the purpose to consider the many-body phenomena in
$f$-electron systems.

Let us generally consider the $f^n$ configuration,
where $n$ is the number of $f$ electrons included on a localized ion.
In the $LS$ coupling scheme, first the spin ${\bm S}$ and angular momentum
${\bm L}$ are formed due to Hund's rules as
${\bm S}$=$\sum_{i=1}^n {\bm s}_i$ and
${\bm L}$=$\sum_{i=1}^n {\bm \ell}_i$,
where ${\bm s}_i$ and ${\bm \ell}_i$ are spin and angular momentum
for $i$-th $f$ electron.
Note that the Hund's rules are based on the Pauli principle and
Coulomb interactions among $f$ electrons.
After forming ${\bm S}$ and ${\bm L}$,
we include the effect of spin-orbit interaction,
given by $\lambda$${\bm L}$$\cdot$${\bm S}$,
where $\lambda$ is the spin-orbit coupling.
We note that $\lambda$$>$0 for $n$$<$7, while $\lambda$$<$0 for $n$$>$7.
Note also that a good quantum number to label such a state is the total
angular momentum ${\bm J}$, given by ${\bm J}$=${\bm L}$+${\bm S}$.
Following from simple algebra, the ground-state level is characterized by
$J$=$|L$$-$$S|$ for $n$$<$7, while $J$=$L$+$S$ for $n$$>$7.

On the other hand, when the spin-orbit interaction becomes
larger than the Coulomb interactions,
it is useful to consider the problem in the $j$-$j$ coupling scheme.
First, we include the spin-orbit coupling so as to define the state
labeled by the total angular momentum ${\bm j}_i$ for the $i$-th electron,
given by ${\bm j}_i$=${\bm s}_i$+${\bm \ell}_i$.
For $f$-orbitals with $\ell$=3, we immediately obtain an octet with
$j$=7/2(=3+1/2) and a sextet with $j$=5/2(=3$-$1/2),
which are well separated by the spin-orbit interaction.
Note here that the level for the octet is higher than that of the sextet.
Then, we take into account the effect of Coulomb interactions
to accommodate $n$ electrons among the sextet and/or octet,
leading to the ground-state level in the $j$-$j$ coupling scheme.

As is easily understood from the above discussion,
the $LS$ coupling scheme works well under the assumption that
the Hund's rule coupling is much larger than the spin-orbit interaction,
since ${\bm S}$ and ${\bm L}$ are formed by the Hund's rule coupling
prior to the inclusion of spin-orbit interaction.
It is considered that this assumption is valid for
insulating compounds with localized $f$ electrons.
However, when the spin-orbit interaction is not small compared with
the Hund's rule coupling, the above assumption is not always satisfied.
In addition, if the $f$ electrons become itinerant owing to hybridization
with the conduction electrons, the effect of Coulomb interactions would
thereby be effectively reduced.
In rough estimation, the effective size of the Coulomb interaction may
be as large as the bandwidth of $f$ electrons, leading to a violation of
the assumption required for the $LS$ coupling scheme.

Furthermore, even in the insulating state, we often encounter
some difficulties to understand the complex magnetic phases
of $f$-electron systems with active {\it multipole} degrees of freedom
from a microscopic viewpoint.
In a phenomenological level, it is possible to analyze
a model for relevant multipoles obtained from the $LS$ coupling
scheme, in order to explain the phenomena of multipole ordering.
However, it is difficult to understand the origin of the
interaction between multipoles in the $LS$ coupling scheme.

From these viewpoints, it seems to be rather useful to exploit
the $j$-$j$ coupling scheme for the purpose to understand
magnetism and superconductivity of $f$-electron materials.
Since individual $f$-electron states is clearly defined,
it is convenient for including many-body effects using the standard
quantum-field theoretical techniques.
However, it is not the reason to validate to use the $j$-$j$
coupling scheme for the model construction.
In order to clarify how the $j$-$j$ coupling scheme works,
it is necessary to step back to the understanding of
the local $f$-electron state.
In the next subsection, let us consider this issue in detail.

%
\subsection{Local f-electron state}

In general, the local $f$-electron term is composed of three parts as
\begin{eqnarray}
   H_{\rm f} = H_{\rm el-el} + H_{\rm so} + H_{\rm CEF},
\end{eqnarray}
where $H_{\rm C}$ is the Coulomb interaction term, written as
\begin{eqnarray}
   H_{\rm el-el} =
   \sum_{\bf i}\sum_{m_1 \sim m_4}\sum_{\sigma_1, \sigma_2}
   I^f_{m_1,m_2,m_3,m_4}
   f_{{\bf i}m_1\sigma_1}^{\dag}f_{{\bf i}m_2\sigma_2}^{\dag}
   f_{{\bf i}m_3\sigma_2}f_{{\bf i}m_4\sigma_1}.
\end{eqnarray}
Here $f_{{\bf i}m\sigma}$ is the annihilation operator for $f$-electron
with spin $\sigma$ and angular momentum $m$(=$-3$,$\cdots$,3)
at a site ${\bf i}$.
Similar to the $d$-electron case, the Coulomb integral
$I^f_{m_1,m_2,m_3,m_4}$ is given by
\begin{eqnarray}
   I^{f}_{m_1,m_2,m_3,m_4}=\delta_{m_1+m_2,m_3+m_4}
   \sum_{k=0}^{6} F^k_f c^{(k)}(m_1,m_4)c^{(k)}(m_2,m_3),
\end{eqnarray}
where the sum on $k$ includes only even values ($k$=0, 2, 4, and 6),
$F^k_f$ is the Slater-Condon parameter for $f$ electrons
including the complex integral of the radial function,
and $c^k$ is the Gaunt coefficient.
It is convenient to express the Slater-Condon parameters as
\begin{eqnarray}
 \begin{array}{l}
    F^0_f=A+15C+9D/7, \\
    F^2_f=225(B-6C/7+D/42), \\
    F^4_f=1089(5C/7+D/77), \\
    F^6_f=(429/5)^2\cdot(D/462),
 \end{array}
\end{eqnarray}
where $A$, $B$, $C$, and $D$ are the Racah parameters
for $f$ electrons \cite{Racah1942}.

The spin-orbit coupling term, $H_{\rm so}$, is given by
\begin{eqnarray}
   H_{\rm so} = \sum_{\bf i} \sum_{m,\sigma,m',\sigma'}
   \lambda_{\rm so} \zeta_{m,\sigma,m',\sigma'}
   f_{{\bf i}m\sigma}^{\dag}f_{{\bf i}m'\sigma'},
\end{eqnarray}
where $\lambda_{\rm so}$ is the spin-orbit interaction
and the matrix elements are explicitly given by
\begin{eqnarray}
  \begin{array}{l}
   \zeta_{m,\sigma,m,\sigma}=m\sigma/2,\\
   \zeta_{m+1,\downarrow,m,\uparrow}=\sqrt{12-m(m+1)}/2,\\
   \zeta_{m-1,\uparrow,m,\downarrow}=\sqrt{12-m(m-1)}/2,
  \end{array}
\end{eqnarray}
and zero for other cases.

The CEF term $H_{\rm CEF}$ is given by
\begin{eqnarray}
  \label{Eq:CEF}
  H_{\rm CEF} = \sum_{{\bf i},m,m',\sigma} A_{m,m'}
  f_{{\bf i}m\sigma}^{\dag}f_{{\bf i}m'\sigma},
\end{eqnarray}
where $A_{m,m'}$ can be evaluated in the same manner as has done
in Sec.~2 for $d$ electrons with $\ell$=2.
However, there is no new information, if we repeat here lengthy
calculations for $f$ electrons with $\ell$=3.
As already mentioned in Sec.~2, it is rather useful and convenient
to consult with the table of Hutchings for angular momentum $J$=3
\cite{Hutchings}.
For cubic symmetry, $A_{m,m'}$ is expressed by using a couple of
CEF parameters, $B_{4}^0$ and $B_{6}^0$, as
\begin{eqnarray}
  \begin{array}{l}
    A_{3,3}=A_{-3,-3}=180B_{4}^0+180B_{6}^0, \\
    A_{2,2}=A_{-2,-2}=-420B_{4}^0-1080B_{6}^0, \\
    A_{1,1}=A_{-1,-1}=60B_{4}^0+2700B_{6}^0, \\
    A_{0,0}=360B_{4}^0-3600B_{6}^0, \\
    A_{3,-1}=A_{-3,1}=60\sqrt{15}(B_{4}^0-21B_{6}^0),\\
    A_{2,-2}=A_{-2,2}=300B_{4}^0+7560B_{6}^0.
  \end{array}
\end{eqnarray}
Following the traditional notation, we define
\begin{eqnarray}
  \begin{array}{l}
    B_{4}^0=Wx/F(4),\\
    B_{6}^0=W(1-|x|)/F(6),\\
  \end{array}
\end{eqnarray}
where $x$ specifies the CEF scheme for $O_{\rm h}$ point group,
while $W$ determines an energy scale for the CEF potential.
Although $F(4)$ and $F(6)$ have not been determined uniquely,
we simply follow the traditional definitions as
$F(4)$=15 and $F(6)$=180 for $J$=3 \cite{Hutchings}.

Here we note that the CEF potential is originally given
by the sum of electrostatic energy from the ligand ions
at the position of $f$-electron ion,
leading to the one-electron potential acting on the charge distribution
of $f$-orbitals, as expressed by Eq.~(\ref{Eq:CEF}).
Thus, in principle, it is not necessary to change the CEF potential,
depending on the $f$-electron number.
As we will see later, the CEF schemes
for $n$=1$\sim$13 are automatically reproduced by diagonalizing
the local $f$-electron term $H_{\rm loc}$,
once we fix the CEF parameters in the form of one-electron
potential Eq.~(\ref{Eq:CEF}).

Now we compare the electronic states of $H_{\rm loc}$
with those of $LS$ and $j$-$j$ coupling schemes.
We believe that it is quite instructive to understand the meanings
of the CEF potential in $f$-electron systems.
We introduce ``$U$'' as an energy scale
for the Racah parameters, $A$, $B$, $C$, and $D$.
In this subsection, $U$ is the energy unit, which is typically
considered to be 1 eV.
In $f$-electron compounds,
the magnitude of the CEF potential is much smaller than
both spin-orbit coupling and Coulomb interactions.
Thus, it is reasonable to consider that $W$ is always
much smaller than $\lambda_{\rm so}$ and $U$.
However, there occur two situations, depending on the order
for taking
the limits of $\lambda_{\rm so}/W$$\rightarrow$$\infty$ and
$U/W$$\rightarrow$$\infty$.
When the limit of $U/W$$\rightarrow$$\infty$ is first taken and
then, we include the effect of the spin-orbit coupling $\lambda_{\rm so}$,
we arrive at the $LS$ coupling scheme.
On the other hand, it is also possible to take first
the infinite limit of $\lambda_{\rm so}/W$.
After that, we include the effect of Coulomb interaction,
leading to the $j$-$j$ coupling scheme.
In the present local $f$-electron term $H_{\rm loc}$,
it is easy to consider two typical situations for
$f$-electron problems,
$|W|$$\ll$$\lambda_{\rm so}$$<$$U$ and
$|W|$$\ll$$U$$<$$\lambda_{\rm so}$,
corresponding to the $LS$ and $j$-$j$ coupling schemes, respectively.

\begin{figure}[t]
\begin{center}
\includegraphics[width=0.8\textwidth]{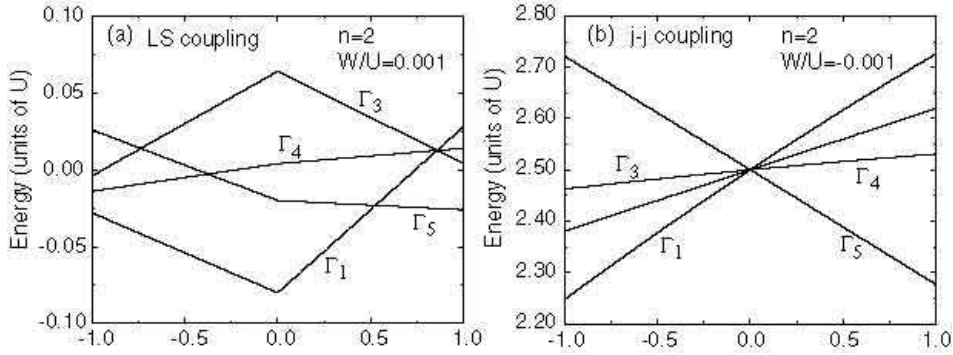}
\caption{%
Energies of $f$ electrons as functions of $x$ for
(a) the $LS$ coupling and (b) the $j$-$j$ coupling schemes for $n$=2.
The magnitude of the CEF potential energy is fixed as $|W|/U$=0.001.
}
\label{fig5-1}
\end{center}
\end{figure}

Let us consider the case of $n$=2 as a typical example of
the comparison between the two schemes.
In the $LS$ coupling scheme for the $f^2$-electron system,
we obtain the ground-state level as $^{3}H$ with $S$=1 and $L$=5
from the Hund's rules, where $S$ and $L$ denote sums of
$f$-electron spin and angular momentum, respectively.
Upon further including the spin-orbit interaction, the ground state
is specified by $J$=4 expressed as $^3H_4$ in the traditional notation.
Note that the total angular momentum $J$ is given by $J$=$|L-S|$ and
$J$=$L+S$ for $n$$<$7 and $n$$>$7, respectively.
In order to consider further the CEF effect, we consult with
the table of Hutchings for the case of $J$=4.
In the $LS$ coupling scheme, $W$ is taken as $W/U$=0.001 and
we set $F(4)$=60 and $F(6)$=1260 for $J$=4
by following the traditional definitions \cite{Hutchings}.
Then, we easily obtain the nine eigen values, including $\Gamma_1$ singlet,
$\Gamma_{3}$ doublet, and two kinds of triplets, $\Gamma_4$ and $\Gamma_5$,
as shown in Fig.~\ref{fig5-1}(a).
Note that for odd $n$, the eigenstate has odd parity,
specified by ``u'' in Mulliken's notation and
``$-$'' in Bethe's notation,
while the even $n$ configuration has even parity,
labeled by ``g'' and ``$+$'' \cite{Burns}.
When we use Bethe's notation to specify the $f$-electron eigenstate,
the ``$+$'' or ``$-$'' superscript is suppressed for convenience.

In the $j$-$j$ coupling scheme, on the other hand,
first we take the infinite limit of $\lambda_{\rm so}$.
Thus, we consider only the $j$=5/2 sextet, where $j$ denotes the total
angular momentum of one $f$ electron.
In the $f^2$-electron system, two electrons are accommodated in
the sextet, leading to fifteen eigen states including
$J$=4 nontet, $J$=2 quintet, and $J$=0 singlet.
Due to the effect of Hund's rule coupling,
$J$=4 nontet should be the ground state.
When we further include the CEF potential, it is necessary to reconsider
the accommodations of two electrons in the $f^1$-electron potential
with $\Gamma_7$ doublet and $\Gamma_8$ quartet.
Thus, in the $j$-$j$ coupling schemes, except for the energy scale $W$,
only relevant CEF parameter is $x$, leading to the level splitting
between $\Gamma_7$ doublet and $\Gamma_8$ quartet.
For the $j$-$j$ coupling scheme,
we set $F(4)$=60 and $W/U$=$-0.001$.
Note that the minus sign in $W$ is added for the purpose of
the comparison with the $LS$ coupling scheme.
As shown in Fig.~\ref{fig5-1}(b), the $J$=4 nontet is split into
$\Gamma_1$ singlet, $\Gamma_{3}$ doublet, $\Gamma_4$ triplet,
and $\Gamma_5$ triplet.
The ground state for $x$$>$0 is $\Gamma_5$ triplet composed of
a couple of $\Gamma_8$ electrons, while for $x$$<$0, it is $\Gamma_1$
singlet which is mainly composed of two $\Gamma_7$ electrons.
Note that for $x$$>$0, the first excited state is $\Gamma_4$ triplet,
composed of $\Gamma_7$ and $\Gamma_8$ electrons.

At the first glance,
the energy levels in the $j$-$j$ coupling scheme seems to be
different from those of the $LS$ coupling scheme.
How do we connect these different results?
In order to answer to this question,
let us directly diagonalize $H_{\rm f}$
by changing $U$ and $\lambda$.
Here it is convenient to introduce a new parameter to connect the
$LS$ and $j$-$j$ coupling schemes as
\begin{equation}
   k=\frac{\lambda_{\rm so}/|W|}{U/|W|+\lambda_{\rm so}/|W|},
\end{equation}
where we explicitly show $|W|$ in this formula,
since both $U$ and $\lambda_{\rm so}$ should be always very large
compared with $|W|$ in actual $f$-electron compounds.
Note that $k$=0 and 1 are corresponding to the limits of
$\lambda_{\rm so}/U$=0 and $\lambda_{\rm so}/U$=$\infty$, respectively.
Then, we can control the change of two schemes by one parameter $k$,
by keeping $U/|W|$$\gg$1 and $\lambda_{\rm so}/|W|$$\gg$1.

\begin{figure}[t]
\begin{center}
\includegraphics[width=0.8\textwidth]{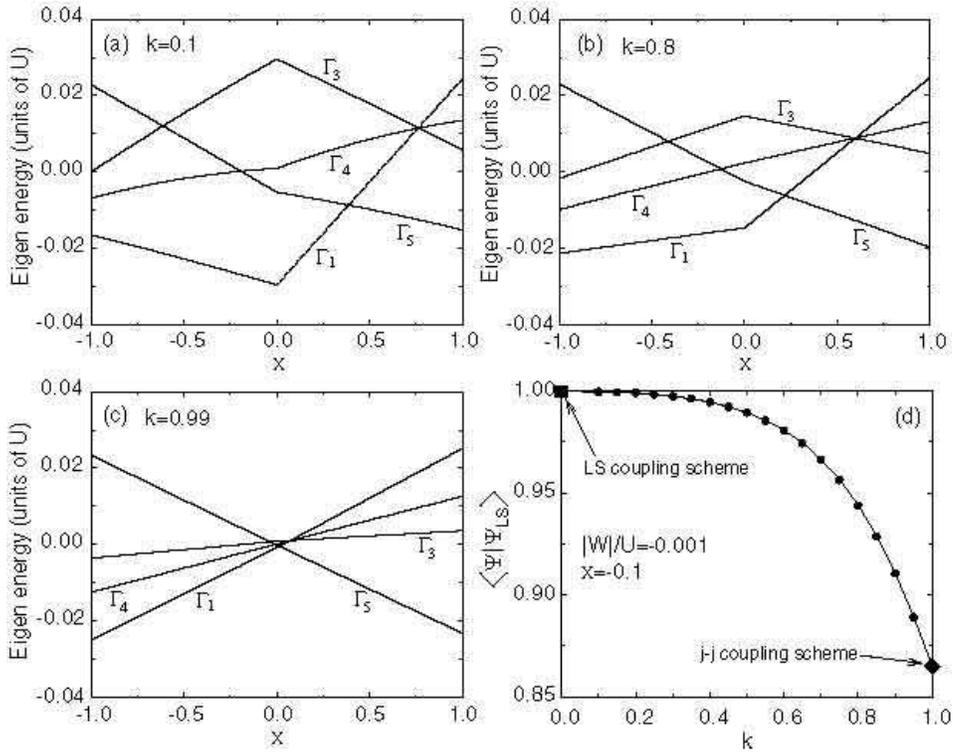}
\end{center}
\caption{%
Eigen energies of $H_{\rm f}$ as functions of $x$ for
(a) $k$=0.1, (b) $k$=0.8, and (c) $k$=0.99.
Racah parameters are set as $A/U$=10, $B/U$=0.3, $C/U$=0.1,
and $D/U$=0.05.
The energy scale for CEF potentials are given by $W/U$=$-0.001$.
(d) Overlap integral between the eigenstate of $H_{\rm f}$ and
that in the $LS$ coupling scheme for the case of
the $\Gamma_1$ ground state.
Solid squares at $k$=0 and 1 are obtained separately from
the $LS$ and $j$-$j$ coupling schemes, respectively.
}
\label{fig5-2}
\end{figure}

In Figs.~\ref{fig5-2}(a)-(d), we show the energy levels of $H_{\rm f}$
for several values of $k$ with both $\lambda_{\rm so}$ and $U$
larger than $|W|$.
Racah parameters are set as $A/U$=10, $B/U$=0.3, $C/U$=0.1,
and $D/U$=0.05 in the units of $U$.
As described above, the CEF potential is always small and
here we set $W/U$=$-0.001$.
In Fig.~\ref{fig5-2}(a), results for $k$=0.1 are shown.
In this case, $\lambda_{\rm so}/U$=0.11, while
the condition $\lambda_{\rm so}/|W|$$\gg$1 is still satisfied.
Without the spin-orbit interaction,
the ground-state level is expressed as $^{3}H$
with $S$=1 and $L$=5 due to the Hund's rules.
When we increase $\lambda_{\rm so}$, the multiplets labeled by $J$
are well separated and the ground-state level is specified by $J$=4,
as expected from the $LS$ coupling scheme.
Then, the energy levels in Fig.~\ref{fig5-2}(a) are quite similar
to those of Fig.~\ref{fig5-1}(a), since we are now in the region
where the $LS$ coupling scheme is appropriate.

Even when $\lambda_{\rm so}$ is further increased and $k$ is equal
to 0.5, the structure of the energy levels is almost the same
as that of the $LS$ coupling scheme (not shown here).
However, when $k$ becomes 0.8, as shown in Fig.~\ref{fig5-2}(b),
the energy level structure is found to be deviated from
that of the $LS$ coupling scheme.
Rather, it becomes similar to the energy level structure of
the $j$-$j$ coupling scheme.
To see the agreement with the $j$-$j$ coupling scheme more clearly,
we consider very large $\lambda$ which gives $k$=0.99.
As shown in Fig.~\ref{fig5-2}(c), we can observe the energy level
structure similar to Fig.~\ref{fig5-1}(b).
In particular, the region of the $\Gamma_3$ ground state becomes
very narrow, as discussed later.
Thus, it is concluded that $H_{\rm f}$ actually reproduces
the energy levels both for the $LS$ and $j$-$j$ coupling schemes.
We also stress that $H_{\rm f}$ provides correct results
in any value of $f$-electron number.

A crucial point is that the structure of energy levels is
continuously changed, as long as $\lambda_{\rm so}$ and $U$ are
large compared with the CEF potential.
Namely, the states both in the $LS$ and $j$-$j$ coupling schemes
are continuously connected in the parameter space.
Thus, depending on the situation to consider the problem,
we are allowed to use either $LS$ or $j$-$j$ coupling scheme.
In order to clarify this point, we evaluate the overlap
$\langle \Psi |\Psi_{\rm LS}\rangle$,
where $|\Psi\rangle$ and $|\Psi_{\rm LS}\rangle$ are
the eigenstate of $H_{\rm f}$ and that in the $LS$ coupling scheme,
respectively.
In Fig.~\ref{fig5-2}(d), we show the overlap for the case of $\Gamma_1$
ground state for $x$=$-0.1$ and $W/U$=$-0.001$.
For $k$=0, $\langle \Psi |\Psi_{\rm LS}\rangle$=1
due to the definition.
The overlap is gradually decreased with the increase of $k$,
but it smoothly converges to the value at $k$=1, i.e.,
the $j$-$j$ coupling scheme.
Note that the overlap between the eigenstates of the
$LS$ and $j$-$j$ coupling schemes is as large as 0.865,
which seems to be larger than readers may naively anticipated
from the clear difference between
Figs.~\ref{fig5-1}(a) and \ref{fig5-1}(b).
It is not surprising, if we are based on the principle of
adiabatic continuation, since the eigenstates of the
$LS$ and $j$-$j$ coupling schemes are continuously connected.

Remark that we can observe the common structure around
at the value of $x$, in which singlet and triplet
ground states are interchanged.
Namely, essential point of the singlet-triplet crossing
can be captured both in the two schemes.
However, the $\Gamma_3$ non-Kramers doublet cannot be the ground state
in the $j$-$j$ coupling scheme, since the doublet in the $J$=4 nontet
is composed of degenerate two singlets
formed by $\Gamma_7$ and $\Gamma_8$ electrons.
As easily understood, such singlets are energetically penalized
by the Hund's rule interaction and the energy for $\Gamma_4$ triplet
composed of $\Gamma_7$ and $\Gamma_8$ electrons is
always lower than that of the singlets.
Thus, in the $j$-$j$ coupling scheme, $\Gamma_3$ non-Kramers doublet
does not appear as the ground state except for $x$=0.

Of course, if $j$=7/2 octet is explicitly included and $\lambda$
is kept finite, it is possible to reproduce $\Gamma_3$ doublet.
Namely, taking account of the effect of $j$=7/2 octet is
equivalent to consider the local $f$-electron term $H_{\rm f}$,
as we have done in this subsection.
If we simply expand the Hilbert space so as to include both
$j$=5/2 sextet and $j$=7/2 octet, we lose the advantage of
the $j$-$j$ coupling scheme considering only $j$=5/2 sextet.
However, for an actual purpose, it is enough to consider
perturbatively such effect in the order of $1/\lambda_{\rm so}$.
In fact, quite recently, Hotta and Harima have shown
that the result of the $LS$ coupling scheme
can be reproduced quantitatively
even in the $j$-$j$ coupling scheme,
when the effect of $j$=7/2 octet is included
as effective one- and two-body potentials up to
the order of $1/\lambda_{\rm so}$ \cite{Hotta2006}.

One may claim that it is possible to reproduce the result of
the $LS$ coupling scheme even within the $j$-$j$ coupling scheme,
just by assuming that the CEF potential for $J$=4 in
the $LS$ coupling scheme also works on the $J$=4 $f^2$-states
composed of a couple of $f$ electrons among $j$=5/2 sextet.
However, such a procedure is $not$ allowed due to
the following two reasons.
First it should be noted that the CEF potential is $not$
determined only by the value of $J$.
For instance, the results of the energy levels for $n$=7 and 13
are apparently different, even though both
of the ground-state multiplets are characterized by $J$=7/2,
since the CEF potential depends also on the values of $L$ and $S$.
Note that for $n$=7, $S$=7/2 and $L$=0, while for $n$=13,
$L$=3 and $S$=1/2.
For the case of $n$=2, even if the $f^2$-state is characterized
by $J$=4 in the $j$-$j$ coupling scheme,
we cannot simply validate the application of the
CEF potential in the $LS$ coupling scheme to the $J$=4
$f^2$-state in the $j$-$j$ coupling scheme.

Second we should note again that
the CEF effect appears only as a one-electron potential.
The CEF potential working on the two-electron state should be
given by the superimposition of the one-electron potential.
Thus, when we use the basis which diagonalizes the spin-orbit interaction,
it is necessary to consider that the CEF potential should work
on the state labeled by the $z$-component of $j$.
This is the only way to define the CEF potential in the $j$-$j$
coupling scheme, even though the $\Gamma_3$ non-Kramers doublet
is not reproduced.
As mentioned in the above paragraph,
in order to 
reproduce the results of the $LS$ coupling scheme
including the non-Kramers doublet,
it is necessary to consider appropriately the effect of $j$=7/2 octet,
leading to the effective potential among $j$=5/2 states.

%
\subsection{Local f-electron term in the j-j coupling scheme}

In the previous subsection, we have shown the relation between
$LS$ and $j$-$j$ coupling schemes on the basis of the local term
including correctly the Coulomb interaction, spin-orbit coupling,
and CEF potential terms.
In order to make further steps to the construction of
a microscopic Hamiltonian,
let us first discuss the local $f$-electron state
on the basis of the $j$-$j$ coupling scheme.
For the purpose, it is necessary to define the one $f$-electron
state, labelled by $\mu$, but in the $j$-$j$ coupling scheme,
the meaning of $\mu$ is clear.
In the case of $n$$<$7, $\mu$ should be the label to specify
the state in the $j$=5/2 sextet, namely,
the $z$-component of the total angular momentum $j$=5/2 and
takes the values of $\mu$=$-5/2$, $-3/2$, $\cdots$, $5/2$.
Note that for 3$<$$n$$<$7, $j$=7/2 octet is $not$ occupied,
since we presume that the effect of spin-orbit interaction is larger
than that of the Hund's rule coupling in the $j$-$j$ coupling scheme.
On the other hand, for the case of $n$$\geq$7, $\mu$ should be
considered to specify the state in the $j$=7/2 octet,
since $j$=5/2 sextet is fully occupied.
Note again that spin-orbit interaction is larger
than that of the Hund's rule coupling.
In this paper, we concentrate only on the case of $n$$<$7.
Thus, in the following, $\mu$ indicates the $z$-component of the
total angular momentum which specifies the state
in the $j$=5/2 sextet.

In the $j$-$j$ coupling scheme,
the local $f$-electron term should be composed of two parts as
\begin{equation}
  H_{\rm loc} = H_{\rm CEF}+H_{\rm el-el},
\end{equation}
where $H_{\rm el-el}$ and $H_{\rm CEF}$ are
Coulomb interactions among $f$ electrons and the CEF term,
respectively.
Note that the spin-orbit interaction has been already included,
when we define the one $f$-electron state in the $j$-$j$ coupling scheme.
In order to express $H_{\rm el-el}$ and $H_{\rm CEF}$,
it is useful to define the annihilation operator of $f$ electron
in the $j$-$j$ coupling scheme, $a_{{\bf i}\mu}$, which is related to
$f_{{\bf i}m\sigma}$ with real-spin $\sigma$ and orbital
$m$ (=$-3$,$\cdots$,3) as
\begin{equation}
  a_{{\bf i}\mu}=\sum_{\sigma} C_{\mu\sigma}
  f_{{\bf i},\mu-\sigma/2,\sigma}.
\end{equation}
where $C_{\mu\sigma}$ is the Clebsch-Gordan coefficient,
given by
\begin{equation}
  C_{\mu\sigma}=-\sigma\sqrt{7/2-\mu\sigma \over 7},
\end{equation}
with $\sigma$=$+1$ ($-1$) for up (down) real spin.

As mentioned in the the previous subsection, in the $j$-$j$ coupling
scheme, we should take into account the CEF effect as the one-electron
potential.
Multi-electron state is obtained by simply accommodating electrons
due to the balance between the Coulomb interaction and
the one-electron potential, as has been done in $d$-electron systems.
Then, the CEF term is given by
\begin{equation}
  \label{H:CEF}
  H_{\rm CEF} = \sum_{{\bf i},\mu,\nu}
  B_{\mu\nu} a_{{\bf i}\mu}^{\dag} a_{{\bf i}\nu},
\end{equation}
where $B_{\mu\nu}$ is expressed by the CEF parameters for $J$=5/2.
For the case of cubic structure, we can easily obtain
\begin{equation}
\label{Eq:cubic-CEF}
\begin{array}{rcl}
  B_{\pm 5/2,\pm 5/2} &=&  60 B_4^0, \\
  B_{\pm 3/2,\pm 3/2} &=& -180 B_4^0, \\
  B_{\pm 1/2,\pm 1/2} &=&  120 B_4^0, \\
  B_{\pm 5/2,\mp 3/2} &=& B_{\mp 3/2,\pm 5/2}=60\sqrt{5} B_4^0,
\end{array}
\end{equation}
and zero for other $\mu$ and $\nu$.
For the case of tetragonal structure, we obtain
\begin{equation}
\begin{array}{rcl}
  B_{\pm 5/2,\pm 5/2} &=&  10 B_2^0 +  60 B_4^0, \\
  B_{\pm 3/2,\pm 3/2} &=&  -2 B_2^0 - 180 B_4^0, \\
  B_{\pm 1/2,\pm 1/2} &=&  -8 B_2^0 + 120 B_4^0, \\
  B_{\pm 5/2,\mp 3/2} &=& B_{\mp 3/2,\pm 5/2}=12\sqrt{5} B_4^4,
\end{array}
\end{equation}
and zero for other $\mu$ and $\nu$.

Note that the coefficients $B_n^m$ are, in actuality, determined
by the fitting of experimental results for physical quantities
such as magnetic susceptibility and specific heat.
Note also that the above formulae have been obtained
from the case of $J$=5/2.
In general, the CEF term is expressed in matrix form,
depending on the value of $J$;
for $J$ larger than 5/2, higher terms in $B_n^m$ should occur.
However, as already mentioned above, since in this paper
the effect of the CEF is considered as a one-electron potential
based on the $j$-$j$ coupling scheme,
it is enough to use the CEF term for $J$=5/2.

Next we consider $H_{\rm el-el}$ in the $j$-$j$ coupling scheme.
It is easy to understand that the Coulomb interaction term is
given in the form of
\begin{equation}
  H_{\rm el-el}=
  {1 \over 2}\sum_{{\bf i},\mu,\nu,\mu',\nu'}
  I(\mu,\nu;\nu',\mu')
  a_{{\bf i}\mu}^{\dag} a_{{\bf i}\nu}^{\dag}
  a_{{\bf i}\nu'} a_{{\bf i}\mu'},
\end{equation}
where $I$ is the Coulomb interactions.
The point here is the calculation of $I$,
which is the sum of two contributions, written as
\begin{equation}
  I(\mu,\nu;\nu',\mu') = K_{\mu\nu,\nu'\mu'}-K_{\mu\nu,\mu'\nu'},
\end{equation}
with the Coulomb integral $K$.
The former indicates the Coulomb term, while the latter
denotes the exchange one.
It should be noted that $I$ vanishes unless $\mu$+$\nu$=$\mu'$+$\nu'$
due to the conservation of $z$-component of total angular momentum.
The matrix element $K_{\mu_1\mu_2,\mu_3\mu_4}$ is explicitly given by
\begin{eqnarray}
  K_{\mu_1 \mu_2,\mu_3 \mu_4} =
  \sum_{\sigma,\sigma'}  C_{\mu_1 \sigma} C_{\mu_2 \sigma'}
  C_{\mu_3 \sigma'}  C_{\mu_4 \sigma}
  I^f_{\mu_1-\sigma/2, \mu_2-\sigma'/2,\mu_3-\sigma'/2,
  \mu_4-\sigma/2},
\end{eqnarray}
where $I^f$ is the Coulomb matrix element among $f$ electrons,
already defined in the previous subsection.

When two electrons are accommodated in the $j$=5/2 sextet,
the allowed values for total angular momentum $J$ are 0, 2, and 4
due to the Pauli principle.
Thus, the Coulomb interaction term should be written
in a 15$\times$15 matrix form.
Note that ``15'' is the sum of the basis numbers for singlet ($J$=0),
quintet ($J$=2), and nontet ($J$=4).
As is easily understood, this 15$\times$15 matrix can be decomposed
into a block-diagonalized form labeled by $J_z$,
including one 3$\times$3 matrix for $J_z$=0,
four 2$\times$2 matrices for $J_z$=$\pm$2 and $\pm$1,
and four 1$\times$1 for $J_z$=$\pm 4$ and $\pm 3$.
We skip the details of tedious calculations for the matrix elements
and here only summarize the results in the following by using
the Racah parameters $E_k$ ($k$=0,1,2) in the $j$-$j$ coupling scheme
\cite{Norman}, which are related to the Slater-Condon parameters $F^k$ as
\begin{eqnarray}
  E_0 &=& F^0-{80 \over 1225}F^2-{12 \over 441}F^4, \\
  E_1 &=& {120 \over 1225}F^2+{18 \over 441}F^4, \\
  E_2 &=& {12 \over 1225}F^2-{1 \over 441}F^4.
\end{eqnarray}
For the sectors of $J_z$=4 and 3, we obtain
\begin{equation}
\label{Eq:Jz4}
  I(5/2,3/2;3/2,5/2)=E_0-5E_2,
\end{equation}
and
\begin{equation}
\label{Eq:Jz3}
  I(5/2,1/2;1/2,5/2)=E_0-5E_2,
\end{equation}
respectively.
For $J_z$=2 and 1, we obtain
\begin{equation}
\label{Eq:Jz2}
\begin{array}{rcl}
  I(3/2,1/2;1/2,3/2) &=& E_0+4E_2, \\
  I(5/2,-1/2;-1/2,5/2) &=& E_0, \\
  I(3/2,1/2;-1/2,5/2) &=& -3\sqrt{5}E_2,
\end{array}
\end{equation}
and
\begin{equation}
\label{Eq:Jz1}
\begin{array}{rcl}
  I(3/2,-1/2;-1/2,3/2) &=& E_0-E_2, \\
  I(5/2,-3/2;-3/2,5/2) &=& E_0+5E_2, \\
  I(3/2,-1/2;-3/2,5/2) &=& -2\sqrt{10}E_2,
\end{array}
\end{equation}
Finally, for $J_z$=0 sector, we obtain
\begin{equation}
\label{Eq:Jz0}
\begin{array}{rcl}
  I(1/2,-1/2;-1/2,1/2) &=& E_0+2E_2+E_1, \\
  I(3/2,-3/2;-3/2,3/2) &=& E_0-3E_2+E_1, \\
  I(5/2,-5/2;-5/2,5/2) &=& E_0+5E_2+E_1, \\
  I(1/2,-1/2;-3/2,3/2) &=& -E_1-3E_2,  \\
  I(1/2,-1/2;-5/2,5/2) &=& E_1-5E_2, \\
  I(3/2,-3/2;-5/2,5/2) &=& -E_1.
\end{array}
\end{equation}
Note here the following relations:
\begin{equation}
 I(\mu,\nu;\nu',\mu')=I(\mu',\nu';\nu,\mu),
\end{equation}
and
\begin{equation}
 I(\mu,\nu;\nu',\mu')=I(-\nu,-\mu;-\mu',-\nu').
\end{equation}
By using these two relations and Eqs.~(\ref{Eq:Jz4}-\ref{Eq:Jz0}), 
we can obtain all the Coulomb matrix elements \cite{Inglis}.

It is instructive to understand how the $f^2$ configuration is
determined by the Coulomb interaction in the $j$-$j$ coupling scheme.
We will discuss later the local $f$-electron state determined by
$H_{\rm CEF}$+$H_{\rm el-el}$.
In the $j$-$j$ coupling scheme, two electrons are accommodated
in the $j$=5/2 sextet.
When we diagonalize the 15$\times$15 matrix for Coulomb interaction terms,
we can easily obtain the eigen energies as
$E_0$$-$$5E_2$ for the $J$=4 nontet,
$E_0$+$9E_2$ for the $J$=2 quintet,
and $E_0$+$3E_1$ for the $J$=0 singlet.
Since the Racah parameters are all positive, the ground state is specified
by $J$=4 in the $j$-$j$ coupling scheme.
In the $LS$ coupling scheme, on the other hand, we obtain the ground-state
level as $^{3}H$ with $S$=1 and $L$=5 from the Hund's rules.
On further inclusion of the spin-orbit interaction,
the ground state becomes characterized by $J$=4,
expressed as $^3H_4$ in the traditional notation.
Note that we are now considering a two-electron problem.
Thus, if we correctly include the effects of Coulomb interactions,
it is concluded that the same quantum number as that in
the $LS$ coupling scheme is obtained in the $j$-$j$ coupling scheme
for the ground-state multiplet.

In order to understand further the physical meaning of Racah parameters,
it is useful to consider a simplified Coulomb interaction term.
In the above discussion, the expressions using Racah parameters are not
convenient, since they depend on the orbitals in a very complicated manner,
even though they keep the correct symmetry required by group theory.
To clarify their meanings, let us step back to the following simplified
interaction form among $\ell=3$ orbitals:
\begin{eqnarray}
  \label{Eq:Hint2}
  H_{\rm int} &&=
  U \sum_{{\bf i}m}\rho_{{\bf i}m\uparrow} \rho_{{\bf i}m\downarrow}
  + U' \sum_{{\bf i},\sigma,\sigma',m>m'}
  \rho_{{\bf i}m\sigma} \rho_{{\bf i}m'\sigma'} \nonumber \\
  && + J \sum_{{\bf i},\sigma,\sigma',m>m'}
  f_{{\bf i}m\sigma}^{\dag} f_{{\bf i}m'\sigma'}^{\dag}
  f_{{\bf i}m\sigma'}f_{{\bf i}m'\sigma},
\end{eqnarray}
where $\rho_{{\bf i}m\sigma}$=$f_{{\bf i}m\sigma}^{\dag} f_{{\bf i}m\sigma}$.
In this equation, we include only three interactions;
intra-orbital Coulomb interaction $U$,
inter-orbital Coulomb interaction $U'$,
and the exchange interaction $J$.
We ignore the pair-hopping $J'$ for simplicity.
Since we set $J'$=0 in the relation of $U$=$U'$+$J$+$J'$,
the relation $U$=$U'$+$J$ holds among Coulomb interactions
to ensure rotational invariance in the orbital space.

By using Clebsch-Gordan coefficients, $f_{{\bf i}m\sigma}$
with real-spin $\sigma$ can be related to $a_{{\bf i}\mu}$ as
\begin{equation}
  f_{{\bf i}m\sigma}=-\sigma\sqrt{3-\sigma m \over 7}a_{{\bf i}m+\sigma/2}.
\end{equation}
Note here that we consider only the $j=5/2$ sextet.
The Coulomb interaction term for $j$=5/2 is given by
\begin{eqnarray}
  H_{\rm int} = U_{\rm eff} \sum_{{\bf i} \mu>\mu'}
  n_{{\bf i}\mu} n_{{\bf i}\mu'}
  -J_{\rm eff} {\bm J}_{\bf i}^2+(35J_{\rm eff}/4)N_{\bf i},
\end{eqnarray}
where $n_{{\bf i}\mu}$=$a_{{\bf i}\mu}^{\dag}a_{{\bf i}\mu}$,
$N_{\bf i}$=$\sum_{\mu}n_{{\bf i}\mu}$,
$U_{\rm eff}$=$U'$$-$$J/2$, $J_{\rm eff}$=$J/49$,
and ${\bm J}_{\bf i}$ is the operator for total angular momentum
with $j$=5/2.
Explicitly, ${\bm J}_{\bf i}^2$ is written as
\begin{eqnarray}
 {\bm J}_{\bf i}^2 &&= \sum_{\mu,\mu'}
 [\mu\mu' n_{{\bf i}\mu}n_{{\bf i}\mu'} \nonumber \\
 && + (\phi^{+}_{\mu}\phi^{-}_{\mu'}
 a_{{\bf i}\mu+1}^{\dag} a_{{\bf i}\mu}
 a_{{\bf i}\mu'-1}^{\dag}a_{{\bf i}\mu'}
 \!+\! \phi^{-}_{\mu}\phi^{+}_{\mu'}
 a_{{\bf i}\mu-1}^{\dag} a_{{\bf i}\mu}
 a_{{\bf i}\mu'+1}^{\dag}a_{{\bf i}\mu'})/2],
\end{eqnarray}
with $\phi_{\mu}^{\pm}=\sqrt{j(j+1)-\mu(\mu\pm 1)}$
=$\sqrt{35/4-\mu(\mu\pm 1)}$.

For two electrons in the $j$=5/2 sextet, based upon
the simplified Coulomb interaction term, we can easily obtain
the energy levels as $U_{\rm eff}$$-$$5J_{\rm eff}/2$ for the $J$=4 nontet,
$U_{\rm eff}$+$23J_{\rm eff}/2$ for the $J$=2 quintet,
and $U_{\rm eff}$+$35J_{\rm eff}/2$ for the $J$=0 singlet.
When we compare these energy levels with the results obtained
using Racah parameters,
we understand the correspondence such as
$E_0$$\sim$$U_{\rm eff}$ and $E_2$$\sim$$J_{\rm eff}$.
Namely, $E_0$ is the effective inter-orbital Coulomb interaction,
while $E_2$ denotes the effective Hund's rule coupling.
Note that $E_1$ does not appear, since it is related to the
pair-hopping interaction which is not included here.

We also note the smallness of $J_{\rm eff}$,
given as $J_{\rm eff}$=$J/49$.
The origin of the large reduction factor $1/49$ is,
in one word, due to the neglect of $j$=7/2 octet.
In the Coulomb interaction term Eq.~(\ref{Eq:Hint2}),
the Hund's rule term is simply written as $-J{\bm S}^2$. 
Note the relation ${\bm S}$=$(g_J-1){\bm J}$ with $g_J$
the Land\'e's $g$-factor.
For $j$=5/2, we easily obtain $g_J$=6/7,
indicating ${\bm S}$=$-(1/7){\bm J}$.
Thus, the original Hund's rule term is simply rewritten
as $-(J/49) {\bm J}^2$.

%
\subsection{Level scheme in the j-j coupling scheme}

Before proceeding to the exhibition of the model Hamiltonian
obtained by further considering the kinetic term of $f$ electrons,
it is instructive to show how the $j$-$j$ coupling scheme works
to reproduce the local level scheme of actual $f$-electron materials.
It is an important point that we can resort to the analogy with
the $d$-electron-like configuration, as discussed in Sec.~2.
As a typical example, here we consider the $f$-electron state
for the case of cubic CEF potential.

After some algebraic calculations, we obtain two degenerate
levels under the cubic CEF.
One is $\Gamma_7$ doublet with Kramers degeneracy and
another is $\Gamma_8$ quartet including two Kramers doublets.
It is quite useful to define new operators with ``orbital''
degrees of freedom to distinguish two Kramers doublets included
in $\Gamma_8$ as
\begin{equation}
  \begin{array}{rcl}
    f_{{\bf i}{\rm a} \uparrow}
    &=& \sqrt{5/6}a_{{\bf i}-5/2}+\sqrt{1/6}a_{{\bf i}3/2}, \\
    f_{{\bf i}{\rm a} \downarrow}
    &=& \sqrt{5/6}a_{{\bf i}5/2}+\sqrt{1/6}a_{{\bf i}-3/2},
  \end{array}
\end{equation}
for ``a''-orbital electrons and 
\begin{equation}
   f_{{\bf i}{\rm b} \uparrow}=a_{{\bf i}-1/2},~~
   f_{{\bf i}{\rm b} \downarrow}=a_{{\bf i}1/2},
\end{equation}
for ``b''-orbital electrons, respectively.
The $\Gamma_7$ state, defined as ``c'' orbital, is characterized by
\begin{equation}
  \begin{array}{rcl}
   f_{{\bf i}{\rm c} \uparrow}
   &=& \sqrt{1/6}a_{{\bf i}-5/2}-\sqrt{5/6}a_{{\bf i}3/2}, \\
   f_{{\bf i}{\rm c} \downarrow}
   &=& \sqrt{1/6}a_{{\bf i}5/2}-\sqrt{5/6}a_{{\bf i}-3/2}.
\end{array}
\end{equation}
For the standard time reversal operator ${\cal K}$=$-i\sigma_y K$,
where $K$ denotes an operator to take the complex conjugate,
we can easily show the relation
\begin{equation}
 {\cal K}f_{{\bf i}\tau \sigma}=\sigma f_{{\bf i}\tau -\sigma}.
\end{equation}
Note that this has the same definition for real spin.

\begin{figure}[t]
\begin{center}
\includegraphics[width=0.5\linewidth]{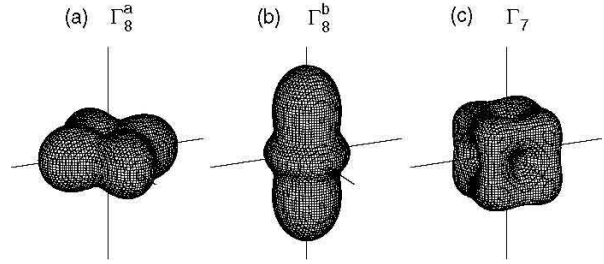}
\end{center}
\caption{
Views for (a) $\Gamma_8^{\rm a}$, (b) $\Gamma_8^{\rm b}$, and
(c) $\Gamma_7$ orbitals.
$\Gamma_{\gamma}$ indicates the irreducible representation
of point group in Bethe's notation \cite{Burns}.
}
\label{fig5-3}
\end{figure}

In Fig.~\ref{fig5-3}, we show the shape of three orbitals.
As intuitively understood from the shape of $\Gamma_7$ orbital,
this keeps the cubic symmetry, indicating A-representation.
In fact, in the group theory, it is characterized by $a_{\rm u}$.
Note that the subscript ``u'' indicates ungerade,
since we consider $f$ electron with $\ell$=3.
On the other hand, degenerate $\Gamma_8^{\rm a}$ and $\Gamma_8^{\rm b}$
orbitals seems to be similar to $x^2-y^2$ and $3z^2-r^2$ orbitals
of $3d$ electrons, respectively, indicating E-representation.
In the group theoretical argument, these are classified into $e_{\rm u}$.
Concerning the similarity between $\Gamma_8$ and $e_{\rm g}$ orbitals,
it is quite natural from a mathematical viewpoint, since
we recall the fact that
$\Gamma_8$ is isomorphic to $\Gamma_3 \times \Gamma_6$,
where $\Gamma_3$ indicates $E$ representation for the orbital part
and $\Gamma_6$ denotes the spin part.
This point is quite impressive when we consider the orbital physics
for $d$- and $f$-electron systems.
Namely, by exploiting this mathematical similarity,
it is possible to understand the complex
$f$-electron phenomena with the use of the microscopic Hamiltonian
in common with that of the $d$-electron multiorbital model.
We will see later this point in the construction of the
model Hamiltonian.

Now we discuss the $f$-electron configuration in the $\Gamma_7$ and
$\Gamma_8$ levels in the manner in which we have considered
the $d$-electron configuration.
First, we pick up the AuCu$_3$-type cubic crystal structure.
A typical AuCu$_3$-type material with one $f$ electron per site
is is CeIn$_3$, in which $\Gamma_7$ and $\Gamma_8$ are the ground
and first excited states, respectively \cite{Knafo}.
If we accommodate one more electron to consider the $f^2$ configuration,
immediately there appear two possibilities,
``low'' and ``high'' spin states, as we have discussed in the
$d$-electron configuration.
When the CEF splitting energy between $\Gamma_7$ and
$\Gamma_8$ levels is smaller than the Hund's rule coupling,
the second electron should be accommodated in the $\Gamma_8$ levels.
In the situation in which one is in the $\Gamma_7$ and the other
in the $\Gamma_8$, a $\Gamma_4$ triplet appears
for the $f^2$ state in the $j$-$j$ coupling scheme.
As mentioned in the previous subsection, $\Gamma_3$ non-Kramers doublet
does not appear in the $j$-$j$ coupling scheme.
On the other hand, if the CEF splitting is larger than
the Hund's rule interaction, then the $f^2$ ground state is formed
from two $\Gamma_7$ electrons, leading to a $\Gamma_1$ singlet state.
When we compare this $\Gamma_1$ state with that in
the $LS$ coupling scheme, we notice that it is given by
a mixture of $J$=0 and $J$=4 states,
but the $J$=4 component is found to be dominant.
Note also that $\Gamma_1$ is the antisymmetric representation of
$\Gamma_7 \times \Gamma_7$.

Since we do not know the exact value of the Hund's rule
interaction in $f$-electron compounds, it is difficult to
determine the $f^2$ state by purely theoretical arguments.
In this case, we have to refer to the data on actual materials.
Fortunately, we have the example of PrIn$_3$,
a typical $f^2$ material with AuCu$_3$-type crystal structure.
From several experimental results, $\Gamma_1$ has been
confirmed to be the ground level in PrIn$_3$
\cite{PrIn3-1,PrIn3-2}.
Thus, the low-spin state should be taken for the AuCu$_3$-type
structure in the $j$-$j$ coupling scheme.

\begin{figure}[t]
\begin{center}
\includegraphics[width=0.4\linewidth]{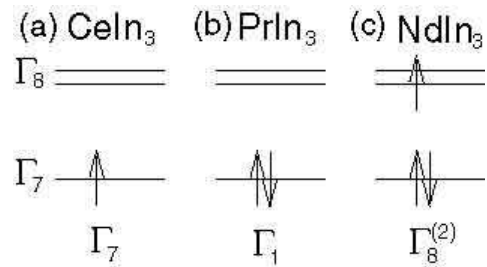}
\end{center}
\caption{
Electron configurations in the $j$-$j$ coupling
scheme for (a) CeIn$_3$, (b) PrIn$_3$, and (c) NdIn$_3$.
$\Gamma_{\gamma}$ denotes the irreducible representation
of point group in Bethe's notation \cite{Burns}.
}
\label{fig5-4}
\end{figure}

Here the reader may pose a naive question:
Is the Hund's rule interaction really so small in $f$-electron systems?
However, we have already discussed this point in the previous subsection.
Namely, the effective Hund's rule interaction $J_{\rm eff}$ is given by
$J_{\rm eff}$=$J/49$ in the $j$-$j$ coupling scheme,
where $J$ is the original Hund's rule interaction among $f$ electrons.
Note again that the magnitude of the Hund's rule interaction is
effectively reduced by the factor 1/49 in the $j$-$j$ coupling scheme.
Even if $J$=1eV, $J_{\rm eff}$ is reduced to be about 200K,
which is comparable with the CEF splitting energy.
Thus, it is possible to have the low-spin state in the $j$-$j$
coupling scheme.

Next, we take a further step to the $f^3$ state by adding one more
$f$ electron.
Since $\Gamma_7$ is fully occupied to form $\Gamma_1$,
the next electron should be placed in the $\Gamma_8$ state,
as shown in Fig.~\ref{fig5-4}(c), clearly indicating that there exists
an active orbital degree of freedom.
The $f^3$ state composed of two $\Gamma_7$ and one $\Gamma_8$
electron is expressed as $\Gamma_8^{(2)}$ in the terminology of
group theory.
When we again consider actual materials,
NdIn$_3$ is found to be a typical $f^3$ material with
the AuCu$_3$-type crystal structure.
In experiments, it has been confirmed that $\Gamma_8^{(2)}$
is the ground level \cite{NdIn3-1,NdIn3-2,NdIn3-3},
as we have found with the present $j$-$j$ coupling scheme.

Let us turn our attention to another crystal structure,
in which $\Gamma_8$ is lower than $\Gamma_7$ in the $f^1$ configuration.
Typical materials are the rare-earth hexaborides RB$_6$
with R=Ce, Pr, and Nd.
As is well known, the ground level of CeB$_6$ is $\Gamma_8$,
indicating that the quadrupolar degree of freedom plays an
active role in this material \cite{CeB6}.
In fact, anomalous behavior related to quadrupolar ordering
has been suggested by several experimental results.

First, we note that the level splitting between $\Gamma_8$ and
$\Gamma_7$ is assumed to be larger than the Hund's rule interaction.
When we accommodate two electrons in $\Gamma_8$ orbitals,
the triplet ($\Gamma_5$), doublet ($\Gamma_3$), and
singlet ($\Gamma_1$) states are allowed.
Among these, owing to the effect of the Hund's rule interaction,
even if it is small, the $\Gamma_5$ triplet should be the ground state.
This has actually been observed in PrB$_6$ \cite{Loewenhaupt,PrB6}.
Further, in order to consider NdB$_6$, another electron is put into
the $\Gamma_8$ orbital, making a total of three.
Alternatively, we may say that there is one hole in the $\Gamma_8$ orbital.
Such a state is found, again, to be characterized by $\Gamma_8^{(2)}$.
Experimental results on NdB$_6$ have actually been reported which lead
to the ground state of $\Gamma_8^{(2)}$
\cite{Loewenhaupt,NdB6-1,NdB6-2,NdB6-3}.
Thus, when $\Gamma_8$ is the ground state for the one $f$-electron
case, we obtain $\Gamma_5$ for the $f^2$ and $\Gamma_8^{(2)}$
for the $f^3$ configurations.

\begin{figure}[t]
\begin{center}
\includegraphics[width=0.4\linewidth]{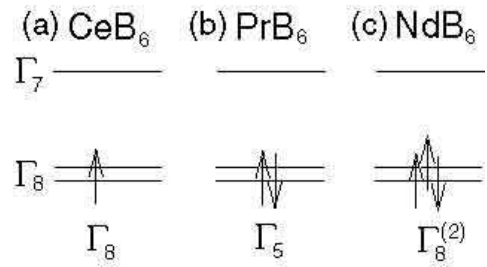}
\end{center}
\caption{
Electron configurations in the $j$-$j$ coupling
scheme for rare-earth hexaborides,
(a) CeB$_6$, (b) PrB$_6$, and (c) NdB$_6$.
$\Gamma_{\gamma}$ is the irreducible representation
of point group in Bethe's notation \cite{Burns}.
}
\label{fig5-5}
\end{figure}

We have shown that the ground states deduced from the $j$-$j$ coupling
scheme are consistent with experimental results.
However, in order to explain the experimental results quantitatively,
it is unavoidable to analyze the CEF levels
using the $LS$ coupling scheme.
As mentioned above, it is possible to reproduce the result of the
$LS$ coupling scheme by considering the effective potentials
from the $j$=7/2 octet,
but what we would like to stress here is that even in a localized system,
the symmetry of the ground level can be understood via the simple
$j$-$j$ coupling scheme.
We need to recognize the limitations of the simple $j$-$j$ coupling scheme
when we treat a local electronic state.
For instance, to consider the $f^3$ state, we simply put
three electrons into the CEF level scheme
which is determined with the $f^1$ configuration.
Thus, the wavefunction of the $f^3$ state is uniquely determined.
However, in an actual situation, the dectet labeled by $J$=9/2
($L$=6 and $S$=3/2) is split into two $\Gamma_8$ and one $\Gamma_6$ orbital.
The ground-state wavefunctions will then depend on the
two CEF parameters $B_4^0$ and $B_6^0$ \cite{LLW}.
In order to explain experimental results on localized $f$-electron materials,
one should analyze the Hamiltonian which also includes the complex
effective potentials from $j$=7/2 octet.
In this paper, however, the electronic states are considered with
an itinerant picture based on the simple $j$-$j$ coupling scheme.
Thus, it is important to check that the local electronic state
formed by $f$ electrons in this way is consistent with the symmetry of
the state obtained with the $LS$ coupling scheme.

In summary, it has been shown that the ground states of the $f^2$
and $f^3$ configurations can be qualitatively reproduced by
accommodating $f$ electrons in the CEF levels of a corresponding
$f^1$ material, provided that the CEF level splitting is larger
than the Hund's rule interaction.
Thus, the $j$-$j$ coupling scheme works even in the localized case.
Accordingly, we believe that a microscopic theory can be developed
in which we discuss the magnetism and superconductivity of $f$-electron
compounds in terms of the $j$-$j$ coupling scheme.

%
\subsection{Model Hamiltonian}

In previous subsections,
we have explained in detail that the $j$-$j$ coupling scheme
works for the local $f$-electron state.
Now let us include the effect of kinetic motion of $f$ electrons
\cite{Hotta2003}.
In this article, we are based on a itinerant picture for
$f$ electrons.
From our experience, this picture seems to be valid, when we
consider actinide compounds, in particular, heavy actinides.
On the other hand, for rare-earth materials,
it has been firmly believed to take 
the localized picture for $f$ electrons.
In particular, in order to consider the heavy fermion behavior,
it is indispensable to consider the system including both
the conduction electron with wide band and the almost localized
$f$ electron, which are hybridized with each other.

It is believed that the hybridization of $f$ electrons
with conduction electron band is
important to understand the magnetism of $f$-electron systems.
In fact, in the traditional prescription, first we derive
the Coqblin-Schrieffer model from the periodic Anderson model
by evaluating the $c$-$f$ exchange interaction $J_{\rm cf}$
within the second-order perturbation in terms of the hybridization
between $f$- and conduction electrons.
Then, we derive the RKKY interactions again using the second-order
perturbation theory with respect to $J_{\rm cf}$.
In general, the RKKY interactions are orbital dependent
and interpreted as multipole interactions.
Such orbital dependence originates from that of the hybridization.
Note that the hybridization should occur only between $f$- and
conduction band with the same symmetry.
Here we emphasize that the symmetry of $f$-electron state is
correctly included in our calculations.
Thus, the structure in the multipole interactions will not be
changed so much, even if we consider the effect of hybridization
with conduction band, as long as we consider correctly
the symmetry of $f$ electron states.

In this paper, we consider the tight-binding approximation
for $f$-electron hopping motion.
The kinetic term can be written as
\begin{equation}
 H_{\rm kin}=\sum_{{\bf i,a},\mu,\nu}
 t_{\mu\nu}^{\bf a} a_{{\bf i}\mu}^{\dag} a_{{\bf i+a}\nu},
\end{equation}
where $t_{\mu\nu}^{\bf a}$ is the overlap integral between
the $\mu$- and $\nu$-states connected by the vector ${\bf a}$.

Let us consider the hopping motion of $f$ electrons
based on the $j$-$j$ coupling scheme.
In order to evaluate $t^{\bf a}_{\mu\nu}$, which is
hopping of $f$ electrons between the $\mu$-state
at ${\bf i}$ site and the $\nu$-state at ${\bf i+a}$ site,
again it is convenient to step back to $f$-electron operators
in the $\ell$=3 multiplet, defined as $f_{{\bf i}m\sigma}$.
Since the real spin should be conserved in the hopping process,
$t^{\bf a}_{\mu\nu}$ is given as
\begin{equation}
  \label{Eq:hopping1}
  t^{\bf a}_{\mu\nu} = \sum_{\sigma}
  C_{\mu\sigma}C_{\nu\sigma}
  T^{\bf a}_{3,\mu-\sigma/2;3,\nu-\sigma/2},
\end{equation}
where $T^{\bf a}_{\ell,m; \ell',m'}$ is the hopping amplitude of
electrons between $(\ell,m)$- and $(\ell',m')$-states along
the ${\bf a}$-direction.

Now the problem is reduced to the evaluation of
$T^{\bf a}_{\ell,m; \ell',m'}$.
Although we can simply consult the paper of
Slater and Koster \cite{Slater1954,Takegahara},
a convenient formula has been obtained by Sharma
for the overlap integral between
two orbitals, $(\ell,m)$ and $(\ell',m')$ \cite{Sharma},
connected by unit vector ${\bf a}$. It is expressed as
\begin{equation}
  \label{Eq:hopping2}
  T^{\bf a}_{\ell,m;\ell',m'}
  = (\ell \ell' \sigma) \sqrt{4\pi \over 2\ell+1} \sqrt{4\pi \over 2\ell'+1}
  Y^{*}_{\ell m}(\theta,\varphi) Y_{\ell' m'}(\theta,\varphi),
\end{equation}
where $(\ell \ell' \sigma)$ denotes Slater's two-center integral
through the $\sigma$ bond, for instance, it is
$(ff\sigma)$ for $\ell$=$\ell'$=3
and $(fp\sigma)$ for $\ell$=3 and $\ell'$=1.
$\theta$ and $\varphi$ are polar and azimuth angles, respectively,
to specify the vector {\bf a} as
\begin{equation}
  {\bf a}=
  (\sin \theta \cos \varphi, \sin \theta \sin \varphi, \cos \theta).
\end{equation}

Here we consider the hopping between $f$ orbitals
in nearest neighbor sites by putting $\ell$=$\ell'$=3.
After some algebraic calculations, we obtain the hopping amplitudes
as follows.
For diagonal elements, we obtain
\begin{equation}
\begin{array}{rcl}
 t^{\bf a}_{\pm 5/2,\pm 5/2} &=& 5t_0 \sin^4 \theta, \\
 t^{\bf a}_{\pm 3/2,\pm 3/2} &=& t_0 \sin^2 \theta (1+15 \cos ^2 \theta), \\
 t^{\bf a}_{\pm 1/2,\pm 1/2} &=& 2t_0 (1-2 \cos ^2 \theta +5 \cos^4 \theta),
\end{array}
\end{equation}
where the energy unit $t_0$ is given by
\begin{equation}
 t_0=(3/56)(ff\sigma).
\end{equation}
Here $(ff\sigma)$ is the Slater-Koster two-center integral
between adjacent $f$ orbitals. Note that $t^{\bf a}_{\mu,-\mu}=0$.
For off-diagonal elements, we obtain
\begin{equation}
\begin{array}{rcl}
t^{\bf a}_{\pm 5/2, \pm 1/2} &=&
-t_0 \sqrt{10}e^{\mp 2i\varphi} \sin^2 \theta (1 - 3 \cos^2 \theta), \\
t^{\bf a}_{\pm 5/2, \mp 3/2} &=& t_0 \sqrt{5}e^{\mp 4i\varphi} \sin^4 \theta,
\\
t^{\bf a}_{\pm 1/2, \mp 3/2} &=& -t_0 \sqrt{2}e^{\mp 2i\varphi} \sin^2 \theta
(1 + 5 \cos^2 \theta),
\end{array}
\end{equation}
and
\begin{equation}
\begin{array}{rcl}
 t^{\bf a}_{5/2,-1/2} &=& -t^{\bf a}_{1/2,-5/2}=
 t_0 \sqrt{10}e^{-3i\varphi} \sin^2 \theta \sin 2\theta, \\
 t^{\bf a}_{5/2, 3/2} &=& -t^{\bf a}_{-3/2, -5/2}=
 -2t_0 \sqrt{5}e^{-i\varphi} \sin^2 \theta \sin 2\theta, \\
 t^{\bf a}_{1/2, 3/2} &=& -t^{\bf a}_{-3/2, -1/2}=
 t_0 \sqrt{2}e^{i\varphi} \sin 2\theta 
 (1 - 5 \cos^2 \theta). 
\end{array}
\end{equation}
Note that $t^{\bf a}_{\nu\mu}$=$t^{{\bf a}*}_{\mu\nu}$.

%
\subsection{$\Gamma_8$ model}

In the previous subsection, we have completed the construction of
a model Hamiltonian, which is expected to be a basic model to
investigate the microscopic aspects of magnetism and superconductivity
of $f$-electron systems.
Since it includes six states per site, i.e., three Kramers doublets,
the analysis may be difficult.
Of course, even if the calculations seem to be tedious,
it is necessary to carry out analytical and/or numerical research
on the basis of such a three orbital model.
However, for practical purposes, it is convenient to simplify
the model, if possible.
In this subsection, as an effective model for actinide compounds,
we introduce a $\Gamma_8$ model, by discarding $\Gamma_7$ orbital
\cite{Hotta2003}.

A simple explanation to validate the ignorance of $\Gamma_7$
is to assume large CEF splitting energy between $\Gamma_7$ and
$\Gamma_8$ levels.
This simplification is motivated by the fact that
the possibility of exotic octupole ordering has been actively
discussed in Ce$_x$La$_{1-x}$B$_6$ and NpO$_2$
with $\Gamma_8$ ground state.
Here readers may be doubtful of the reality of our assumption,
since the Coulomb interaction among $f$ electrons is naively thought
to be larger than the CEF level splitting in any case.
However, it should be noted again that we are now considering
the $f$-electron state in the $j$-$j$ coupling scheme,
not in the original $f$-electron state with angular momentum $\ell$=3.
As already mentioned,
the Hund's rule interaction in the $j$-$j$ coupling scheme is
effectively reduced to be 1/49 of the original Hund's rule coupling.
Even when the original Hund's rule coupling among $f$ electrons is
1 eV, it is reduced to 200 K in the $j$-$j$ coupling scheme.
For instance, the CEF level splitting in actinide dioxides
is considered to be larger than 1000 K.
We also recall that the CEF level splitting in CeB$_6$
is as large as 500~K.
Thus, we safely conclude that our present assumption is correctly
related to some actual materials.
Of course, in order to achieve quantitative agreement with experimental
results, it is necessary to include also $\Gamma_7$ level,
since the magnitude of the CEF splitting is always finite,
even if it is large compared with the effective Hund's rule interaction.
However, we strongly believe that it is possible to grasp
microscopic origin of unconventional superconductivity
as well as spin and orbital, i.e., multipole, ordering
in $f$-electron systems on the basis of the $\Gamma_8$ model,
since this model is considered to be connected adiabatically
from the realistic situation.
It is one of future tasks to develop more general theory
to include all the $j$=5/2 sextet in future.

\begin{figure}[t]
\begin{center}
\includegraphics[width=0.6\linewidth]{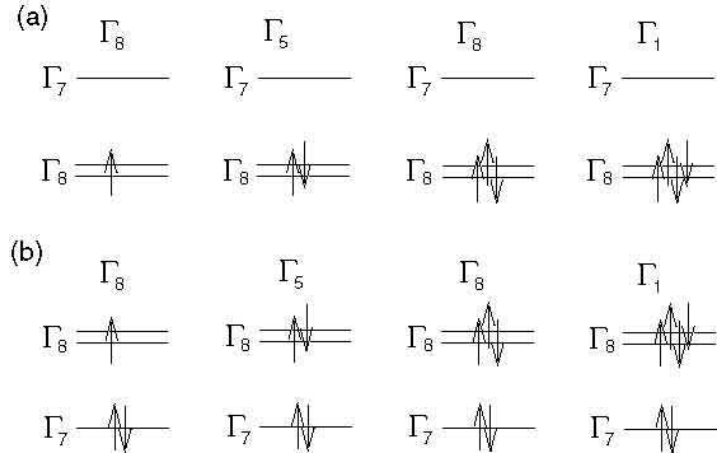}
\end{center}
\caption{
Electron configurations in the $j$-$j$ coupling scheme
to which the $\Gamma_8$ model is applicable.
(a) $f^1 \sim f^4$ configuration when $\Gamma_8$ is lower.
(b) $f^3 \sim f^6$ configuration when $\Gamma_7$ is lower,
}
\label{fig5-6}
\end{figure}

Concerning the $f$-electron number, typically we treat
the case with one $f$ electron in the $\Gamma_8$ multiplet per site.
However, this restriction does not
simply indicate that we consider only the Ce-based compound.
In the $j$-$j$ coupling scheme, in order to consider $f^n$-electron
systems, where $n$ indicates local $f$ electron number per site,
we accommodate $f$ electrons in the one-electron CEF levels
due to the balance between Coulomb interactions and CEF level
splitting energy, just as in the case of $d$-electron systems.
Thus, as shown in Fig.~\ref{fig5-6}(a),
the $\Gamma_8$ model is applicable to the cases for
$n=1 \sim 4$ in the $\Gamma_8$-$\Gamma_7$ system,
where $\Gamma_x$-$\Gamma_y$ symbolically denotes the situation
with $\Gamma_x$ ground and $\Gamma_y$ excited states.
Furthermore, we should note that due to the electron-hole symmetry
in the $\Gamma_8$ subspace, the $\Gamma_8$ model
is also applicable to the cases for
$n=3 \sim 6$ in the $\Gamma_7$-$\Gamma_8$ system,
as shown in Fig.~\ref{fig5-6}(b).

Now let us define the $\Gamma_8$ model.
First we consider the hopping part.
For simplicity, here we set the cubic-based lattice.
Later we will discuss the hopping amplitude of other lattice structures.
We include the hopping in the $xy$ plane and along the $z$-axis
for ${\bf a}$=${\bf x}$=[1,0,0], ${\bf y}$=[0,1,0], and 
${\bf z}$=[0,0,1], respectively.
To evaluate the hopping amplitude,
we simply set ($\theta$, $\varphi$) to be ($\pi/2$,0), ($\pi/2$,$\pi/2$),
and (0,0) for x, y, and z directions.
Then, by using the general results in the previous section,
we easily obtain $t^{\bf a}_{\mu\nu}$ between neighboring $f$ orbitals
in the $xy$ plane and along the $z$ axis.
Further we transform the basis by the above definitions
for $\Gamma_8$ operators with orbital degrees of freedom.
The results are given as
\begin{equation}
  \label{Eq:tx}
  t_{\tau\tau'}^{\bf x} = t
  \left(
   \begin{array}{cc}
     3/4 & -\sqrt{3}/4 \\
     -\sqrt{3}/4 & 1/4 \\
   \end{array}
  \right),
\end{equation}
for the ${\bf x}$-direction,
\begin{equation}
  \label{Eq:ty}
  t_{\tau\tau'}^{\bf y} = t
  \left(
   \begin{array}{cc}
     3/4 & \sqrt{3}/4 \\
     \sqrt{3}/4 & 1/4 \\
   \end{array}
  \right),
\end{equation}
for the ${\bf y}$ direction, and
\begin{equation}
  \label{Eq:tz}
  t_{\tau\tau'}^{\bf z} = t
  \left(
    \begin{array}{cc}
     0 & 0 \\
     0 & 1 \\
    \end{array}
  \right),
\end{equation}
for the ${\bf z}$ direction.
Note that $t$=$8t_0$=(3/7)$(ff\sigma)$.
Here we stress that the hopping amplitudes among $\Gamma_8$ orbitals
are just the same as those for the $e_{\rm g}$ orbitals of $3d$ electrons.
Intuitively, readers can understand this point due to the
shapes of $\Gamma_8$ orbitals as shown in Fig.~\ref{fig5-3},
which are similar to $e_{\rm g}$ orbitals.

After transforming the basis of the Coulomb interaction terms,
the Hamiltonian for $\Gamma_8$ orbitals is given by
\begin{eqnarray}
 \label{H8}
 H_{\Gamma_8} &=&
 \sum_{{\bf i,a},\sigma,\tau,\tau'} t^{\bf a}_{\tau\tau'}
 f^{\dag}_{{\bf i}\tau\sigma} f_{{\bf i+a}\tau'\sigma}
 + U \sum_{{\bf i},\tau}\rho_{{\bf i}\tau\uparrow}
 \rho_{{\bf i}\tau\downarrow}
 + U' \sum_{\bf i} \rho_{{\bf i}a} \rho_{{\bf i}b} \nonumber \\
 &+& J \sum_{{\bf i},\sigma,\sigma'}
 f_{{\bf i}a\sigma}^{\dag} f_{{\bf i}b\sigma'}^{\dag}
 f_{{\bf i}a\sigma'} f_{{\bf i}b\sigma}
 + J' \sum_{{\bf i},\tau \ne \tau'}
 f_{{\bf i}\tau\uparrow}^{\dag} f_{{\bf i}\tau\downarrow}^{\dag}
 f_{{\bf i}\tau'\downarrow} f_{{\bf i}\tau'\uparrow},
\end{eqnarray}
where
$\rho_{{\bf i}\tau\sigma}$=
$f_{{\bf i}\tau\sigma}^{\dag} f_{{\bf i}\tau\sigma}$
and
$\rho_{{\bf i}\tau}$=$\sum_{\sigma}\rho_{{\bf i}\sigma\tau}$.
In the Coulomb interaction terms, $U$, $U'$, $J$, and $J'$ denote
intra-orbital, inter-orbital, exchange, and pair-hopping interactions
among $\Gamma_8$ electrons, respectively,
expressed by using the Racah parameters $E_k$ as
\begin{equation}
 \begin{array}{rcl}
  U &=& E_0+E_1+2E_2, \\
  U'&=& E_0+(2/3)E_2, \\
  J &=& 5E_2, \\
  J'&=& E_1-(11/3)E_2.
 \end{array}
\end{equation}
Note that the relation $U$=$U'$+$J$+$J'$ holds, ensuring rotational
invariance in pseudo-orbital space for the interaction part.
For $d$-electron systems, one also has another relation $J$=$J'$,
as mentioned in Sec.~2.
When the electronic wavefunction is real, this relation is
easily demonstrated from the definition of the Coulomb integral.
However, in the $j$-$j$ coupling scheme, the wavefunction is
complex, and $J$ is not equal to $J'$, in general.
For simplicity, we shall assume here that $J$=$J'$,
noting that essential results are not affected.
Since double occupancy of the same orbital is suppressed
owing to the large value of $U$, pair-hopping processes
are irrelevant in the present case.

We believe that this $\Gamma_8$ Hamiltonian provides a simple,
but non-trivial model to consider superconductivity and magnetism
in $f$-electron systems.
Note again that it is essentially the same as the model for
$e_{\rm g}$ electron systems such as manganites,
although the coupling with Jahn-Teller distortion is not included
in the present model.
Due to the complex interplay and competition among charge, spin,
and orbital degrees of freedom, a rich phase diagram has been
obtained for manganites.
Thus, it is definitely expected that a similar richness will also
be unveiled for $f$-electron systems
based on the $\Gamma_8$ model Hamiltonian.

%
%
\section{Orbital physics in $f$-electron systems}

We have constructed a microscopic model Hamiltonian for $f$-electron
systems in the previous subsection.
In particular, we could obtain the $\Gamma_8$ orbital degenerate model
as an effective Hamiltonian for actinide compounds.
In this section, we review the theoretical results of
the spin and orbital structure,
i.e., multipole order, based on the $\Gamma_8$ model.

\begin{figure}[t]
\begin{center}
\includegraphics[width=0.6\linewidth]{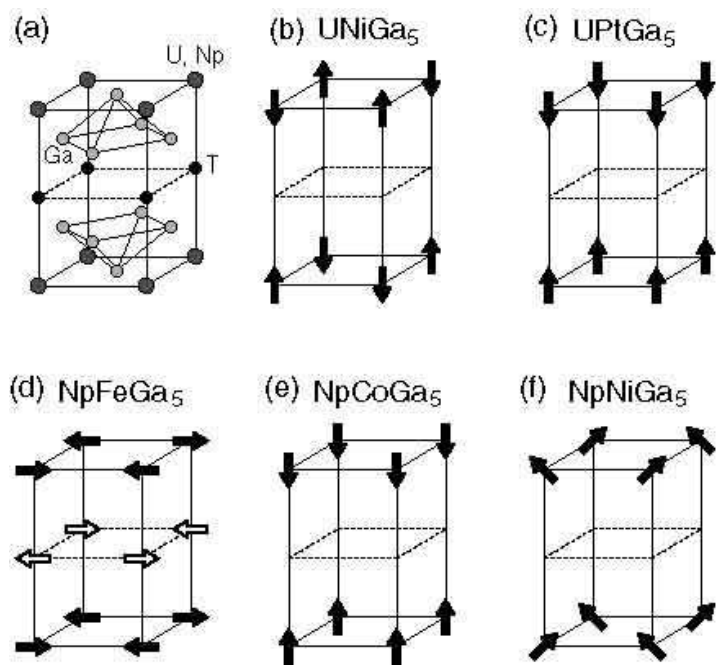}
\end{center}
\caption{(a) Crystal structure of AnTGa$_5$.
Schematic views of magnetic structures at low temperatures,
composed of magnetic moments of U and Np ions for
(b) UNiGa$_5$, (c) UPtGa$_5$,
(d) NpFeGa$_5$, (e) NpCoGa$_5$, and (f) NpNiGa$_5$.
For NpFeGa$_5$, magnetic moments at Fe sites are also depicted.}
\label{fig6-1}
\end{figure}

%
\subsection{Spin and orbital structure of actinide compounds}

In this subsection, we review the theoretical effort to
understand the magnetic structure
of uranium and neptunium compounds with HoCoGa$_5$-type
tetragonal crystal structure \cite{Hotta2004,Onishi2004}.
The crystal structure is shown in Fig.~\ref{fig6-1}(a).
The varieties of magnetic structure of U-115 and Np-115
explained in Sec.~1 are summarized
in Figs.~\ref{fig6-1}(b)-\ref{fig6-1}(f)
In order to set up the microscopic model for actinide 115 materials,
it is useful to consider UGa$_3$ and NpGa$_3$,
which are the mother compounds of U-115 and Np-115.
Among them, it has been reported that UGa$_3$ exhibits a G-type AF
metallic phase in the low-temperature region \cite{UGa3},
but a ``hidden'' ordering different from the magnetic one has been
suggested by resonant X-ray scattering measurements \cite{Mannix}.
Unfortunately, orbital ordering in UGa$_3$ is not yet confirmed
experimentally, but it may be an interesting possibility
to understand the result of resonant X-ray scattering experiment
on UGa$_3$ based on the orbital-ordering scenario.

\begin{figure}[t]
\begin{center}
\includegraphics[width=0.5\textwidth]{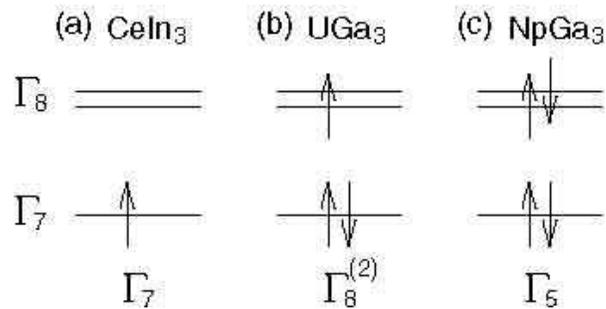}
\end{center}
\caption{
Level schemes for (a) CeIn$_3$,
(b) UGa$_3$, and (c) NpGa$_3$ based on the $j$-$j$ coupling scheme.
Here we assume trivalent actinide ions as
U$^{3+}$ ($5f^3$) and Np$^{3+}$ ($5f^4$).
It should be noted that up and down arrows denote pseudospins to
distinguish the states in the Kramers doublet.
Note also that for NpGa$_3$, a couple of electrons in $\Gamma_8$
orbitals form a local triplet, leading to $\Gamma_5$.
}
\label{fig6-2}
\end{figure}

Although it is difficult to determine the valence of actinide
ions in the solid state, for the time being, we assume that
the valence is U$^{3+}$ or Np$^{3+}$,
including three or four $f$ electrons per ion.
By considering the CEF potential and Coulomb interactions,
we then assign three or four electrons
to the states in the $j$=5/2 sextet.
In order to proceed with the discussion, it is necessary to
know which is lower, $\Gamma_7$ or $\Gamma_8$,
in the one $f$-electron picture.
For some crystal structures it is possible to determine
the level scheme from intuitive discussions of
$f$-electron wavefunctions and the positions of ligand ions.
However, this is not the case for the AuCu$_3$-type crystal structure.
For this case, we again invoke experimental results on CeIn$_3$,
a typical AuCu$_3$-type Ce-based compound,
where $\Gamma_7$ and $\Gamma_8$ have been reported as ground and excited
states, respectively, with an energy difference of 12meV \cite{Knafo}.
Thus, we take $\Gamma_7$ to be lower for the present considerations,
as shown in Fig.~\ref{fig6-2}(a).

In the $j$-$j$ coupling scheme for UGa$_3$ and NpGa$_3$,
we accommodate three or four electrons in the one-electron energy states
$\Gamma_7$ and $\Gamma_8$.
We immediately notice that there are two possibilities, i.e.,
low- and high-spin states,
depending on the Hund's rule interaction
and the splitting between the $\Gamma_7$ and $\Gamma_8$ levels.
As discussed in the previous subsection,
the effective Hund's rule interaction can be small
in the $j$-$j$ coupling scheme and thus,
the low-spin state should be realized,
as shown in Figs.~\ref{fig6-2}(b) and (c).
We emphasize that this low-spin state is consistent with
the $LS$ coupling scheme.
In fact, for NpGa$_3$, the observed magnetic moment at Np ion
has been found to be consistent with $\Gamma_5$ triplet \cite{NpGa3}.

In the electron configuration shown in Figs.~\ref{fig6-2}(b) and (c),
the $\Gamma_7$ level is fully occupied to form a singlet.
If this $\Gamma_7$ level is located well below the $\Gamma_8$,
the occupying electrons will not contribute to the magnetic properties.
Thus, we can ignore the $\Gamma_7$ electrons for our present purposes.
In order to validate this simplification,
it is useful to introduce the results of band-structure calculations
for CeIn$_3$ \cite{Betsuyaku} and UGa$_3$ \cite{Harima}.
Note that both results have been obtained assuming
the system is in the paramagnetic state.
In order to focus on the $f$ electron components of the energy band,
we concentrate on the bands around the $\Gamma$ point
near the Fermi level.
For CeIn$_3$, the energy band dominated by $\Gamma_7$ character
is found to be lower than the $\Gamma_8$-dominated band,
consistent with the local level scheme in Fig.~\ref{fig6-2}(a).
An important point is that the Fermi level intersects the
$\Gamma_7$-dominant band, indicating that the Fermi surface
is mainly composed of $\Gamma_7$ electrons hybridized with
Ga-ion $p$ electrons.
On the other hand, for UGa$_3$, the $\Gamma_7$ band is
also lower than the $\Gamma_8$
band, but the Fermi level crosses the $\Gamma_8$ band.
Thus, the $\Gamma_7$ band appears to be fully occupied,
consistent with the $j$-$j$ coupling level scheme,
as shown in Fig.~\ref{fig6-2}(b).
Since the main contribution to the Fermi surface comes from
$\Gamma_8$ electrons, it is natural to dwell on the $\Gamma_8$
bands and ignore the occupied $\Gamma_7$ bands
in giving further consideration to many-body effects.

So far, we have considered the model in the cubic system,
but as mentioned before, 115 materials exhibit
tetragonal crystal structure.
To include the effect of tetragonality, here we introduce
two ingredients into the model Hamiltonian.
One is non-zero $\Delta$, which is the level splitting
between two orbitals.
Under the tetragonal CEF, the local electronic levels are
given by two $\Gamma_7$ and one $\Gamma_6$ states.
Among them, $\Gamma_6$ is just equal to $\Gamma_8^b$
in the cubic system.
Two $\Gamma_7$ states are given by the linear combinations of
$a^{\dag}_{{\bf i}\pm3/2} |0 \rangle$ and
$a^{\dag}_{{\bf i}\mp5/2} |0 \rangle$,
which can be expressed also by the mixture of
$\Gamma_7$ and $\Gamma_8^a$.
Here for simplicity, we introduce $\Delta$, splitting energy
between $\Gamma_8$ orbitals,
by ignoring the change of wavefunctions from cubic to tetragonal systems.
Another is the change in the hopping amplitude along the $z$-axis.
In AnTGa$_5$ (An=U and Np), AnGa$_3$ layer is sandwiched by
TGa$_2$ blocks, as shown in Fig.~\ref{fig6-1}(a),
indicating that the hopping of $f$-electron along the $z$-axis
should be reduced from that in AnGa$_3$.
However, it is difficult to estimate the reduction quantitatively,
since it is necessary to include correctly the hybridization with
$d$-electrons in transition metal ions and $p$-electrons in Ga ions.
Thus, we change the hopping $t$ as $t_z$
in the definition of $t_{\tau\tau'}^{\bf z}$.

Then, the Hamiltonian is the sum of
$H_{\Gamma_8}$ and the level splitting term,
given by
\begin{eqnarray}
   H=H_{\Gamma_8} 
   -\Delta \sum_{\bf i}(\rho_{{\bf i}a}-\rho_{{\bf i}b})/2.
\end{eqnarray}
Concerning the hopping amplitudes in the xy plane,
they are given by Eqs.~(\ref{Eq:tx}) and (\ref{Eq:ty}),
but along the z-axis, it is necessary to include
the change from cubic to tetragonal case.
Namely, the effective hopping along the $z$ axis is expressed as
\begin{equation}
  t_{\tau\tau'}^{\bf z} = t_z
  \left(
    \begin{array}{cc}
     0 & 0 \\
     0 & 1 \\
    \end{array}
  \right),
\end{equation}
where $t_z$ is the reduced hopping amplitude along the $z$-axis.
The ratio $t_z/t$ is less than unity.
We note that in actuality, $\Delta$ should be related to
the value of $t_z$, since both quantities depend on
the lattice constant along the $z$ axis.
However, the relation between $t_z$ and $\Delta$
is out of the scope at present and thus,
here we simply treat them as independent parameters.

Among several methods to analyze the microscopic model,
we resort to an exact diagonalization technique
on a 2$\times$2$\times$2 lattice.
Although there is a demerit that it is difficult to enlarge the
system size, we take a clear advantage that it is possible to deduce
the magnetic structure by including the effect of electron correlation.
In order to discuss the ground-state properties,
it is useful to measure the spin and orbital correlations,
which are, respectively, defined by
\begin{equation}
  S(\bm{q}) = (1/N)\sum_{{\bf i,j}}
  \langle \sigma^{z}_{\bf i} \sigma^{z}_{\bf j} \rangle
  e^{i\bm{q}\cdot({\bf i}-{\bf j})},
\end{equation}
with $\sigma^{z}_{\bf i}$=%
$\sum_{\tau}(n_{{\bf i}\tau\uparrow}$$-$$n_{{\bf i}\tau\downarrow})/2$,
and
\begin{equation}
  T(\bm{q}) = (1/N)\sum_{{\bf i,j}}
  \langle {\bf \tau}^{z}_{\bf i}{\bf \tau}^{z}_{\bf j} \rangle
  e^{i\bm{q}\cdot({\bf i}-{\bf j})},
\end{equation}
with $\tau^{z}_{\bf i}$=%
$\sum_{\sigma}(n_{{\bf i}a\sigma}$$-$$n_{{\bf i}b\sigma})/2$.
Here $N$ is the number of sites.

\begin{figure}[t]
\begin{center}
\includegraphics[width=1.0\textwidth]{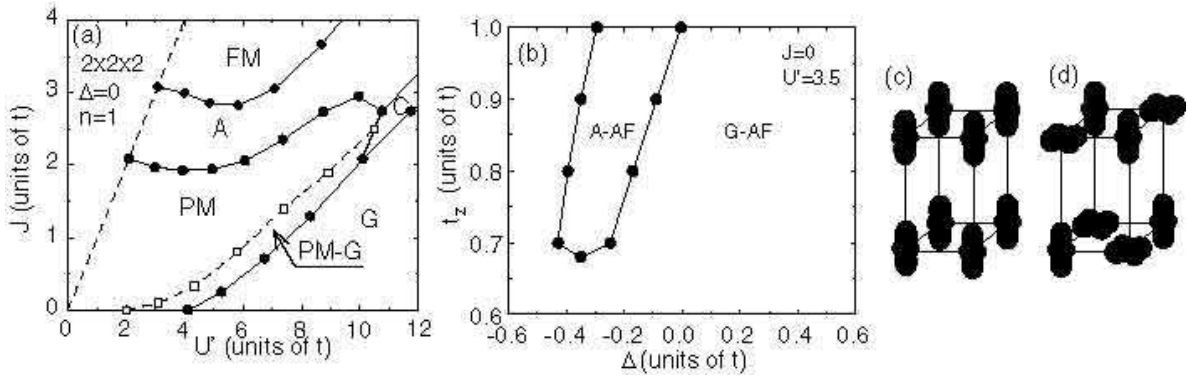}
\end{center}
\caption{
(a) Phase diagram for UGa$_3$ obtained by the exact
diagonalization.
The region of $J$$>$$U'$ is ignored, since it is unphysical.
See Fig.~\ref{fig7}(a) for the definitions of abbreviations.
Here ``PM-G'' indicates the PM phase with
enhanced $(\pi, \pi, \pi)$ spin correlation.
(b) Phase diagram of the magnetic structure
in the $(\Delta, t_z)$ plane for $J$=0 and $U'=3.5$.
(c) Ferro orbital pattern in the A-type AF phase.
(d) Antiferro orbital pattern in the G-type AF phase.
}
\label{fig6-3}
\end{figure}

\subsubsection{U-115}

Let us review the results for $n$=1 \cite{Hotta2004},
in which one electron is accommodated in $\Gamma_8$ orbital,
corresponding to uranium compounds.
First we consider the cubic case ($\Delta$=0 and $t_z/t$=1).
After we evaluate spin and orbital correlations for several parameter
sets, the ground-state phase diagram is completed on the $(U', J)$ plane,
as shown in Fig.~\ref{fig6-3}(a).
In the small-$J$ region, the paramagnetic (PM) phase exists
for large parameter space and
in the boundary region between PM and G-type AF states, we can see
the PM phase with dominant $(\pi, \pi, \pi)$ spin correlation.
Note that such a PM-G region is not specific to the case of $J$=0,
since it appears even when we increase the Hund's rule interaction.

Here we briefly discuss the phases in the large-$J$ region.
We observe an interesting similarity with the phase diagram for
undoped manganites RMnO$_3$,
in which mobile $e_{\rm g}$-electrons are tightly coupled with
the Jahn-Teller distortions and the background $t_{\rm 2g}$ spins.
Note that the present Hamiltonian is just equal to the $e_{\rm g}$
electron part of the model for manganites \cite{Hotta2003}.
In the so-called double-exchange system with large Hund's rule
coupling between $e_{\rm g}$ and $t_{\rm 2g}$ electrons,
the Jahn-Teller distortion suppresses the probability of
double occupancy and it plays a similar role as
the interorbital Coulomb interaction $U'$.
The AF coupling among $t_{\rm 2g}$ spins, $J_{\rm AF}$,
controls the FM tendency in the $e_{\rm g}$-electron phases.
Roughly speaking, large (small) $J_{\rm AF}$ denotes small (large) $J$.
Then, we see an interesting similarity between Fig.~\ref{fig6-3}(a)
and the phase diagram for manganites, except for the PM region.
See Figs.~\ref{fig7}(b) and \ref{fig8}(a).
In particular, a chain of the transition,
FM$\rightarrow$A-AF$\rightarrow$C-AF$\rightarrow$G-AF,
occurs with decreasing $J$ (increasing $J_{\rm AF}$).
Again we stress that the present $\Gamma_8$ model for
$f$-electron systems is essentially the same as
the $e_{\rm g}$ orbital model in the $d$-electron systems.
It is interesting to observe common phenomena
concerning orbital degree of freedom in $f$-electron systems.

Now we include the effect of tetragonality.
Since the mother compound UGa$_3$ is AF metal, in our phase diagram,
it is reasonable to set the parameter region corresponding to ``PM-G''.
Then, we choose $U'$=3.5 and $J$=0.
Again we evaluate spin and orbital correlations
by changing $\Delta$ and $t_z$, and
obtain the phase diagram in the $(\Delta, t_z)$ plane,
as shown in Fig.~\ref{fig6-3}(b) for $J$=0 and $U'$=3.5.
Note that the ground state for $\Delta$=0 and $t_z$=1
is magnetic metallic, as seen in Fig.~\ref{fig6-3}(a).
It is found that an A-type AF phase appears in the negative
$\Delta$ region for $t_z$$>$0.68.
Note that the appearance of the A-AF phase is not sensitive to $t_z$
as long as $t_z$$>$0.68.
Rather, $\Delta$ seems to play a key role in controlling the change of
the magnetic phase.
Here we recall the experimental fact that
UNiGa$_5$ exhibits a G-type AF phase,
while UPtGa$_5$ shows an A-type.
Thus, it is necessary to relate the effect of $\Delta$ to
the difference in magnetic structure
found between UNiGa$_5$ and UPtGa$_5$.
Although $t_z$ may differ among U-115 compounds,
we focus here on the effect of $\Delta$.

From the orbital correlation, we obtain the ferro-orbital
and anti-ferro orbital patterns for A-AF and G-AF phases,
respectively, as shown in
Figs.~\ref{fig6-3}(c) and \ref{fig6-3}(d).
Let us now discuss the reasons for the appearance of an A-AF phase.
For negative values of $\Delta$, we easily obtain ferro-orbital
(FO) pattern composed of $\Gamma_8^{b}$ orbitals,
as illustrated in Fig.~\ref{fig6-3}(c).
For electrons to gain kinetic energy of motion along the $z$-axis,
it is necessary to place the AF spin arrangement along this same axis.
In the FM spin configuration, electrons cannot move along the $z$-axis
due to the Pauli principle, since hopping occurs $only$ between
$\Gamma_8^{b}$ orbitals along the $z$-axis.
On the other hand, in the $xy$ plane $b$-orbital electrons
can hop to neighboring $a$-orbitals with a significant amplitude,
which is larger than that between neighboring $b$-orbitals.
Thus, in order to gain kinetic energy, electrons tend to occupy
$a$-orbitals even in the FO state composed of $b$-orbitals,
as long as $|\Delta|$ is not so large.
When we explicitly include the effects of the Hund's rule
interaction $J$, electron spins should have FM alignment between
neighboring sites in order to gain energy
in hopping processes from $b$- to $a$-orbitals.
Consequently, a FM spin configuration is favored in the $xy$ plane.
In the case with antiferro orbital correlations,
spin correlation tends in general to be FM,
as has been widely recognized in orbitally degenerate systems.

Here we mention a relation of $\Delta$ to
the magnetic anisotropy in U-115 materials.
For UPtGa$_5$ with the A-AF phase, $\chi_a$ is larger than $\chi_c$,
whereas this anisotropy is not pronounced in UNiGa$_5$
with the G-AF phase \cite{U115-8}.
An analysis for the high-temperature region based on $LS$ coupling
yields the $J_z$=$\pm$1/2 Kramers doublet as the ground state
among the dectet of $J$=9/2 ($L$=6 and $S$=3/2).
The states with $J_z$=$\pm$1/2 in the $LS$ coupling scheme
have significant overlap with
$f^{\dag}_{{\bf i}b \uparrow}f^{\dag}_{{\bf i}c \uparrow}
f^{\dag}_{{\bf i}c \downarrow}|0 \rangle$ and
$f^{\dag}_{{\bf i}b \downarrow}f^{\dag}_{{\bf i}c \uparrow}
f^{\dag}_{{\bf i}c \downarrow}|0 \rangle$
in the $j$-$j$ coupling scheme.
Accordingly, by the present definition $\Delta$ should be
negative to place $\Gamma_8^{b}$ below $\Gamma_8^{a}$.
If the absolute value of $\Delta$($<$0) becomes large,
$\Gamma_8^{b}$ is well separated from $\Gamma_8^{a}$
and the magnetic anisotropy will consequently become large.
Thus, a change from G- to A-type AF phase is consistent with
the trends of magnetic anisotropy in UNiGa$_5$ and UPtGa$_5$.

Finally, we make a brief comment about the effect of $t_{z}$.
Following the above discussion, the A-AF phase should
appear even for small $t_{z}$.  However, in the present
calculation it disappears for $t_{z}$$<$0.68,
a critical value which seems to be rather large.
Such a quantitative point depends on the system size, and
we note that it is necessary to perform the calculation
in the thermodynamic limit.

While such investigations are just beginning, we already see
a number of opportunities for future work along this path.
Concerning issues directly related to the present context,
it is highly recommended that calculations be carried out
in the thermodynamic limit, in order
to confirm the present exact diagonalization results.
For instance, the magnetic susceptibility should be evaluated
in the random phase approximation or fluctuation-exchange method.
With such an approach, the magnetic structure can be discussed by detecting
the divergence in the magnetic susceptibility.
This is one of our future tasks.
Another problem is how to establish the effective reduction of
$t_{z}$ in considering the case of UTGa$_5$.
In such systems, TGa$_2$ blocks are interspersed between UGa$_3$ layers,
but the main process may occur through the Ga ions.
To analyze this, it is necessary to treat
a three-dimensional $f$-$p$ model with
explicit consideration of U and Ga ions.
This is another problem for future investigation.

\subsubsection{Np-115}

Now we review the theoretical results for magnetic structure
of Np-115 \cite{Onishi2004}.
First we consider the case of $n$=2, which is corresponding
to the trivalent Np ion.
At $t_z$=1 and $\Delta$=0 (cubic case),
local triplet composed of a couple of $f$ electrons is formed
at each site and the G-type AF structure is stabilized
due to the so-called superexchange interaction.
Even when $\Delta$ is introduced as the tetragonal CEF effect,
the G-AF structure remains robust for $|\Delta|<1$.
When $|\Delta|$ is larger than unity, two electrons simultaneously
occupy the lower orbital, leading to the non-magnetic state
composed of local $\Gamma_1$, irrelevant to the present study
to consider the magnetic phase.
When we change $t_z$ for $\Delta$=0,
again the G-type AF structure is stabilized,
but we find that the spin correlation of ${\bm q}$=$(\pi,\pi,0)$
comes to be equivalent to that of ${\bm q}$=$(\pi,\pi,\pi)$
with the decrease of $t_z$,
since the AF structure is stabilized in each $xy$ plane
due to superexchange interaction
and the planes are decoupled for small $t_z$.

At the first glance, it seems difficult to understand the variety
of magnetic phases observed in NpTGa$_5$ even in a qualitative level,
when we consider only the trivalent Np ion.
However, there is no {\it a priori} reason to fix the valence as
Np$^{3+}$.
In NpTGa$_5$, $d$-electron band originating from transition metal
ions may significantly affect the valence of Np ion \cite{Maehira2004}.
In addition, we also stress that the actual compounds exhibit
AF metallic behavior.
In the band-structure calculation, the average number of $f$ electrons
at Np ion is easily decreased from four.
Thus, we treat the local $f$-electron number as a parameter.

We may consider another reason to decrease effectively the number
of $f$ electron from $n$=2 in NpGa$_3$.
In the present two-orbital model, the G-AF structure is robust,
which is natural from the theoretical viewpoint within the model.
However, in the experimental result on NpGa$_3$,
the low-temperature ground state is ferromagnetic, although
the AF phase has been observed around at $T$$\sim$60K.
In order to understand the occurrence of the FM phase in the
two-orbital model, it is necessary to inject some amount of
``hole'' in the AF phase, since the double-exchange mechanism works
to maximize the kinetic motion of electrons.
It is difficult to determine the amount of doped holes
to obtain the FM phase, but at least qualitatively,
the effective decrease of $n$ seems to be physically meaningful
in NpGa$_3$ as well as NpTGa$_5$.

\begin{figure}[t]
\begin{center}
\includegraphics[width=0.8\textwidth]{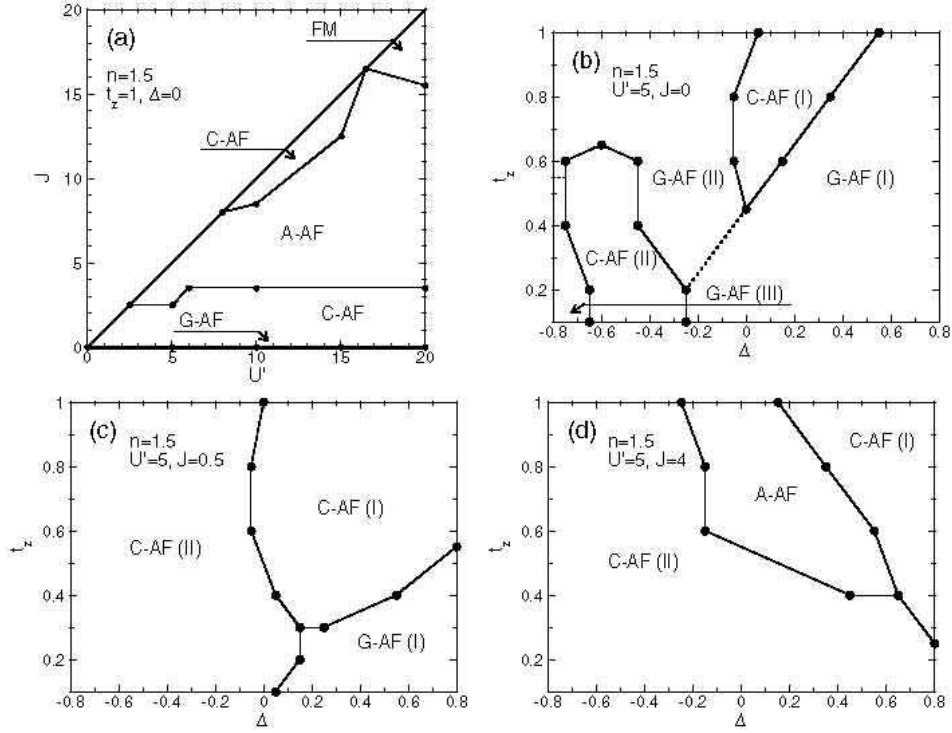}
\end{center}
\caption{%
(a) Ground-state phase diagram in the $(U',J)$ plane
for $n$=1.5, $t_z$=1, and $\Delta$=0.
Ground-state phase diagrams of the magnetic structure
in the $(\Delta, t_z)$ plane for $n$=1.5, $U'$=5,
(b) $J$=0, (c) $J$=0.5, and (d) $J$=4.
}
\label{fig6-4}
\end{figure}

Then, we consider the case of $n$=1.5.
In Fig.~\ref{fig6-4}(a), we show the ground-state phase diagram
in the $(U',J)$ plane at $n$=1.5 for the cubic case
with $t_z$=1 and $\Delta$=0.
At $J$=0, a G-type AF structure is stabilized due to
superexchange interaction in the same way as the case of $n$=2.
However, the G-AF structure is immediately changed to a C-AF structure
only by a small value of the Hund's rule interaction.
With increasing $J$, the magnetic phase changes
in the order of G-AF, C-AF, A-AF, and FM phases,
except for the C-AF phase in the large $J$ region.
Concerning the spin structures, see Fig.~\ref{fig7}(a).
This result is quite natural, since we are now considering
the magnetic structure based on the two-orbital model,
in which the FM tendency is due to the optimization
of kinetic motion of electrons.

After calculations of the spin and orbital correlations
for several parameter sets, we obtain the ground-state phase diagram
in the $(\Delta,t_z)$ plane,
as shown in Figs.~\ref{fig6-4}(b), for $U'$=5 and $J$=0.
In the region of large positive $\Delta$, we find that G-AF(I) phase
with ferro-type arrangement of $\Gamma_8^a$ orbital extends
in a wide range of the phase diagram.
It is found that the C-AF(I) phase appears in the region for
small positive $\Delta$ and 0.5$<$$t_z$$<$1,
The C-AF(I) phase exhibits the dominant component of $(\pi,\pi,0)$
in the spin correlation.
When we further decrease $\Delta$, we find G-AF(II) phase,
which may be considered as orbital disordered,
since there is no dominant component in the orbital correlation.
For small $t_z$ and small negative $\Delta$,
we find another C-AF phase, which we call C-AF(II), in which
the spin correlation of $(\pi,0,\pi)$ and $(0,\pi,\pi)$ are dominant.
For small $t_z$ and large negative $\Delta$, there appears
yet another G-AF phase, called G-AF(III)
with ferro-type arrangement of $\Gamma_8^b$ orbital.
In any case, for $J$=0, we can observe several kinds of C- and G-AF
phases, but A-AF phase does not occur.

Although we increase the value of $J$ as $J$=0.5, no new phases
appear in the phase diagram for $n$=1.5,
as shown in Fig.~\ref{fig6-4}(c).
There are three phases, but they are two C-AF and one G-AF states.
As labeled explicitly in the phase diagrams, C-AF(I), C-AF(II),
and G-AF(I) are the same as those in the phase diagram
of Fig.~\ref{fig6-4}(b).
Due to the effect of $J$, G-AF(II) and G-AF(III) disappear,
since the number of FM bond should be increased to gain
the kinetic energy.
As shown in Fig.~\ref{fig6-4}(d),
when we further increase the value of $J$ as $J$=4,
the G-AF phase completely disappears and instead,
we observe the A-AF phase sandwiched by two C-AF phases.
By analogy with the various phases of manganites, the A-AF phase is
considered to appear due to the double-exchange mechanism
in the two-orbital model, when $J$ is increased.

In the experiments for NpTGa$_5$,
C-, A-, and G-AF magnetic phases have been found in
NpFeGa$_5$, NpCoGa$_5$, and NpNiGa$_5$.
Here we have a naive question: What is a key parameter
to understand the change of the magnetic structure?
In the case of UTGa$_5$, it has been claimed that
the level splitting $\Delta$ is important
to explain the difference in magnetic structure as
well as the magnetic anisotropy for a fixed value of $n$=1.
Roughly speaking, $\Delta$ is positive for T=Fe,
small positive for T=Co, and negative for T=Ni.
Among UTGa$_5$ with T=Ni, Pd, and Pt, when we assume that
the absolute value of $\Delta$ is increased in the order
of Ni, Pd, and Pt, it is possible to understand qualitatively
the change in the magnetic anisotropy,
in addition to the change in the magnetic structure of
G-AF for T=Ni and A-AF for T=Pd and Pt.
It has been found that the value of $t_z$ is not so crucial
to explain qualitatively the magnetic properties of
U-115 based on the two-orbital model for $n$=1.

For $n$=2, we always obtain the G-AF phase.
However, for $n$=1.5, we have observed three kinds of AF
magnetic structure in the phase diagrams.
Let us summarize the change in the magnetic structure
for a fixed value of $t_z$=0.8.
Note that this value is larger than $t_z$=0.1, which we have
considered to reproduce two kinds of cylindrical Fermi-surface
sheets of Np-115.
However, in the small-sized cluster calculations,
it is difficult to compare directly with the values in
the thermodynamic limit.
Thus, we do not discuss further the quantitative point on
the values of $t_z$.
As shown in Fig.~\ref{fig6-4}(b), for $J$=0 and $t_z$=0.8,
we see the change in the magnetic structure
as G-AF ($\Delta$$<$0), C-AF(0$<$$\Delta$$<$0.4),
and G-AF ($\Delta$$>$0.4).
For $J$=0.5 and $t_z$=0.8, as shown in Fig.~\ref{fig6-4}(c),
the C-AF phases are always observed, but they have
different orbital structures.
Finally, for $J$=4 and $t_z$=0.8, we observe
C-AF ($\Delta<-0.15$), A-AF($-0.15<\Delta<0.3$),
and C-AF ($\Delta$$>$0.3), as shown in Fig.~\ref{fig6-4}(d).

In order to understand the appearance of three types of
the AF phases, we may consider an explanation due to
the combination of the changes in $\Delta$ and $n$.
For instance, by assuming that $J$=4 for NpTGa$_5$
and the change in $\Delta$ for NpTGa$_5$ is just
the same as that for UTGa$_5$,
we consider that $n$$\approx$2 with $\Delta$$<$0 for T=Ni,
$n$$\approx$1.5 with $\Delta$$\approx$0 for T=Co,
and $n$$\approx$1.5 with $\Delta$$>0$ for T=Fe.
Then, it seems to be possible to relate our theoretical AF phases
with the experimental observations in NpTGa$_5$.
However, it is difficult to claim that the above parameter assignment
for three Np-115 materials is the best explanation for
the magnetic structure of Np-115,
since in actual compounds, there are other important ingredients
which have not been included in the present model.
For instance, we have never discussed the direction of the
magnetic moment of Np ion.
In particular,
the canted AF structure cannot be considered at all
for the G-AF phase of NpNiGa$_5$.
Thus, we need to recognize some distance between the actual magnetic
states and the theoretically obtained phases.
Our theory should be improved by taking into account
other realistic ingredients of 115 structure.

Finally, let us remark a possible future direction of the research.
In Fig.~\ref{fig6-4}(b), we have explained the G-AF(II) as
the orbital disordered phase, which is considered to be related
to the metallic phase.
Of course, we cannot conclude the matallicity only from the present
calculations for small-size cluster.
However, it seems to be an interesting concept that
the $f$-electron state is controlled by orbital degeneracy.
Quite recently, Onishi and Hotta has pointed out the orbital
incommensurate state appearing between two kinds of localized AF phases,
by using the density matrix renormalization group method
to the model of the $j$-$j$ coupling scheme \cite{Onishi2006}.
We may throw new lights on the long-standing issue
concerning the competition between localized and itinerant nature
of $f$-electrons, when such competition is controlled
by orbital degree of freedom.

%
\subsection{Multipole ordering}

In the previous subsection, we have discussed the spin and orbital
structure of 115 materials on the basis of the $\Gamma_8$ model.
Again we note that such ``spin'' and ``orbital'' are not real
spin and orbital, but pseudo spin and orbital.
In principle, in $f$-electron systems, real spin and orbital
are not independent degrees of freedom,
since they are tightly coupled with each other
due to the strong spin-orbit interaction.
Then, in order to describe such a complicated spin-orbital
coupled system, it is rather appropriate to
represent the $f$-electron state in terms of
``multipole'' degree of freedom,
rather than using spin and orbital degrees of freedom
as in $d$-electron systems.
We show a table to summarize the multipole operators up to rank 3.
The relation of the irreducible representation
with pseudo spin and orbital are also shown.
When we use the term of multipole degree of freedom,
pseudo spin $\sigma$ and pseudo orbital $\tau$ are
$\Gamma_{\rm 4u}$ dipole and $\Gamma_{\rm 3g}$ quadrupole,
respectively, from Table 3.
Thus, in the previous section, we have considered the ordering
tendencies of dipole and quadrupole degree of freedom.
Note that it is important to show explicitly the parity of
the multipole operator to complete its symmetry.
In this section, for convenience, we use the subscripts ``u'' and ``g''
for odd and even parity, respectively.

\begin{table}
\caption{
\label{table:multipole_operators}
Multipole operators in the $\Gamma_8$ subspace, shown
in Ref.~\cite{Shiina}.
The first, second, and third lines denote
$\gamma$ of the irreducible representation $\Gamma_{\gamma}$,
multipole operator $X^{\Gamma_{\gamma}}$,
and pseudospin representation, respectively.
The multipole operators are represented by pseudospin operators as
$\hat{\bm{\tau}}=\sum_{\tau, \tau^{\prime}, \sigma}
c^{\dagger}_{\tau \sigma}\bm{\sigma}_{\tau \tau^{\prime}}
c_{\tau^\prime \sigma}$
and
$\hat{\bm{\sigma}}=\sum_{\tau, \sigma, \sigma^{\prime}}
c^{\dagger}_{\tau \sigma}\bm{\sigma}_{\sigma \sigma^{\prime}}
c_{\tau \sigma^\prime}$,
where $\bm{\sigma}$ are the Pauli matrices.
We use notations
$\hat{\eta}^{\pm}=(\pm\sqrt{3} \hat{\tau}^x-\hat{\tau}^z)/2$
and
$\hat{\zeta}^{\pm}=-(\hat{\tau}^x \pm \sqrt{3}\hat{\tau}^z)/2$.
For simplicity, we suppress the site label $\bm{r}$ in this table.}
\begin{flushleft}
\begin{tabular}{|c|cc|ccc|ccc|}
\hline
        $2u$   & $3gu$  & $3gv$
      & $4u1x$ & $4u1y$ & $4u1z$
      & $4u2x$ & $4u2y$ & $4u2z$ \\
\hline
        $T_{xyz}$   & $O^0_2$     & $O^2_2$
      & $J^{4u1}_x$ & $J^{4u1}_y$ & $J^{4u1}_z$
      & $J^{4u2}_x$ & $J^{4u2}_y$ & $J^{4u2}_z$ \\
\hline
        $\hat{\tau}^y$
      & $\hat{\tau}^z$
      & $\hat{\tau}^x$

      & $\hat{\sigma}^x$
      & $\hat{\sigma}^y$
      & $\hat{\sigma}^z$

      & $\hat{\eta}^+ \hat{\sigma}^x$
      & $\hat{\eta}^- \hat{\sigma}^y$
      & $\hat{\tau}^z \hat{\sigma}^z$ \\
\hline
\end{tabular}
\end{flushleft}
\begin{flushleft}
\begin{tabular}{|ccc|ccc|}
\hline
        $5ux$  & $5uy$  & $5uz$
      & $5gx$  & $5gy$  & $5gz$ \\
\hline
       $T^{5u}_x$  & $T^{5u}_y$ & $T^{5u}_z$
       & $O_{yz}$  & $O_{zx}$   & $O_{xy}$ \\
\hline
        $\hat{\zeta}^+ \hat{\sigma}^x$
      & $\hat{\zeta}^- \hat{\sigma}^y$
      & $\hat{\tau}^x  \hat{\sigma}^z$

      & $\hat{\tau}^y \hat{\sigma}^x$
      & $\hat{\tau}^y \hat{\sigma}^y$
      & $\hat{\tau}^y \hat{\sigma}^z$ \\
\hline
\end{tabular}
\end{flushleft}
\end{table}

In this article, we pick up octupole ordering in NpO$_2$.
First we briefly discuss the level scheme for actinide dioxides
with CaF$_2$ cubic crystal structure.
Due to the CEF effect,
the sextet is split into $\Gamma_8$ quartet and $\Gamma_7$ doublet.
In this case, due to the intuitive discussion on the direction
of the extension of orbital and the position of oxygen ions,
the $\Gamma_7$ state should be higher than the $\Gamma_8$ level.
Here we define the splitting energy as $\Delta$.
As discussed in Sec.~5, in order to make the low-spin state,
we accommodate two, three, and four electrons in the $\Gamma_8$ level.
Then, the ground states are $\Gamma_5$, $\Gamma^{(2)}_8$,
and $\Gamma_1$, respectively \cite{Hotta2003},
consistent with the CEF ground states of
UO$_2$ \cite{UO2-1,UO2-2}, NpO$_2$ \cite{NpO2-1,NpO2-2},
and PuO$_2$ \cite{PuO2-1,PuO2-2}, respectively.
Then, $\Delta$ is estimated from the CEF excitation energy
in PuO$_2$, experimentally found to be 123 meV \cite{PuO2-1,PuO2-2}.
On the other hand, as mentioned repeatedly in this article,
the Hund's rule coupling $J_{\rm H}$ between
$\Gamma_8$ and $\Gamma_7$ levels is 1/49 of the original Hund's
rule interaction among $f$ orbitals.
Namely, $J_{\rm H}$ is as large as a few hundred Kelvins.
Thus, we validate our assumption that
the $\Gamma_7$ state is simply ignored.
From the qualitative viewpoint, unfortunately, this simplification
is not appropriate to reproduce experimental results,
since the ground-state wave-function is not exactly reproduced
in the $\Gamma_8$ model.
However, we believe that this approximation provides a qualitatively
correct approach, in order to understand the complex multipole state
from the microscopic viewpoint.

Let us here review the recent theoretical result on
octupole order of NpO$_2$
\cite{Kubo2005a,Kubo2005b,Kubo2005c,Kubo2005d,Kubo2005e}.
We set the Hamiltonian for actinide dioxides as
the $\Gamma_8$ model Eq.~(\ref{H8}) on an fcc lattice.
The form of the Hamiltonian is already given, but the hopping
should be estimated on the fcc lattice.
For instance, the hopping amplitudes between $f$-orbitals at $(0,0,0)$
and $(a/2,a/2,0)$ ($a$ is the lattice constant) are given by
\begin{equation}
  t^{(a/2, a/2, 0)}_{\tau \uparrow; \tau^{\prime} \uparrow}=
  t^{(a/2, a/2, 0) *}_{\tau \downarrow; \tau^{\prime} \downarrow}=
  t \left(
  \begin{array}{cc}
    4 &  2\sqrt{3}i \\
    -2\sqrt{3}i & 3
  \end{array}
  \right).
\end{equation}
and
\begin{equation}
  t^{(a/2, a/2, 0)}_{\tau \uparrow; \tau^{\prime} \downarrow}=
  t^{(a/2, a/2, 0)}_{\tau \downarrow; \tau^{\prime} \uparrow}= 0,
\end{equation}
where $t=(ff\sigma)/28$ and
$t^{-\bm{\mu}}_{\tau \sigma; \tau^{\prime} \sigma^{\prime}}$
=$t^{\bm{\mu}}_{\tau \sigma; \tau^{\prime} \sigma^{\prime}}$.
Note that the hopping integrals depend on $\bm{\mu}$ and
they are intrinsically complex numbers in the fcc lattice.
It is in sharp contrast to the previous model
on a simple cubic lattice,
in which the hopping integrals are real and
the same form as for $e_{\rm g}$ orbitals of $d$ electrons.
In other words, there is no difference
between the $\Gamma_8$ model for $f$ electrons
and the $e_{\rm g}$ orbital model for $d$ electrons
on the simple cubic lattice in the absence of a magnetic field.
For $e_{\rm g}$ orbitals, we can always set the hopping integrals
to be real irrespective of the lattice type,
by selecting appropriate basis wave-functions,
while $\Gamma_8$ orbitals on the fcc lattice
appear to be complex in nature, specific to $f$-electron systems
with strong spin-orbit coupling.

In order to discuss multipole ordering,
we derive an effective multipole model in the strong-coupling limit
using standard second-order perturbation theory with respect to $t$,
which is exactly the same as that used in the derivation
of the Heisenberg model from the Hubbard model in the
research field of transition metal oxides.
It is one of advantages of the $f$-electron model on the basis of
the $j$-$j$ coupling scheme that we can exploit the technique
developed in the $d$-electron research.
We consider the case of one electron per $f$ ion in the $\Gamma_8$
orbitals, but the effective model is the same for the one-hole case,
which corresponds to NpO$_2$, due to an electron-hole transformation.
Among the intermediate $f^2$-states in the perturbation theory,
we consider only the lowest-energy $\Gamma_5$ triplet states,
in which the two electrons occupy different orbitals,
assuming that other excited states are
well separated from the $f^2$ ground states.
In fact, the excitation energy from the $\Gamma_5$ ground state
of $f^2$ in UO$_2$ is 152 meV \cite{UO2-1,UO2-2}.
Note that the CEF excitation energy is considered to be larger
than the triplet excitation one, since the Hund's rule interaction
is effectively reduced in the $j$-$j$ coupling scheme.
Thus, it is reasonable to take only the $\Gamma_5$ states
as the intermediate states.

\begin{table}
  \caption{\label{table:coupling_constants}
    Coupling constants in the effective model.
    The energy unit is $(1/16)t^2/(U^{\prime}-J)$.
  }
    \begin{tabular}{ccccccccc}
      $a_1$       & $a_3$        & $a_4$ &
      $b_8$       & $b_9$        & $b_{10}$ &
      $b^{(1)}_1$ & $b^{(1)}_2$  & $b^{(1)}_3$ \\
      12          & $64\sqrt{3}$ & 192 &
      195         & $-336$       & 576 &
      $-196$      & $-4$         & 0           \\
      \hline
      $b^{(1)}_4$   & $b^{(1)}_5$ & $b^{(1)}_6$ &
      $b^{(2)}_1$   & $b^{(2)}_2$ & $b^{(2)}_3$ &
      $b^{(2)}_4$   & $b^{(2)}_5$ & $b^{(2)}_6$ \\
      $224\sqrt{3}$ & 0           & 0           &
      4             & 193         & $-336$      &
      $64\sqrt{3}$  & $2\sqrt{3}$ & $112\sqrt{3}$
    \end{tabular}
\end{table}

After straightforward, but tedious calculations \cite{Kubo2005a,Kubo2005b},
we arrive at an effective model in the form of
\begin{equation}
  H_{\rm eff}=
  \sum_{\bm{q}} (H_{1 \bm{q}}
  +H_{2 \bm{q}}+\mathcal{H}_{4u1 \bm{q}}+H_{4u2 \bm{q}}),
  \label{eq:effectiveH}
\end{equation}
where $\bm{q}$ is the wave vector.
$H_{1 \bm{q}}$ denotes
the interactions between quadrupole moments, given by
\begin{eqnarray}
  H_{1 \bm{q}}
 &=& a_1(O^0_{2, -\bm{q}} O^0_{2, \bm{q}} c_x c_y + {\rm c.p.}) \nonumber \\
 &+& a_3(O^0_{2, -\bm{q}} O_{xy, \bm{q}} s_x s_y + {\rm c.p.}) \nonumber \\
 &+& a_4(O_{xy, -\bm{q}} O_{xy, \bm{q}} c_x c_y + {\rm c.p.}),
\end{eqnarray}
where c.p. denotes cyclic permutations,
$c_{\nu}=\cos(q_{\nu} a/2)$,
and $s_{\nu}=\sin(q_{\nu} a/2)$ ($\nu=x$, $y$, or $z$).
The definitions of the multipole operators
and values of the coupling constants $a_i$ are
given in Tables~\ref{table:multipole_operators}
and \ref{table:coupling_constants}, respectively.
Note that
$O^0_{2 \bm{q}}$ transforms to $(\sqrt{3}O^2_{2 \bm{q}}-O^0_{2 \bm{q}})/2$
and $(-\sqrt{3}O^2_{2 \bm{q}}-O^0_{2 \bm{q}})/2$
under c.p. $(x,y,z)\rightarrow(y,z,x)$ and $(x,y,z)\rightarrow(z,x,y)$,
respectively.
$\mathcal{H}_{2 \bm{q}}$ and ${H}_{4un \bm{q}}$ ($n=1$ or 2)
are the interactions between dipole and octupole moments, given by
\begin{eqnarray}
  H_{2 \bm{q}}
  &=& b_8[T^{5u}_{z, -\bm{q}} T^{5u}_{z, \bm{q}} (c_y c_z + c_z c_x)
    + {\rm c.p.}] \nonumber \\
    &+&b_9[T^{5u}_{x, -\bm{q}} T^{5u}_{y, \bm{q}} s_x s_y + {\rm c.p.}]
 \nonumber \\
    &+&b_{10} T_{xyz, -\bm{q}} T_{xyz, \bm{q}} (c_x c_y + {\rm c.p.}),
\end{eqnarray}
and
\begin{eqnarray}
  H_{4u n \bm{q}}
  &=& b^{(n)}_1 [J^{4u n}_{z -\bm{q}} J^{4u n}_{z \bm{q}} c_x c_y
      +{\rm c.p.}] \nonumber \\
  &+& b^{(n)}_2 [J^{4u n}_{z -\bm{q}} J^{4u n}_{z \bm{q}} (c_y c_z +c_z c_x)
      +{\rm c.p.}] \nonumber \\
  &+& b^{(n)}_3 [J^{4u n}_{x -\bm{q}} J^{4u n}_{y \bm{q}} s_x s_y
      +{\rm c.p.}] \nonumber \\
  &+& b^{(n)}_4
    [T_{xyz -\bm{q}} (J^{4u n}_{z \bm{q}} s_x s_y+{\rm c.p.})] \nonumber \\
  &+& b^{(n)}_5 [T^{5u}_{z -\bm{q}} J^{4u n}_{z \bm{q}} c_z(c_x-c_y)
      +{\rm c.p.})] \nonumber \\
  &+& b^{(n)}_6 [T^{5u}_{z -\bm{q}}
      (-J^{4u n}_{x \bm{q}}s_z s_x+J^{4u n}_{y \bm{q}}s_y s_z) +{\rm c.p.}],
  \label{eq:effectiveH_4u}
\end{eqnarray}
where values of the coupling constants $b_i$ and $b^{(n)}_i$
are shown in Table~\ref{table:coupling_constants}.
The above Eqs.~(\ref{eq:effectiveH})--(\ref{eq:effectiveH_4u})
are consistent with the general form of multipole interactions
on the fcc lattice derived by Sakai et al. \cite{Sakai1}.
We follow the notation in Ref.~\cite{Sakai1} for convenience.

\begin{figure}
\begin{center}
  \includegraphics[width=0.8\linewidth]{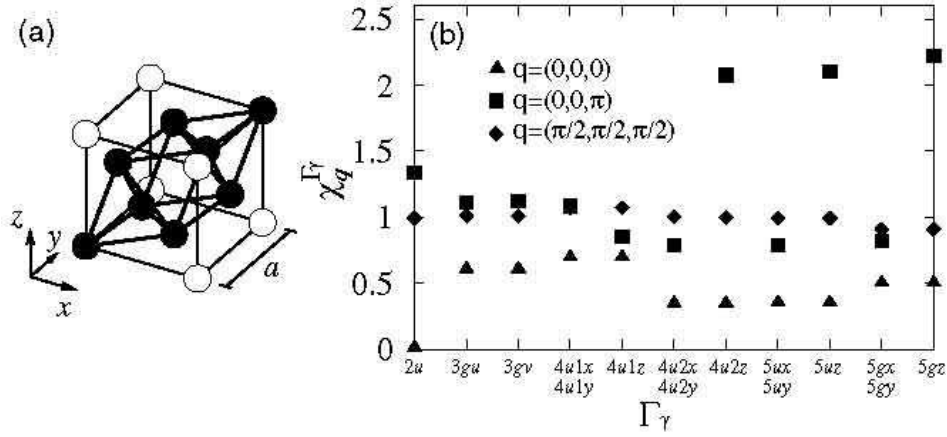}
\end{center}
\caption{
(a) The 8-site fcc cluster, shown by solid spheres,
taken in the calculation of exact diagonalization.
(b) Correlation functions for the 8-site cluster for
$\bm{q}=(0,0,0)$ (triangles),
$\bm{q}=(0,0,1)$ (squares), and
$\bm{q}=(1/2,1/2,1/2)$ (diamonds) in units of $2\pi/a$.
}
\label{figure:correlation_function}
\end{figure}

When a mean-field theory is applied to the effective model,
due care should be taken,
since in an fcc lattice with geometrical frustration,
the effect of fluctuations may be strong enough
to destroy the state obtained within the mean-field theory.
Thus, we first evaluate the correlation function in the ground state
using an unbiased method such as
exact diagonalization on the $N$-site lattice.
Here we set $N=8$, as shown Fig.~\ref{figure:correlation_function}(a).
The correlation function of the multipole operators is defined by
\begin{equation}
  \chi^{\Gamma_{\gamma}}_{\bm{q}}=(1/N)
  \sum_{\bm{r},\bm{r}^{\prime}} e^{i \bm{q} \cdot (\bm{r}-\bm{r}^{\prime})}
  \langle X^{\Gamma_{\gamma}}_{\bm{r}}
  X^{\Gamma_{\gamma}}_{\bm{r}^{\prime}} \rangle,
\end{equation}
where $\langle \cdots \rangle$ denotes
the expectation value by the ground-state wave-function.
Figure~\ref{figure:correlation_function}(b) exhibits
the results for correlation functions.
As shown in the Table 4, the absolute value of the interaction between
$\Gamma_{2u}$ moments ($b_{10}$) is the largest among multipole interactions,
but the correlation function of the $\Gamma_{2u}$ moment is not so enhanced,
suggesting that the frustration effect is significant for an Ising-like moment
such as $\Gamma_{2u}$.
Rather, large values of correlation functions are found
for $J^{4u2}_{z}$, $T^{5u}_z$, and $O_{xy}$ moments
at $\bm{q}=(0,0,1)$ in units of $2\pi/a$.
Note that the effective model does not include the term
which stabilizes $O_{xy}$ quadrupole order at $\bm{q}=(0,0,1)$.
The enhancement of this correlation function is due to
an induced quadrupole moment
in $\Gamma_{4u2}$ or $\Gamma_{5u}$ moment ordered states.
Thus, the relevant interactions of the system should be
$b^{(2)}_2$ and $b_8$, which stabilize the $J^{4u2}_z$ and $T^{5u}_z$
order, respectively, at $\bm{q}=(0,0,1)$.
In the following, then, we consider a simplified multipole model
which includes only $b^{(2)}_2$ and $b_8$.

\begin{figure}
\begin{center}
\includegraphics[width=0.5\textwidth]{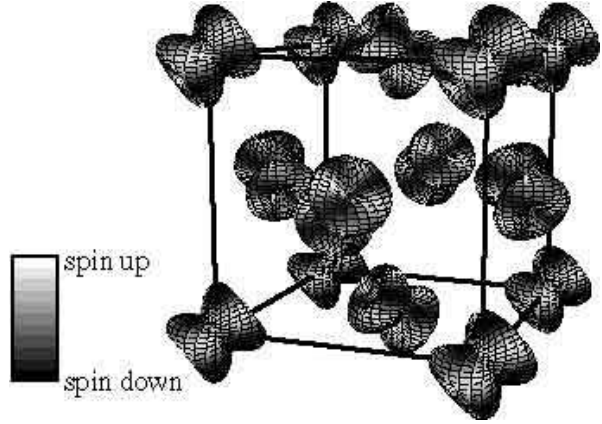}
\end{center}
\caption{
The triple-$\bm{q}$ $\Gamma_{5u}$ octupole state in the fcc lattice.}
\label{fcc}
\end{figure}

Now we apply mean-field theory to the simplified model to specify
the ordered state.
As easily understood, the coupling constant $b_8$ is slightly
larger than $b^{(2)}_2$, indicating that $\Gamma_{5u}$ order has
lower energy than $\Gamma_{4u2}$ order.
The interaction $b_8$ stabilizes longitudinal ordering of
the $\Gamma_{5u}$ moments, but their directions are not entirely
determined by the form of the interaction.
Here we point out that in the $\Gamma_8$ subspace,
the $\Gamma_{5u}$ moment has an easy axis along [111]
\cite{Kubo2003,Kubo2004,Kiss}.
Thus, taking the moment at each site
along [111] or other equivalent directions,
we find that a triple-$\bm{q}$ state is favored,
since it gains interaction energy in all the directions.
In fact, as shown in Fig.~\ref{fcc}, the ground state has
longitudinal triple-$\bm{q}$ $\Gamma_{5u}$ octupole order
with four sublattices, i.e.,
$(\langle T^{5u}_{x \bm{r}} \rangle,
\langle T^{5u}_{y \bm{r}} \rangle,
\langle T^{5u}_{z \bm{r}} \rangle) \propto
(\exp[i 2 \pi x/a], \exp[i 2 \pi y/a], \exp[i 2 \pi z/a])$.
Note that this triple-$\bm{q}$ structure
does not have frustration even in the fcc lattice.
The ground state energy is $-4b_8$ per site,
and the transition temperature $T_0$ is given by $k_{\rm B} T_0=4b_8$.
Another important message of Fig.~\ref{fcc} is
that both up- and down-spin densities are
anisotropic with different distribution.
It may be natural, if we consider octupole as the combined
degree of freedom of spin and orbital, but such an intuitive
explanation of octupole becomes possible from the
microscopic viewpoint on the basis of the $j$-$j$ coupling scheme.

Let us briefly introduce new experimental evidence of
octupole ordering in NpO$_2$.
Quite recently, in order to clarify the nature of the ordered
phase of NpO$_2$, Tokunaga et al. have performed $^{17}$O-NMR
measurement of NpO$_2$ below $T_0$ \cite{Tokunaga}.
Tokunaga et al. have observed the occurrence of two inequivalent oxygen
sites below $T_0$ from the $^{17}$O-NMR spectrum.
They have also found that the hyperfine interaction at the oxygen sites
are explained by invoking a hyperfine interaction
with field-induced AF moments due to the longitudinal
triple-$\bm{q}$ octupole order.
Thus, the NMR results strongly support the occurrence of
the longitudinal triple-$\bm{q}$ multipole structure in NpO$_2$.

Here we briefly mention the physical properties of the phase
in the high-temperature region.
The magnetic susceptibility of UO$_2$ follows the standard
Curie-Weiss law \cite{Arrott}, while that of NpO$_2$ is
significantly deviated from the Curie-Weiss behavior
well above the transition temperature $T_0$ \cite{Ross,Erdos}.
As pointed out by Kubo and Hotta \cite{Kubo2006},
the difference in the temperature dependence of magnetic susceptibilities
in these materials is naturally explained due to the fact that
dipole $and$ octupole moments coexist in NpO$_2$
while only dipole exists in UO$_2$.
When the magnetic moment consists of two independent moments,
it is intuitively understood that in such a situation,
the magnetic susceptibility is given by the sum of two different Curie-Weiss
relations, leading to non-Curie-Weiss behavior.
It is one of characteristic issues of the system with active
orbital degree of freedom.

We have not included the effect of oxygen in the model,
but it has been already shown that the conclusion does not change
\cite{Kubo2005c}, even if we analyze the so-called $f$-$p$ model
which include both $5f$
electrons of Np ion and $2p$ electrons of O ions.
It is also possible to perform similar analysis for another
lattice structure.
In fact, we have found a $\Gamma_{3g}$
antiferro-quadrupole transition for the simple cubic lattice
and a $\Gamma_{2u}$ antiferro-octupole transition for the bcc lattice
\cite{Kubo2005b,Kubo2005d}.

%
%
\section{Conclusions}

We have reviewed orbital ordering phenomena in $d$- and
$f$-electron compounds starting from a basic level
for the construction of the model.
Since this subject includes so many kinds of materials
such as transition metal oxides, rare-earth compounds,
and actinide materials, it is almost impossible to cover
all the results concerning the orbital-related phenomena
in these compounds.
Thus, we have picked up some typical materials and
attempted to explain how and why the orbital ordering occurs.

In order to summarize this article, we would like to emphasize
three points.
One is the understanding of the orbital ordering from
a band-insulating picture, which beautifully explains
the appearance of E- and CE-type spin structure.
The heart of the explanation is the interference effect of
electron phase originating from the anisotropic orbital.
By including further the realistic effect of Jahn-Teller
distortions and/or Coulomb interactions,
it is possible to obtain spin, charge, and orbital ordering
of transition metal oxides.

Second is the similarity between $d$- and $f$-electron orbital,
when we consider the $f$-electron model on the basis
of the $j$-$j$ coupling scheme.
In particular, we have remarked significant correspondence between
$e_{\rm g}$ orbital degenerate model for $d$ electrons
and $\Gamma_8$ model for $f$ electrons.
The large variety of manganite-like magnetic structure observed
in U-115 and Np-115 has been understood qualitatively
on the basis of the $\Gamma_8$ model.

The third point is the understanding of multipole order
in a microscopic level on the basis of the $j$-$j$ coupling model.
As a typical example, we have reviewed the octupole ordering
in NpO$_2$.
By applying the $d$-electron-like analysis on the $\Gamma_8$
model on the fcc lattice, it is possible to understand
naturally the stability of octupole ordering.
It is also remarkable that octupole is clearly understood
by the anisotropic spin-dependent charge distribution.

Of course, there still remain more kinds of
orbital ordering in $d$- and $f$-electron compounds,
which cannot be explained in this review article.
We could not cite lots of important papers of other authors
on the issue of orbital ordering phenomena.
However, it is not the main purpose of this article
to introduce orbital order in $d$- and $f$-electron
systems with complete references.
We would like to convey the unique viewpoint that orbital
ordering (including multipole ordering)
is the common phenomenon in $d$- and $f$-electron systems.
By developing further microscopic theory on orbital ordering
using the orbital degenerate model, we hope that it is
possible to understand complex magnetism among
transition metal oxides, rare-earth compounds,
and actinide materials, from a unified viewpoint.

%
%
\section*{Acknowledgments}

The author would like to thank E. Dagotto, A. Feiguin,
H. Koizumi, A. Malvezzi, M. Mayr, E. Moraghebi,
A. Moreo, H. Onishi, Y. Takada, J. C. Xavier, and S. Yunoki
for collaborations on orbital ordering
in transition metal oxides.
He is also grateful to  K. Kubo, T. Maehira, H. Onishi,
T. Takimoto, and K. Ueda
for collaborations on novel magnetism and unconventional
superconductivity of $f$-electron systems.
The author expresses sincere thanks to M. Kubota for
his kind offer of original figures used in Fig.~\ref{fig22}.
He also thanks D. Aoki, T. Fujimoto, T. Ito, Y. Haga, H. Harima,
R. H. Heffner, W. Higemoto, Y. Homma, F. Honda, S. Ikeda, S. Jonen,
S. Kambe, K. Kaneko, Y. Kuramoto, T. D. Matsuda, N. Metoki,
A. Nakamura, Y. \=Onuki, H. Sakai, Y. Shiokawa, N. Tateiwa,
Y. Tokunaga, R. E. Walstedt, E. Yamamoto, and H. Yasuoka
for useful discussions on $f$-electron materials.

The work on manganites was supported by the Ministry of Education,
Science, Sports, and Culture of Japan during my stay
in the National High Magnetic Field Laboratory, Florida State University.
The author was supported by the Grant-in-Aid for Encouragement of Young
Scientists under the contract Nos.~12740230 and 14740219
from the Ministry of Education, Culture, Sports, Science, and
Technology (MEXT) of Japan.
The work on $f$-electron systems has been also supported
by a Grany-in-Aid for Scientific Research in Priority Area
``Skutterudites'' under the contract Nos.~16037217 and 18027016
from the MEXT.
The author has been partly supported by a Grant-in-Aid for
Scientific Research (C)(2) under the contract No.~16540316
from Japan Society for the Promotion of Science.

%
%

\end{document}